\documentclass[times,twocolumn,final]{elsarticle}

\usepackage{medimamod}

\usepackage{framed,multirow}

\usepackage{amssymb}
\usepackage{latexsym}

\usepackage[hyphens]{url}
\usepackage{xcolor}

\usepackage{hyperref}

\definecolor{newcolor}{rgb}{.8,.349,.1}

\journal{Medical Image Analysis}

\usepackage{soul}

\usepackage[
    left = `,% 
    right = ',% 
    leftsub = `,% 
    rightsub = ' %
]{dirtytalk}

\newcommand{\smalltable}{\small}

\AtBeginEnvironment{table}{\smalltable}
\AtBeginEnvironment{longtable}{\smalltable}

\usepackage{calc}
\usepackage{pdflscape}
\usepackage{longtable}
\usepackage{amsmath}
\usepackage[capitalize]{cleveref}
\usepackage{subcaption}
\usepackage{booktabs}
\usepackage[textsize=tiny]{todonotes}
\crefname{subappendix}{Section}{sections}
\crefname{subsubappendix}{Subsection}{subsections}
\crefname{paragraph}{Paragraph}{paragraphs}

\renewcommand{\appendix}{\par
  \setcounter{section}{0}
  \setcounter{subsection}{0}
  \gdef\thesection{\Alph{section}}
}

\usepackage{makecell}

\usepackage{etoolbox}
\newbool{insideTabular}
\setbool{insideTabular}{false}
\BeforeBeginEnvironment{tabular}{\setbool{insideTabular}{true}}
\AfterEndEnvironment{tabular}{\setbool{insideTabular}{false}}
\BeforeBeginEnvironment{longtable}{\setbool{insideTabular}{true}}
\AfterEndEnvironment{longtable}{\setbool{insideTabular}{false}}

\usepackage[acronyms,shortcuts,acronymlists={hidden}]{glossaries}
\newglossary[algh]{hidden}{acrh}{acnh}{Hidden Acronyms}
\newacronym[type=hidden]{CXR}{CXR}{chest X-ray}
\newacronym[type=hidden]{CNN}{CNN}{convolutional neural network}
\newacronym[type=hidden]{LLM}{LLM}{large language model}
\newacronym[type=hidden]{MAE}{MAE}{mean absolute error}
\newacronym[type=hidden]{AUC}{AUROC}{area under the ROC curve}
\newacronym{MAPLEZ}{MAPLEZ}{``Medical report Annotations with Privacy-preserving Large language model using Expeditious Zero shot answers"}

\begin{document}

\verso{Ricardo Bigolin Lanfredi \textit{et~al.}}

\begin{frontmatter}

\title{Enhancing chest X-ray datasets with privacy-preserving large language models and multi-type annotations: a data-driven approach for improved classification}

\author[1]{Ricardo \snm{Bigolin Lanfredi}}
\author[1]{Pritam \snm{Mukherjee}}
\author[1]{Ronald \snm{Summers}\corref{cor1}}
% \fntext[fn1]{\printglossary[nonumberlist,type=\acronymtype]}
\fntext[fn1]{Acronyms: \textbf{MAPLEZ}: Medical report Annotations with Privacy-preserving Large language model using Expeditious Zero shot answers.}
% \ead{rsummers@cc.nih.gov}
\cortext[cor1]{Corresponding author: 
  rsummers@cc.nih.gov}

\address[1]{Imaging Biomarkers and Computer-Aided Diagnosis Laboratory,\\ Department of Radiology and Imaging Sciences,\\ National Institutes of Health Clinical Center\\
Bldg 10, Room 1C224D, 10 Center Dr, Bethesda, MD 20892-1182, USA}

% \received{1 May 2013}
% \finalform{10 May 2013}
% \accepted{13 May 2013}
% \availableonline{15 May 2013}
% \communicated{S. Sarkar}

\begin{abstract}
%%%
In chest X-ray (CXR) image analysis, rule-based systems are usually employed to extract labels from reports for dataset releases. However, there is still room for improvement in label quality. These labelers typically output only presence labels, sometimes with binary uncertainty indicators, which limits their usefulness. Supervised deep learning models have also been developed for report labeling but lack adaptability, similar to rule-based systems. In this work, we present MAPLEZ (Medical report Annotations with Privacy-preserving Large language model using Expeditious Zero shot answers), a novel approach leveraging a locally executable Large Language Model (LLM) to extract and enhance findings labels on CXR reports. MAPLEZ extracts not only binary labels indicating the presence or absence of a finding but also the location, severity, and radiologists' uncertainty about the finding. Over eight abnormalities from five test sets, we show that our method can extract these annotations with an increase of 3.6 percentage points (pp) in macro F1 score for categorical presence annotations and more than 20 pp increase in F1 score for the location annotations over competing labelers. Additionally, using the combination of improved annotations and multi-type annotations in classification supervision, we demonstrate substantial advancements in model quality, with an increase of 1.1 pp in AUROC over models trained with annotations from the best alternative approach. We share code and annotations.
%%%%
\end{abstract}


\end{frontmatter}

%\linenumbers

\section{Introduction}
\label{sec:intro}
Multi-label classification of \gls*{CXR} images has been widely explored in the computer vision literature. Publicly available large datasets, including  CheXpert~\citep{chexpert}, NIH ChestXray14~\citep{nihdataset} and MIMIC-CXR~\citep{mimiccxr}, provide \gls*{CXR} images as well as the corresponding labels for several common findings or diagnoses. Given the scale of the datasets, the labels used for training the  \gls*{CXR} classifiers are typically extracted from radiology reports using either traditional natural language processing tools, such as the CheXpert labeler~\citep{chexpert} and the Medical-Diff-VQA labeler~\citep{prevqa,vqa}, or deep learning based tools, such as CheXbert~\citep{chexbert}. As demonstrated by the results of this paper, these tools still have room for enhancement, and improving them leads to direct advancements in the models that utilize these labels during development.

Recently, general-purpose pre-trained \gls*{LLM} such as GPT4~\citep{gpt4}, Llama~\citep{llama2} or Vicuna~\citep{vicuna} have been shown to be effective at labeling radiology reports~\citep{template,vicunanih,gpt4reports}. A key advantage of these \gls*{LLM} tools is that they have good performance without additional training or finetuning. Additionally, using \glspl*{LLM} with publicly available weights, such as Llama or Vicuna, allows one to run these models locally (on-premise) without risking patients' privacy. It is also not always possible to include anonymized data provided by public datasets in prompts to cloud-based \gls*{LLM}. For example, to comply with the MIMIC-CXR data use agreement, the use of cloud \glspl*{LLM} with the reports of that dataset is limited to a particular cloud service setup that might not be available to every researcher~\citep{mimicgptlimitation}. Finally, the United States government has shown that it considers the development of privacy-preserving data analytics tools a priority~\citep{national}.

Another improvement that can be made to labelers is to extract more detailed information from the reports than just whether a given finding is present, absent, or not mentioned in a report. For instance, in a concurrent work, the rule-based Medical-Diff-VQA  labeler~\citep{prevqa,vqa} has been used to extract multi-type annotations: presence mentions, relative changes from previous reports, location, uncertainty, and severity of abnormalities from \gls*{CXR} reports. ~\citet{prevqa} and \citet{vqa} showed an advantage of using some of these types of annotations when training classifiers. 

Our hypothesis in this paper is twofold: 
\begin{itemize}
\item the extraction of several types of annotations from \gls*{CXR} reports can be improved by developing a labeler employing a locally-run \gls*{LLM};
\item these annotations can train a \gls*{CXR} image classifier that outperforms models trained using competing labelers.
\end{itemize}

\begin{figure*}[!htb]
  \centering
   \includegraphics[width=0.99\linewidth]{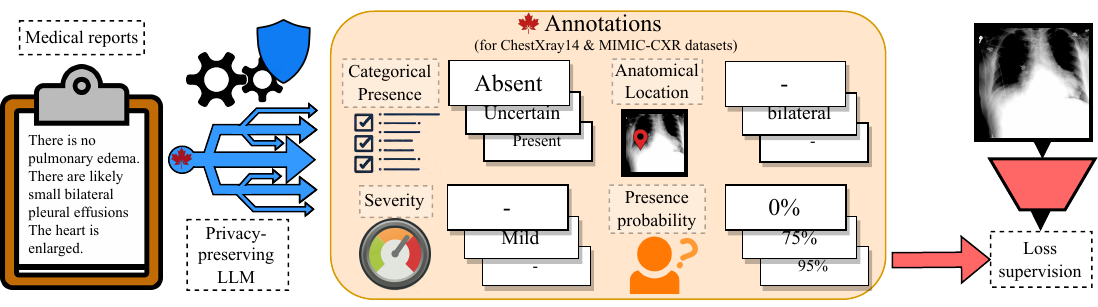}
   \caption{Representation of the \gls*{MAPLEZ} project. A knowledge-driven decision tree prompt system for a privacy-preserving \gls*{LLM} labels medical reports for several abnormalities with four types of annotations: categorical presence, presence probability, severity, and anatomical location. We used MAPLEZ to annotate two publicly available \gls*{CXR} datasets, making annotations with higher quality than competing labelers. We then showed the advantages of employing the annotations for training a \gls*{CXR} classifier.}
   \label{fig:fig1}
\end{figure*}

Therefore, we propose the \gls*{MAPLEZ} labeler based on the SOLAR-0-70b-16bit \gls*{LLM}. It uses a knowledge-driven decision tree prompt system to process medical reports and produce four types of abnormality annotations: presence, probability of presence, anatomical location, and severity. A representation of the \gls*{MAPLEZ} method is presented in \ref{fig:fig1}. We produce and share \gls*{MAPLEZ}'s annotations for the MIMIC-CXR~\citep{mimiccxr} and NIH ChestXray14~\citep{nihdataset}. We show the superiority of the \gls*{MAPLEZ} method against competing labelers in three annotation types. In addition to those datasets, we evaluate the prompt system for a limited set of reports from other medical imaging modalities. We propose using the new annotations and a multi-task loss to supervise a classification model of a \gls*{CXR}. The \gls*{MAPLEZ} annotations translated into a statistically significant improvement of the classifier (\textit{P}=.002; \textit{P}\textless.001) over the best competing labeler. The supervision by the probability of presence and anatomical location provided small but statistically significant classification improvements in weighted \gls*{AUC} (\textit{P}\textless.001, \textit{P}\textless.001). The severity annotations did not provide any classification improvements in the validation set.

\textbf{Key Contributions} 
\begin{itemize}
\item Providing a zero-shot fast prompt system for annotation extraction in medical reports in the form of prompts and open-source code, which other researchers can adapt to their research needs.
\item Providing improved and extended annotations (presence, probability, location, and severity) for two of the largest \gls*{CXR} datasets, MIMIC-CXR~\citep{mimiccxr} and NIH ChestXray14~\citep{nihdataset}. These annotations could be used, for example, in detection and visual question-answering tasks.
\item Performing extensive evaluation of the annotations to show their superior quality against other commonly employed report labelers.
\item Showing that the method can be easily adapted to reports from other medical imaging modalities: we present high F1 scores of the proposed labeler on PET, CT, and MRI reports.
\item Showing statistically significant (\textit{P}\textless.001) improvements in a downstream classification task when employing the annotations of our method.
\end{itemize}

\subsection{Related works}

\subsubsection{Using large language models to extract report labels}

Few works employ privacy-preserving \glspl*{LLM} to extract medical report labels.~\citet{cmimi} performed a small-scale experiment to show that a privacy-preserving \gls*{LLM} could provide good labeling for one specific abnormality from CT reports.~\citet{vicunanih} showed that a privacy-preserving \glspl*{LLM} performed on par with rule-based labelers for \gls*{CXR} reports. Our work develops a more complex prompt system and performs extensive experiments to show that a privacy-preserving \gls*{LLM} can actually perform better than rule-based labelers.~\citet{opentemplate} developed an \gls*{LLM} prompt system for labeling CXR reports with open-source \glspl*{LLM} parallel to our project. The authors showed that their labeler performed better than the CheXbert labeler. We failed to reproduce this result in our experiments and showed that the MAPLEZ prompt was the only one that could perform better than the CheXbert labeler~\citep{chexbert}.

~\citet{template} did a preliminary study showing that GPT-4 provided abnormality category labels on par with a state-of-the-art deep learning tool, whereas ~\citep{gpt4reports} showed a better performance by the GPT family of \glspl*{LLM}. Both works only processed hundreds of reports for their experiments without having to deal with making the prompt system tractable for several abnormalities, annotation types, and hundreds of thousands of reports.

A recent concurrent work by~\citet{chexgpt} used GPT-4 to label 50,000 reports for 13 abnormalities from the MIMIC-CXR dataset. They then trained a deep learning model on those automated ground truths for classifying the presence or absence of the findings based on the reports. We judge that the F1 scores in their results in the categorical label annotation are similar to what we present in our paper. However, their method is less quickly adaptable to new labels and modalities since their prompt is not zero-shot, and producing a new labeler requires tens of thousands of example reports to be evaluated by GPT-4 and another round of training for CheXbert. In contrast, our method requires only the replacement of abnormality names in the prompt system. 

With the exception of the works from \citet{vicunanih} and \citet{gpt4reports}, the extraction of categorical abnormality presence employing \glspl*{LLM} has been limited to binary presence or absence. In contrast, the more fuzzy approach of our method can highlight the uncertainties of the noisy labels extracted directly from reports. Furthermore, to our knowledge, we are the first to demonstrate significant (\textit{P}\textless.001;\textit{P}\textless.001) downstream classification task improvements with labels from an LLM-based labeler compared to employing annotations from previous state-of-the-art labeler tools.

\subsubsection{Other methods to extract information from large datasets of CXR reports}

Rule-based systems have been extensively used in CXR dataset releases. The CheXpert ~\citep{chexpert} and the NegBio~\citep{negbio} labelers have a similar approach. They search for keywords related to each concept defined by experts and employ a rule-based search of negative phrases to exclude such cases. Metamap~\citep{metamap} uses a sequence of traditional natural language processing tools to map text mentions to medical concept IDs. This type of tool performs worse than other methods~\citep{chexbert}, partially because they depend on rule-based parsing and might have difficulties with more complex or poorly written sentence structures. They are also very dependent on their list of keywords, which might miss several less common ways of mentioning a concept and need experts to be defined.

Supervised machine learning has been successfully employed to label medical reports. Dnorm~\citep{dnorm} employed 793 hand-annotated PubMed abstracts to train a conditional random field~\citep{crf} named entity recognizer model, named BANNER~\citep{banner}, to map keywords to medical concept IDs. CheXbert~\citep{chexbert} employed 1,687 manually labeled reports to train a BERT~\citep{bert} model to extract the presence of abnormalities. The PadChest dataset~\citep{padchest} employed 27,593 labeled reports to train an RNN on the task. These tools usually need a more extensive training set than the 100 validation reports we employed for the prompt engineering of \gls*{MAPLEZ}. They might also be less adaptable to a change of domain or type of report since they were trained for a particular task.

Other approaches employing deep learning methods do not require any annotations for reports. For example, the CheXzero paper~\citep{chexzero} used a CLIP-based model~\cite{clip}, which employs trainable deep learning text and image encoders to bring both data types into the same embedding space and employs a contrasting loss over pairs of \glspl*{CXR} and reports to learn its weights.hese models might perform worse than supervised models. In our paper, we employed CheXzero as one of the three possible choices of initializing weights of our network, as reported in \cref{sec:trainingexplained}, but models trained with those weights performed worse than the EfficientNetV2-M~\citep{efficientnet} trained on ImageNet~\citep{imagenet}. This result might be a sign of the superiority of supervised models or that the input resolution of the CheXzero model is too low.

\subsubsection{Extracting structured multi-type annotations from reports}

LesaNet~\citep{lesanet} was trained on a CT dataset that contained 171 labels, several of which characterized the location or severity of the abnormality. These labels were extracted with a rule-based labeler only from sentences that contained lesion bookmarks, probably causing several false negatives from attributes present in other sentences of the report. ~\citet{vqa} used a \gls*{CXR} report rule-based labeler to extract the same four types of annotations as we propose to extract with \gls*{MAPLEZ}: categorical presence, probability of presence, severity, and location of abnormalities. They also extracted labels characterizing the comparison of previous reports, which we did not extract. However, we show in our results that \glspl*{LLM} perform significantly better (all \textit{P}\textless.001) than their rule-based labeler, in the aggregated scores of the tasks of categorical presence, probability and location extraction, and argue that our proposed method is much more adaptable than a rule-based system, which requires a list of all the possible wording of mentions of each type of abnormality. To our knowledge, our work is the first to use \glspl*{LLM} to generate report annotations for numerical probability, severity, and location of abnormalities. 

\subsubsection{Extracting labels from PET, CT, and MRI reports}

~\citet{mriml1} (MRI), ~\citet{mriml2} (MRI), ~\citet{mriml3} (MRI), ~\citet{ctml} (CT), ~\citet{ctml2} (CT), ~\citet{ctml3} (CT), ~\citet{ctml4} (CT), ~\citet{ctml5} (CT), ~\citet{ctml6} (CT), and ~\citet{ctmriruleml} (CT, MRI) developed supervised machine learning systems for labeling binary presence of specific types of abnormalities in medical reports after manually labeling thousands of reports for supervision. ~\citet{ctrule1} (CT) extracted  ~\citet{ctrule2} (CT), ~\citet{lesanet} (CT), ~\citep{ctmriruleml} (CT, MRI), and ~\citet{petrule} (PET) developed rule-based systems for extracting presence of abnormalities from reports. The existence of so many labeling systems with their own annotated supervision data or set of rules suggests that there is no single established tool for extracting abnormality labels from PET, CT, or MRI reports. These reports contain a more diverse set of abnormalities reported than \gls*{CXR} reports, so the lack of easy adaptability of rule-based and supervised machine learning systems probably impacts the use of these developed tools in subsequent research. 

~\citet{ctmripetgeneric} provided a more generic approach to labeling CT, MRI, and PET reports: they extracted sentences containing reference images (e.g., \say{(Series 1001, image 32)}) from hundreds of thousands of reports and processed those sentences with an unsupervised natural language processing clustering method to create 80 classes of abnormality presents in the reports. In opposition to this approach, \gls*{MAPLEZ} does not require a vast corpus of reports to be developed and does not have the restriction of only working for sentences with reference images. ~\citet{cmimi} presented a small-scale experiment to show that \glspl*{LLM} could be employed to extract the presence of one specific abnormality from CT reports. Unlike their method, our method provides fuzzy and multi-type annotations, and our paper performs a more extensive analysis of the adaptability and the quality of the labels generated by \glspl*{LLM}. 

\section{Methods}

\subsection{A Prompt system for automatic annotation of CXR reports}
\label{sec:methods}
To enhance the quality of annotations derived from \gls*{CXR} reports, we use the SOLAR-0-70b-16bit \gls*{LLM}~\citep{solar,huggingface}, which is accessible to the public under the CC BY-NC-4.0 license. This model adapts the Llama 2 model~\citep{llama2}, further finetuned on two unspecified instruction datasets similar to broadly-employed datasets~\citep{openorca,orca,flan,alpaca}. We chose this model after, at the start of the project (from April 2023 to August 2023), we tested several openly available \glspl*{LLM}: BioGPT~\citep{biogpt}, Koala~\citep{koala}, Vicuna~\citep{vicuna}, Alpaca~\citep{alpaca}, Bloom~\citep{bloom}, Stable Vicuna~\citep{stablevicuna}, Galpaca~\citep{galpaca}, GPT4All~\citep{gpt4all}, Guanaco~\citep{guanaco}, Llama 2~\citep{llama2}. When we found one model that, according to our preliminary tests, could improve the report labeling task (SOLAR-0-70b-16bit), we decided to freeze the choice of \gls*{LLM} and focus on improving the prompt system for that specific \gls*{LLM}. 

We did not modify or finetune the \gls*{LLM}, employing it in a zero-shot manner. Our custom-designed prompt system takes a radiologist's report and generates annotations for the 13 abnormalities listed in \cref{sec:labelprompts}. These 13 abnormality labels were selected from CheXpert~\citep{chexpert}, the most known baseline. A new set of labels would complicate comparison experiments, leading to additional label translations. However, as shown in this paper, our method is easily adaptable to other abnormality labels.

We focused on extracting four specific annotation categories for each abnormality: categorical presence, probability of presence, severity, and anatomical location. To improve our prompts, we experimented on a validation set with 100 manually annotated reports from the NIH ChestXray14 dataset~\citep{nihdataset}. Initial testing with tailored prompts revealed that querying the \gls*{LLM} about specific abnormalities yielded more precise responses than multiple abnormalities simultaneously. Furthermore, we observed that chain-of-thought prompts~\citep{cot}  were impractical for processing 227,827 \glspl*{CXR} reports from the MIMIC-CXR dataset~\citep{mimiccxr} for 13 types of abnormalities and four annotation categories due to computation time. To enhance efficiency, we used prompts that demanded brief responses, typically up to four tokens in length, reducing computational demands.

\begin{figure*}[!htb]
  \centering
  \begin{subfigure}{0.35\linewidth}
       \includegraphics[width=0.99\linewidth]{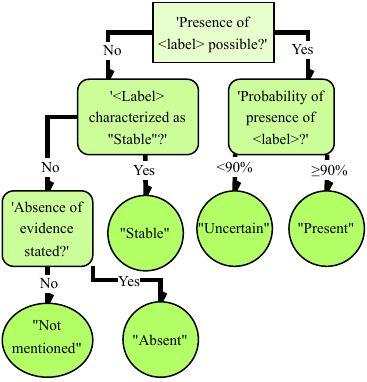}
    \caption{Categorical presence decision tree}
    \label{fig:labels}
  \end{subfigure}
  \hfill
  \begin{subfigure}{0.19\linewidth}
    \includegraphics[width=0.99\linewidth]{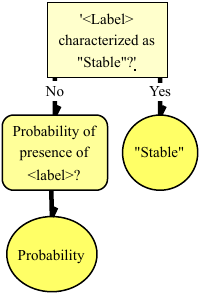}
    \caption{Probability decision tree}
    \label{fig:probability}
  \end{subfigure}
    \hfill
  \begin{subfigure}{0.25\linewidth}
    \includegraphics[width=0.99\linewidth]{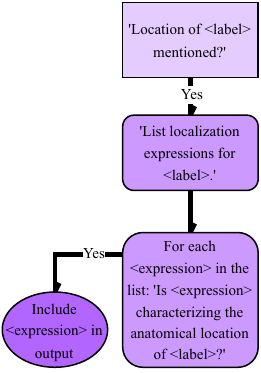}
    \caption{Location decision tree}
    \label{fig:severity}
  \end{subfigure}
    \hfill
  \begin{subfigure}{0.132\linewidth}
    \includegraphics[width=0.99\linewidth]{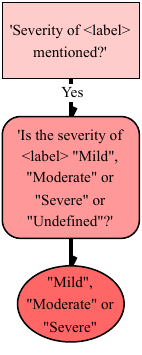}
    \caption{Severity decision tree}
    \label{fig:location}
  \end{subfigure}
  \caption{Simplified representation of the knowledge-driven decision tree prompt system and its possible outputs. The location and severity knowledge-driven decision trees are only run when the abnormality is labeled as ``Uncertain" or ``Present". }
  \label{fig:flowchart}
\end{figure*}

We implemented a knowledge-based decision tree prompt system to guide the annotation process. Simplified visual representations of this system can be found in \cref{fig:flowchart}, with the full prompts detailed in Section \cref{sec:prompts}. The five possible outputs for the categorical presence of an abnormality were ``Present", ``Absent", ``Not mentioned", ``Uncertain", i.e.,  the radiologist expresses uncertainty about the abnormality being present (radiologist uncertainty), or ``Stable", i.e., the radiologist compares the state of an abnormality to the state from a previous report without indicating its presence or not (text uncertainty). We diverged from the CheXpert labeler's practice of assigning the same category to ``Uncertain" and ``Stable" cases, opting for distinct categories to allow for their separate handling. Upon establishing the possible presence of an abnormality, the \gls*{LLM} checks if the report includes details on location or severity. For severity annotations, the \gls*{LLM} outputs one of three categories: ``Mild", ``Moderate", or ``Severe". For location annotations, the \gls*{LLM} is instructed to enumerate the characterizing locations, separating them with semi-colons. Listing the locations is the only prompt requiring a lengthy response in our labeler. Subsequently, the \gls*{LLM} verifies each listed location to ensure they describe anatomical locations. For the probability annotations, the model either categorizes the abnormality as ``Stable" or outputs a probability between 0\% and 100\%.

We observed an improvement in performance on our prompt experimentation set when we expanded abnormality denominations -- the way different abnormalities are mentioned in our prompts -- to include synonyms and subtypes, which were based on the rules of the CheXpert labeler~\citep{chexpertdefinitions}. The text we used as abnormality denominations in the prompts can be found in \cref{sec:labelprompts}.
\subsection{Merging abnormality subtypes}
\label{sec:merging}
Some abnormalities are a merge of subtypes of that abnormality. Specifically, we categorize ``Consolidation" as a blend of ``Consolidation" and ``Pneumonia", while ``Lung opacity" merges several conditions: ``Atelectasis", ``Consolidation", ``Pneumonia", ``Edema", ``Lung lesion", and ``Lung opacity" itself. When combining these conditions, we define ``Consolidation" and ``Lung opacity" as primary labels. Our methodology for integrating the diagnostic labels varies by the type of label:
\begin{itemize}
\item for presence labels, we prioritize them as follows: ``Present" is prioritized over ``Uncertain," which is prioritized over ``Stable," which in turn is prioritized over ``Absent" (if it is a primary label), followed by ``Not mentioned," and lastly ``Absent." (if it is not a primary label);
\item for probability labels, we combine abnormalities by using the highest probability, with ``Stable" prioritized over probabilities less than or equal to 50\%;
\item for severity labels, we apply a similar highest-value approach, treating the absence of abnormality or severity as the lowest priority severity level;
\item for location labels, we concatenate the different locations into one list, ensuring there are no duplicates.
\end{itemize}

\subsubsection{Adapting the system to other medical imaging modalities}
\label{sec:adapt}
To test its adaptability, we made minor modifications to run MAPLEZ with other types of medical text: CT, MRI, and PET reports. We changed the selection of abnormalities to contain the ones usually mentioned in those reports, and we did not validate the abnormality denominations. The abnormality labels and denominations are presented in \cref{sec:labelprompts}. The rest of the prompt system was the same, except for a mention of the modality of the report, replacing \say{Given the (complete/full) report} excerpts with \say{Given the (complete/full) \textless modality\textgreater~ report}, where \textless modality\textgreater~ was \say{ct}, \say{mri}, or \say{pet}.

\subsection{Employing the new annotations}
\label{sec:methods2}

We trained a \gls*{CNN} \gls*{CXR} classifier using the annotations obtained with the \gls*{LLM}. Out of the 13 available abnormalities, we focused on eight to standardize the outputs of our approach and baseline methods. The classifier was trained to detect abnormalities with supervision from categorical or probabilistic annotations. The binary cross-entropy loss was employed in both cases, and the probabilities were used as soft labels. Additionally, we explored leveraging severity and anatomical location annotations as additional supervision in a multi-task loss.

We selected anatomically meaningful keywords with non-overlapping meanings to allow the classifier to learn from the location annotations. Our selection involved analyzing the most frequent n-grams in the annotations across different abnormality groups. The chosen keywords, such as \say{right}, \say{left}, and \say{lower}, are listed in \cref{sec:classifierwords}. We also identified terms in the annotations that were synonymous with some keywords and created a replacement-word list; for instance, the presence of \say{bilateral} in an annotation would indicate both \say{left} and \say{right} keyword labels were positive. The complete list of these replacement words is in \cref{sec:classifierwords}.

Determining negative keyword labels for each case posed a challenge. We decided on a rule: a label is negative only if mutually exclusive with a positive label. For example, if \say{right} is positive, \say{left} becomes negative, but \say{lower} remains unaffected since an abnormality in the left lung might be in the lower left lung. Additionally, anatomically adjacent terms prevented each other from being labeled as negative. For example, if \say{lower} and \say{upper} are positive, \say{base} will not be negative, even though it is mutually exclusive with \say{upper}: \say{lower} being anatomically close to \say{base} will prevent that. Similarly,  if there is abnormality on both sides, for example, \say{pneumonia on the right and left lungs}, both \say{left} and \say{right} are positive since a positive mention has priority over the rule for negative labels. Full details of these relationships are shown in \cref{sec:classifierwords}. Any keywords not categorized as positive or negative for an abnormality are treated as unlabeled and ignored in the location loss calculation.

Our model's architecture included distinct logit outputs for each selected location keyword of each abnormality. We calculated the location loss using binary cross-entropy for each logit, integrating it into the overall loss function by multiplying it with a weighting factor, $\lambda_{loc}$, and adding it to the presence classification loss.

Lastly, we experimented with the severity labels from Medical-Diff-VQA and MAPLEZ labelers. We only applied a scaled multi-class cross-entropy loss when a severity annotation was available. However, this modification did not yield improvements in the \gls*{AUC} on our validation set, so we do not detail this aspect further.

\section{Results}
\subsection{Labeler evaluation}
\label{sec:labelereval}

For evaluating the \gls*{LLM} annotations in \gls*{CXR} reports, we hand-labeled categorical presence, severity and location of abnormalities for 350 reports from the MIMIC-CXR dataset~\citep{physionet,mimiccxr,mimiccxrphysionet} and 200 reports from the NIH ChestXray14 dataset~\citep{nihdataset}. Details about these hand annotations are given in \cref{sec:handannotations}. We also used public datasets that were labeled by radiologists directly from the \gls*{CXR} images: 
\begin{itemize}
\item \textit{REFLACX}~\citep{reflacx,reflacxphysionet,physionet}:  phase 3 of the \textit{REFLACX} dataset has 2,507 frontal \glspl*{CXR} from the MIMIC-CXR dataset~\citep{mimiccxr} labeled for several abnormalities by a single radiologist among five radiologists. Phases 1 and 2 of the same dataset have 109 frontal \glspl*{CXR}, each labeled by five radiologists. We used these two phases for the inter-observer scores for \cref{tab:humans}. The dataset also includes probabilities assigned by radiologists to each abnormality. The probability labels are annotated using five categories, and we convert them to probability intervals: ``Absent": $[0\%,5\%]$; ``Unlikely (\textless 10\%)": $[0\%,15\%]$; ``Less Likely (25\%)": $[10\%,40\%]$; ``Possibly (50\%)": $[35\%,65\%]$; ``Suspicious for/Probably (75\%)": $[60\%,90\%]$; ``Consistent with (\textgreater 90\%)": $[85\%,100\%]$. We selected five abnormalities from this dataset that were equivalent to the ones from the CheXpert labeler~\citep{chexpert}. One of the five abnormalities was a merge from several labels, as shown in \cref{sec:reflacxchanges}, following the merging rules from \cref{sec:merging}.
\item \textit{Pneumonia}: the RSNA Pneumonia Detection Challenge (2018) dataset~\citep{pneumonia} contains 26,684 \gls*{CXR} studies from the NIH ChestXray14 dataset~\citep{nihdataset}, each labeled for presence of pneumonia/consolidation by one to three radiologists from a set of 18 radiologists.
\item \textit{Pneumothorax}: the dataset from the National University Hospital from Singapore~\citep{pneumothorax} contains 24,709 studies from the NIH ChestXray14 dataset~\citep{nihdataset} labeled for the presence of pneumothorax by one of four radiologists.
\end{itemize}

\begin{table*}[!htb]
  \caption{F1 scores ($\uparrow$) for seven labelers for the categorical presence annotation. MAPLEZ is our proposed method, and CheXpert~\citep{chexpert}, Vicuna~\citep{vicunanih}, Medical-Diff-VQA (VQA)~\citep{vqa}, CheXbert~\citep{chexbert} and Template~\citep{template} are the competing methods. These symbols represent the p-values (\textit{P}) from the two-sided permutation hypothesis test for the difference between the score of MAPLEZ and the other scores in that specific table row: \(^{\scriptscriptstyle{*}\scriptscriptstyle{*}\scriptscriptstyle{*}}\) for \textit{P}$\textless$.001,  \(^{\scriptscriptstyle{*}\scriptscriptstyle{*}}\) for \textit{P}$\textless$.01, \(^{\scriptscriptstyle{*}}\) for \textit{P}$\textless$.05, and \(^{ns}\) for not significant. Precision, recall, confidence intervals, and aggregations by abnormality are presented in \cref{tab:labelsfull}. NIH=NIH ChestXray14 dataset~\citep{nihdataset}; MIMIC=MIMIC-CXR dataset~\citep{mimiccxr}; \textit{RFL-3}= phase 3 of the \textit{REFLACX} dataset~\citep{reflacx}; \textit{PNA}=\textit{Pneumonia} dataset~\citep{pneumonia}; \textit{PTX}=\textit{Pneumothorax} dataset~\citep{pneumothorax}; Abn.=Abnormality; Atel.=``Atelectasis"; Card.=``Cardiomegaly"; Cons.=``Consolidation"; Fract.=``Fracture"; Opac.=``Lung opacity"; Effus.=``Pleural Effusion"; PTX=``Pneumothorax"; N=total labeled test cases; N$^+$=number of positive cases for that abnormality; W=weight used in the aggregation of dataset scores (All-W), related to the variance of the score; MAPLEZ-G=MAPLEZ-Generic is the model that uses simpler abnormality denominations in its prompts; \textit{Human}=aggregation for all datasets labeled by radiologists straight from the \glspl*{CXR}, which are marked with italic; All-M: macro scores, calculated using a simple average over abnormality scores. }
  \label{tab:labels}
  \centering
  \begin{tabular}{@{}llrrcccccccccc@{}}
\toprule
Data & Abn. & $N$ & $N^+$ & W & \footnotesize{CheXpert} & \footnotesize{Vicuna} & \footnotesize{VQA} & \footnotesize{CheXbert} & \footnotesize{Template} & \makecell{\footnotesize{MAPLEZ}\\\footnotesize{Generic}} & \makecell{\footnotesize{MAPLEZ}\\\footnotesize{(Ours)}}\\
\midrule
NIH & Atel. & 200 & 27 & 0.009 & \makebox[\widthof{\(^{\scriptscriptstyle{*}\scriptscriptstyle{*}\scriptscriptstyle{*}}\)}][c]{}0.906\makebox[\widthof{\(^{\scriptscriptstyle{*}\scriptscriptstyle{*}\scriptscriptstyle{*}}\)}][l]{\(^{ns}\)} & \makebox[\widthof{\(^{\scriptscriptstyle{*}\scriptscriptstyle{*}\scriptscriptstyle{*}}\)}][c]{}0.881\makebox[\widthof{\(^{\scriptscriptstyle{*}\scriptscriptstyle{*}\scriptscriptstyle{*}}\)}][l]{\(^{ns}\)} & - & \makebox[\widthof{\(^{\scriptscriptstyle{*}\scriptscriptstyle{*}\scriptscriptstyle{*}}\)}][c]{}\textbf{0.964}\makebox[\widthof{\(^{\scriptscriptstyle{*}\scriptscriptstyle{*}\scriptscriptstyle{*}}\)}][l]{\(^{\scriptscriptstyle{*}\scriptscriptstyle{*}}\)} & \makebox[\widthof{\(^{\scriptscriptstyle{*}\scriptscriptstyle{*}\scriptscriptstyle{*}}\)}][c]{}0.857\makebox[\widthof{\(^{\scriptscriptstyle{*}\scriptscriptstyle{*}\scriptscriptstyle{*}}\)}][l]{\(^{ns}\)} & \makebox[\widthof{\(^{\scriptscriptstyle{*}\scriptscriptstyle{*}\scriptscriptstyle{*}}\)}][c]{}0.871\makebox[\widthof{\(^{\scriptscriptstyle{*}\scriptscriptstyle{*}\scriptscriptstyle{*}}\)}][l]{\(^{ns}\)} & 0.818 \\
MIMIC & Atel. & 350 & 104 & 0.065 & \makebox[\widthof{\(^{\scriptscriptstyle{*}\scriptscriptstyle{*}\scriptscriptstyle{*}}\)}][c]{}0.832\(^{\scriptscriptstyle{*}\scriptscriptstyle{*}\scriptscriptstyle{*}}\) & \makebox[\widthof{\(^{\scriptscriptstyle{*}\scriptscriptstyle{*}\scriptscriptstyle{*}}\)}][c]{}0.959\makebox[\widthof{\(^{\scriptscriptstyle{*}\scriptscriptstyle{*}\scriptscriptstyle{*}}\)}][l]{\(^{ns}\)} & \makebox[\widthof{\(^{\scriptscriptstyle{*}\scriptscriptstyle{*}\scriptscriptstyle{*}}\)}][c]{}0.949\makebox[\widthof{\(^{\scriptscriptstyle{*}\scriptscriptstyle{*}\scriptscriptstyle{*}}\)}][l]{\(^{ns}\)} & \makebox[\widthof{\(^{\scriptscriptstyle{*}\scriptscriptstyle{*}\scriptscriptstyle{*}}\)}][c]{}\textbf{0.963}\makebox[\widthof{\(^{\scriptscriptstyle{*}\scriptscriptstyle{*}\scriptscriptstyle{*}}\)}][l]{\(^{ns}\)} & \makebox[\widthof{\(^{\scriptscriptstyle{*}\scriptscriptstyle{*}\scriptscriptstyle{*}}\)}][c]{}0.927\makebox[\widthof{\(^{\scriptscriptstyle{*}\scriptscriptstyle{*}\scriptscriptstyle{*}}\)}][l]{\(^{ns}\)} & \makebox[\widthof{\(^{\scriptscriptstyle{*}\scriptscriptstyle{*}\scriptscriptstyle{*}}\)}][c]{}0.941\makebox[\widthof{\(^{\scriptscriptstyle{*}\scriptscriptstyle{*}\scriptscriptstyle{*}}\)}][l]{\(^{ns}\)} & 0.950 \\
NIH & Card. & 200 & 21 & 0.003 & \makebox[\widthof{\(^{\scriptscriptstyle{*}\scriptscriptstyle{*}\scriptscriptstyle{*}}\)}][c]{}0.649\(^{\scriptscriptstyle{*}\scriptscriptstyle{*}\scriptscriptstyle{*}}\) & \makebox[\widthof{\(^{\scriptscriptstyle{*}\scriptscriptstyle{*}\scriptscriptstyle{*}}\)}][c]{}0.833\makebox[\widthof{\(^{\scriptscriptstyle{*}\scriptscriptstyle{*}\scriptscriptstyle{*}}\)}][l]{\(^{ns}\)} & - & \makebox[\widthof{\(^{\scriptscriptstyle{*}\scriptscriptstyle{*}\scriptscriptstyle{*}}\)}][c]{}0.690\(^{\scriptscriptstyle{*}\scriptscriptstyle{*}\scriptscriptstyle{*}}\) & \makebox[\widthof{\(^{\scriptscriptstyle{*}\scriptscriptstyle{*}\scriptscriptstyle{*}}\)}][c]{}0.878\makebox[\widthof{\(^{\scriptscriptstyle{*}\scriptscriptstyle{*}\scriptscriptstyle{*}}\)}][l]{\(^{ns}\)} & \makebox[\widthof{\(^{\scriptscriptstyle{*}\scriptscriptstyle{*}\scriptscriptstyle{*}}\)}][c]{}0.833\makebox[\widthof{\(^{\scriptscriptstyle{*}\scriptscriptstyle{*}\scriptscriptstyle{*}}\)}][l]{\(^{ns}\)} & \textbf{0.950} \\
MIMIC & Card. & 350 & 132 & 0.032 & \makebox[\widthof{\(^{\scriptscriptstyle{*}\scriptscriptstyle{*}\scriptscriptstyle{*}}\)}][c]{}0.737\(^{\scriptscriptstyle{*}\scriptscriptstyle{*}\scriptscriptstyle{*}}\) & \makebox[\widthof{\(^{\scriptscriptstyle{*}\scriptscriptstyle{*}\scriptscriptstyle{*}}\)}][c]{}0.913\makebox[\widthof{\(^{\scriptscriptstyle{*}\scriptscriptstyle{*}\scriptscriptstyle{*}}\)}][l]{\(^{ns}\)} & \makebox[\widthof{\(^{\scriptscriptstyle{*}\scriptscriptstyle{*}\scriptscriptstyle{*}}\)}][c]{}0.757\(^{\scriptscriptstyle{*}\scriptscriptstyle{*}\scriptscriptstyle{*}}\) & \makebox[\widthof{\(^{\scriptscriptstyle{*}\scriptscriptstyle{*}\scriptscriptstyle{*}}\)}][c]{}0.912\makebox[\widthof{\(^{\scriptscriptstyle{*}\scriptscriptstyle{*}\scriptscriptstyle{*}}\)}][l]{\(^{ns}\)} & \makebox[\widthof{\(^{\scriptscriptstyle{*}\scriptscriptstyle{*}\scriptscriptstyle{*}}\)}][c]{}0.923\makebox[\widthof{\(^{\scriptscriptstyle{*}\scriptscriptstyle{*}\scriptscriptstyle{*}}\)}][l]{\(^{ns}\)} & \makebox[\widthof{\(^{\scriptscriptstyle{*}\scriptscriptstyle{*}\scriptscriptstyle{*}}\)}][c]{}\textbf{0.938}\makebox[\widthof{\(^{\scriptscriptstyle{*}\scriptscriptstyle{*}\scriptscriptstyle{*}}\)}][l]{\(^{ns}\)} & 0.931 \\
\textit{RFL-3} & Card. & 506 & 171 & 0.007 & \makebox[\widthof{\(^{\scriptscriptstyle{*}\scriptscriptstyle{*}\scriptscriptstyle{*}}\)}][c]{}0.448\(^{\scriptscriptstyle{*}\scriptscriptstyle{*}\scriptscriptstyle{*}}\) & \makebox[\widthof{\(^{\scriptscriptstyle{*}\scriptscriptstyle{*}\scriptscriptstyle{*}}\)}][c]{}0.598\makebox[\widthof{\(^{\scriptscriptstyle{*}\scriptscriptstyle{*}\scriptscriptstyle{*}}\)}][l]{\(^{ns}\)} & \makebox[\widthof{\(^{\scriptscriptstyle{*}\scriptscriptstyle{*}\scriptscriptstyle{*}}\)}][c]{}0.464\(^{\scriptscriptstyle{*}\scriptscriptstyle{*}\scriptscriptstyle{*}}\) & \makebox[\widthof{\(^{\scriptscriptstyle{*}\scriptscriptstyle{*}\scriptscriptstyle{*}}\)}][c]{}0.613\makebox[\widthof{\(^{\scriptscriptstyle{*}\scriptscriptstyle{*}\scriptscriptstyle{*}}\)}][l]{\(^{ns}\)} & \makebox[\widthof{\(^{\scriptscriptstyle{*}\scriptscriptstyle{*}\scriptscriptstyle{*}}\)}][c]{}0.610\makebox[\widthof{\(^{\scriptscriptstyle{*}\scriptscriptstyle{*}\scriptscriptstyle{*}}\)}][l]{\(^{ns}\)} & \makebox[\widthof{\(^{\scriptscriptstyle{*}\scriptscriptstyle{*}\scriptscriptstyle{*}}\)}][c]{}0.616\makebox[\widthof{\(^{\scriptscriptstyle{*}\scriptscriptstyle{*}\scriptscriptstyle{*}}\)}][l]{\(^{ns}\)} & \textbf{0.618} \\
NIH & Cons. & 200 & 70 & 0.003 & \makebox[\widthof{\(^{\scriptscriptstyle{*}\scriptscriptstyle{*}\scriptscriptstyle{*}}\)}][c]{}0.531\(^{\scriptscriptstyle{*}\scriptscriptstyle{*}\scriptscriptstyle{*}}\) & \makebox[\widthof{\(^{\scriptscriptstyle{*}\scriptscriptstyle{*}\scriptscriptstyle{*}}\)}][c]{}0.479\(^{\scriptscriptstyle{*}\scriptscriptstyle{*}\scriptscriptstyle{*}}\) & - & \makebox[\widthof{\(^{\scriptscriptstyle{*}\scriptscriptstyle{*}\scriptscriptstyle{*}}\)}][c]{}0.551\(^{\scriptscriptstyle{*}\scriptscriptstyle{*}\scriptscriptstyle{*}}\) & \makebox[\widthof{\(^{\scriptscriptstyle{*}\scriptscriptstyle{*}\scriptscriptstyle{*}}\)}][c]{}0.611\(^{\scriptscriptstyle{*}\scriptscriptstyle{*}\scriptscriptstyle{*}}\) & \makebox[\widthof{\(^{\scriptscriptstyle{*}\scriptscriptstyle{*}\scriptscriptstyle{*}}\)}][c]{}0.604\(^{\scriptscriptstyle{*}\scriptscriptstyle{*}\scriptscriptstyle{*}}\) & \textbf{0.951} \\
MIMIC & Cons. & 350 & 90 & 0.017 & \makebox[\widthof{\(^{\scriptscriptstyle{*}\scriptscriptstyle{*}\scriptscriptstyle{*}}\)}][c]{}0.838\makebox[\widthof{\(^{\scriptscriptstyle{*}\scriptscriptstyle{*}\scriptscriptstyle{*}}\)}][l]{\(^{ns}\)} & \makebox[\widthof{\(^{\scriptscriptstyle{*}\scriptscriptstyle{*}\scriptscriptstyle{*}}\)}][c]{}0.738\makebox[\widthof{\(^{\scriptscriptstyle{*}\scriptscriptstyle{*}\scriptscriptstyle{*}}\)}][l]{\(^{\scriptscriptstyle{*}\scriptscriptstyle{*}}\)} & \makebox[\widthof{\(^{\scriptscriptstyle{*}\scriptscriptstyle{*}\scriptscriptstyle{*}}\)}][c]{}0.845\makebox[\widthof{\(^{\scriptscriptstyle{*}\scriptscriptstyle{*}\scriptscriptstyle{*}}\)}][l]{\(^{ns}\)} & \makebox[\widthof{\(^{\scriptscriptstyle{*}\scriptscriptstyle{*}\scriptscriptstyle{*}}\)}][c]{}\textbf{0.892}\makebox[\widthof{\(^{\scriptscriptstyle{*}\scriptscriptstyle{*}\scriptscriptstyle{*}}\)}][l]{\(^{ns}\)} & \makebox[\widthof{\(^{\scriptscriptstyle{*}\scriptscriptstyle{*}\scriptscriptstyle{*}}\)}][c]{}0.775\makebox[\widthof{\(^{\scriptscriptstyle{*}\scriptscriptstyle{*}\scriptscriptstyle{*}}\)}][l]{\(^{\scriptscriptstyle{*}\scriptscriptstyle{*}}\)} & \makebox[\widthof{\(^{\scriptscriptstyle{*}\scriptscriptstyle{*}\scriptscriptstyle{*}}\)}][c]{}0.843\makebox[\widthof{\(^{\scriptscriptstyle{*}\scriptscriptstyle{*}\scriptscriptstyle{*}}\)}][l]{\(^{ns}\)} & 0.881 \\
\textit{RFL-3} & Cons. & 506 & 154 & 0.004 & \makebox[\widthof{\(^{\scriptscriptstyle{*}\scriptscriptstyle{*}\scriptscriptstyle{*}}\)}][c]{}0.454\makebox[\widthof{\(^{\scriptscriptstyle{*}\scriptscriptstyle{*}\scriptscriptstyle{*}}\)}][l]{\(^{\scriptscriptstyle{*}}\)} & \makebox[\widthof{\(^{\scriptscriptstyle{*}\scriptscriptstyle{*}\scriptscriptstyle{*}}\)}][c]{}0.406\(^{\scriptscriptstyle{*}\scriptscriptstyle{*}\scriptscriptstyle{*}}\) & \makebox[\widthof{\(^{\scriptscriptstyle{*}\scriptscriptstyle{*}\scriptscriptstyle{*}}\)}][c]{}0.441\makebox[\widthof{\(^{\scriptscriptstyle{*}\scriptscriptstyle{*}\scriptscriptstyle{*}}\)}][l]{\(^{\scriptscriptstyle{*}}\)} & \makebox[\widthof{\(^{\scriptscriptstyle{*}\scriptscriptstyle{*}\scriptscriptstyle{*}}\)}][c]{}0.474\makebox[\widthof{\(^{\scriptscriptstyle{*}\scriptscriptstyle{*}\scriptscriptstyle{*}}\)}][l]{\(^{ns}\)} & \makebox[\widthof{\(^{\scriptscriptstyle{*}\scriptscriptstyle{*}\scriptscriptstyle{*}}\)}][c]{}0.502\makebox[\widthof{\(^{\scriptscriptstyle{*}\scriptscriptstyle{*}\scriptscriptstyle{*}}\)}][l]{\(^{ns}\)} & \makebox[\widthof{\(^{\scriptscriptstyle{*}\scriptscriptstyle{*}\scriptscriptstyle{*}}\)}][c]{}0.466\makebox[\widthof{\(^{\scriptscriptstyle{*}\scriptscriptstyle{*}\scriptscriptstyle{*}}\)}][l]{\(^{\scriptscriptstyle{*}}\)} & \textbf{0.508} \\
\textit{PNA} & Cons. & 7186 & 2589 & 0.062 & \makebox[\widthof{\(^{\scriptscriptstyle{*}\scriptscriptstyle{*}\scriptscriptstyle{*}}\)}][c]{}0.381\(^{\scriptscriptstyle{*}\scriptscriptstyle{*}\scriptscriptstyle{*}}\) & \makebox[\widthof{\(^{\scriptscriptstyle{*}\scriptscriptstyle{*}\scriptscriptstyle{*}}\)}][c]{}0.429\(^{\scriptscriptstyle{*}\scriptscriptstyle{*}\scriptscriptstyle{*}}\) & - & \makebox[\widthof{\(^{\scriptscriptstyle{*}\scriptscriptstyle{*}\scriptscriptstyle{*}}\)}][c]{}0.384\(^{\scriptscriptstyle{*}\scriptscriptstyle{*}\scriptscriptstyle{*}}\) & \makebox[\widthof{\(^{\scriptscriptstyle{*}\scriptscriptstyle{*}\scriptscriptstyle{*}}\)}][c]{}0.505\(^{\scriptscriptstyle{*}\scriptscriptstyle{*}\scriptscriptstyle{*}}\) & \makebox[\widthof{\(^{\scriptscriptstyle{*}\scriptscriptstyle{*}\scriptscriptstyle{*}}\)}][c]{}0.516\(^{\scriptscriptstyle{*}\scriptscriptstyle{*}\scriptscriptstyle{*}}\) & \textbf{0.633} \\
NIH & Edema & 200 & 15 & 0.001 & \makebox[\widthof{\(^{\scriptscriptstyle{*}\scriptscriptstyle{*}\scriptscriptstyle{*}}\)}][c]{}\textbf{0.786}\makebox[\widthof{\(^{\scriptscriptstyle{*}\scriptscriptstyle{*}\scriptscriptstyle{*}}\)}][l]{\(^{ns}\)} & \makebox[\widthof{\(^{\scriptscriptstyle{*}\scriptscriptstyle{*}\scriptscriptstyle{*}}\)}][c]{}0.692\makebox[\widthof{\(^{\scriptscriptstyle{*}\scriptscriptstyle{*}\scriptscriptstyle{*}}\)}][l]{\(^{ns}\)} & - & \makebox[\widthof{\(^{\scriptscriptstyle{*}\scriptscriptstyle{*}\scriptscriptstyle{*}}\)}][c]{}0.774\makebox[\widthof{\(^{\scriptscriptstyle{*}\scriptscriptstyle{*}\scriptscriptstyle{*}}\)}][l]{\(^{ns}\)} & \makebox[\widthof{\(^{\scriptscriptstyle{*}\scriptscriptstyle{*}\scriptscriptstyle{*}}\)}][c]{}0.759\makebox[\widthof{\(^{\scriptscriptstyle{*}\scriptscriptstyle{*}\scriptscriptstyle{*}}\)}][l]{\(^{ns}\)} & \makebox[\widthof{\(^{\scriptscriptstyle{*}\scriptscriptstyle{*}\scriptscriptstyle{*}}\)}][c]{}0.524\makebox[\widthof{\(^{\scriptscriptstyle{*}\scriptscriptstyle{*}\scriptscriptstyle{*}}\)}][l]{\(^{\scriptscriptstyle{*}\scriptscriptstyle{*}}\)} & 0.688 \\
MIMIC & Edema & 350 & 111 & 0.021 & \makebox[\widthof{\(^{\scriptscriptstyle{*}\scriptscriptstyle{*}\scriptscriptstyle{*}}\)}][c]{}0.849\makebox[\widthof{\(^{\scriptscriptstyle{*}\scriptscriptstyle{*}\scriptscriptstyle{*}}\)}][l]{\(^{ns}\)} & \makebox[\widthof{\(^{\scriptscriptstyle{*}\scriptscriptstyle{*}\scriptscriptstyle{*}}\)}][c]{}0.769\makebox[\widthof{\(^{\scriptscriptstyle{*}\scriptscriptstyle{*}\scriptscriptstyle{*}}\)}][l]{\(^{\scriptscriptstyle{*}\scriptscriptstyle{*}}\)} & \makebox[\widthof{\(^{\scriptscriptstyle{*}\scriptscriptstyle{*}\scriptscriptstyle{*}}\)}][c]{}0.844\makebox[\widthof{\(^{\scriptscriptstyle{*}\scriptscriptstyle{*}\scriptscriptstyle{*}}\)}][l]{\(^{ns}\)} & \makebox[\widthof{\(^{\scriptscriptstyle{*}\scriptscriptstyle{*}\scriptscriptstyle{*}}\)}][c]{}\textbf{0.882}\makebox[\widthof{\(^{\scriptscriptstyle{*}\scriptscriptstyle{*}\scriptscriptstyle{*}}\)}][l]{\(^{ns}\)} & \makebox[\widthof{\(^{\scriptscriptstyle{*}\scriptscriptstyle{*}\scriptscriptstyle{*}}\)}][c]{}0.776\makebox[\widthof{\(^{\scriptscriptstyle{*}\scriptscriptstyle{*}\scriptscriptstyle{*}}\)}][l]{\(^{\scriptscriptstyle{*}}\)} & \makebox[\widthof{\(^{\scriptscriptstyle{*}\scriptscriptstyle{*}\scriptscriptstyle{*}}\)}][c]{}0.815\makebox[\widthof{\(^{\scriptscriptstyle{*}\scriptscriptstyle{*}\scriptscriptstyle{*}}\)}][l]{\(^{ns}\)} & 0.839 \\
\textit{RFL-3} & Edema & 506 & 115 & 0.005 & \makebox[\widthof{\(^{\scriptscriptstyle{*}\scriptscriptstyle{*}\scriptscriptstyle{*}}\)}][c]{}0.471\makebox[\widthof{\(^{\scriptscriptstyle{*}\scriptscriptstyle{*}\scriptscriptstyle{*}}\)}][l]{\(^{\scriptscriptstyle{*}\scriptscriptstyle{*}}\)} & \makebox[\widthof{\(^{\scriptscriptstyle{*}\scriptscriptstyle{*}\scriptscriptstyle{*}}\)}][c]{}\textbf{0.561}\makebox[\widthof{\(^{\scriptscriptstyle{*}\scriptscriptstyle{*}\scriptscriptstyle{*}}\)}][l]{\(^{ns}\)} & \makebox[\widthof{\(^{\scriptscriptstyle{*}\scriptscriptstyle{*}\scriptscriptstyle{*}}\)}][c]{}0.498\makebox[\widthof{\(^{\scriptscriptstyle{*}\scriptscriptstyle{*}\scriptscriptstyle{*}}\)}][l]{\(^{ns}\)} & \makebox[\widthof{\(^{\scriptscriptstyle{*}\scriptscriptstyle{*}\scriptscriptstyle{*}}\)}][c]{}0.527\makebox[\widthof{\(^{\scriptscriptstyle{*}\scriptscriptstyle{*}\scriptscriptstyle{*}}\)}][l]{\(^{ns}\)} & \makebox[\widthof{\(^{\scriptscriptstyle{*}\scriptscriptstyle{*}\scriptscriptstyle{*}}\)}][c]{}0.536\makebox[\widthof{\(^{\scriptscriptstyle{*}\scriptscriptstyle{*}\scriptscriptstyle{*}}\)}][l]{\(^{ns}\)} & \makebox[\widthof{\(^{\scriptscriptstyle{*}\scriptscriptstyle{*}\scriptscriptstyle{*}}\)}][c]{}0.549\makebox[\widthof{\(^{\scriptscriptstyle{*}\scriptscriptstyle{*}\scriptscriptstyle{*}}\)}][l]{\(^{ns}\)} & 0.551 \\
MIMIC & Fract. & 350 & 20 & 0.007 & \makebox[\widthof{\(^{\scriptscriptstyle{*}\scriptscriptstyle{*}\scriptscriptstyle{*}}\)}][c]{}0.519\makebox[\widthof{\(^{\scriptscriptstyle{*}\scriptscriptstyle{*}\scriptscriptstyle{*}}\)}][l]{\(^{\scriptscriptstyle{*}\scriptscriptstyle{*}}\)} & \makebox[\widthof{\(^{\scriptscriptstyle{*}\scriptscriptstyle{*}\scriptscriptstyle{*}}\)}][c]{}0.842\makebox[\widthof{\(^{\scriptscriptstyle{*}\scriptscriptstyle{*}\scriptscriptstyle{*}}\)}][l]{\(^{ns}\)} & \makebox[\widthof{\(^{\scriptscriptstyle{*}\scriptscriptstyle{*}\scriptscriptstyle{*}}\)}][c]{}0.833\makebox[\widthof{\(^{\scriptscriptstyle{*}\scriptscriptstyle{*}\scriptscriptstyle{*}}\)}][l]{\(^{ns}\)} & \makebox[\widthof{\(^{\scriptscriptstyle{*}\scriptscriptstyle{*}\scriptscriptstyle{*}}\)}][c]{}\textbf{0.872}\makebox[\widthof{\(^{\scriptscriptstyle{*}\scriptscriptstyle{*}\scriptscriptstyle{*}}\)}][l]{\(^{ns}\)} & \makebox[\widthof{\(^{\scriptscriptstyle{*}\scriptscriptstyle{*}\scriptscriptstyle{*}}\)}][c]{}0.800\makebox[\widthof{\(^{\scriptscriptstyle{*}\scriptscriptstyle{*}\scriptscriptstyle{*}}\)}][l]{\(^{ns}\)} & \makebox[\widthof{\(^{\scriptscriptstyle{*}\scriptscriptstyle{*}\scriptscriptstyle{*}}\)}][c]{}\textbf{0.872}\makebox[\widthof{\(^{\scriptscriptstyle{*}\scriptscriptstyle{*}\scriptscriptstyle{*}}\)}][l]{\(^{ns}\)} & 0.842 \\
NIH & Opac. & 200 & 122 & 0.037 & \makebox[\widthof{\(^{\scriptscriptstyle{*}\scriptscriptstyle{*}\scriptscriptstyle{*}}\)}][c]{}0.893\makebox[\widthof{\(^{\scriptscriptstyle{*}\scriptscriptstyle{*}\scriptscriptstyle{*}}\)}][l]{\(^{ns}\)} & \makebox[\widthof{\(^{\scriptscriptstyle{*}\scriptscriptstyle{*}\scriptscriptstyle{*}}\)}][c]{}0.822\(^{\scriptscriptstyle{*}\scriptscriptstyle{*}\scriptscriptstyle{*}}\) & - & \makebox[\widthof{\(^{\scriptscriptstyle{*}\scriptscriptstyle{*}\scriptscriptstyle{*}}\)}][c]{}0.902\makebox[\widthof{\(^{\scriptscriptstyle{*}\scriptscriptstyle{*}\scriptscriptstyle{*}}\)}][l]{\(^{ns}\)} & \makebox[\widthof{\(^{\scriptscriptstyle{*}\scriptscriptstyle{*}\scriptscriptstyle{*}}\)}][c]{}0.832\(^{\scriptscriptstyle{*}\scriptscriptstyle{*}\scriptscriptstyle{*}}\) & \makebox[\widthof{\(^{\scriptscriptstyle{*}\scriptscriptstyle{*}\scriptscriptstyle{*}}\)}][c]{}0.895\makebox[\widthof{\(^{\scriptscriptstyle{*}\scriptscriptstyle{*}\scriptscriptstyle{*}}\)}][l]{\(^{ns}\)} & \textbf{0.923} \\
MIMIC & Opac. & 350 & 262 & 0.122 & \makebox[\widthof{\(^{\scriptscriptstyle{*}\scriptscriptstyle{*}\scriptscriptstyle{*}}\)}][c]{}0.877\(^{\scriptscriptstyle{*}\scriptscriptstyle{*}\scriptscriptstyle{*}}\) & \makebox[\widthof{\(^{\scriptscriptstyle{*}\scriptscriptstyle{*}\scriptscriptstyle{*}}\)}][c]{}0.906\makebox[\widthof{\(^{\scriptscriptstyle{*}\scriptscriptstyle{*}\scriptscriptstyle{*}}\)}][l]{\(^{\scriptscriptstyle{*}\scriptscriptstyle{*}}\)} & \makebox[\widthof{\(^{\scriptscriptstyle{*}\scriptscriptstyle{*}\scriptscriptstyle{*}}\)}][c]{}0.898\makebox[\widthof{\(^{\scriptscriptstyle{*}\scriptscriptstyle{*}\scriptscriptstyle{*}}\)}][l]{\(^{\scriptscriptstyle{*}\scriptscriptstyle{*}}\)} & \makebox[\widthof{\(^{\scriptscriptstyle{*}\scriptscriptstyle{*}\scriptscriptstyle{*}}\)}][c]{}\textbf{0.938}\makebox[\widthof{\(^{\scriptscriptstyle{*}\scriptscriptstyle{*}\scriptscriptstyle{*}}\)}][l]{\(^{ns}\)} & \makebox[\widthof{\(^{\scriptscriptstyle{*}\scriptscriptstyle{*}\scriptscriptstyle{*}}\)}][c]{}0.900\makebox[\widthof{\(^{\scriptscriptstyle{*}\scriptscriptstyle{*}\scriptscriptstyle{*}}\)}][l]{\(^{\scriptscriptstyle{*}\scriptscriptstyle{*}}\)} & \makebox[\widthof{\(^{\scriptscriptstyle{*}\scriptscriptstyle{*}\scriptscriptstyle{*}}\)}][c]{}0.917\makebox[\widthof{\(^{\scriptscriptstyle{*}\scriptscriptstyle{*}\scriptscriptstyle{*}}\)}][l]{\(^{\scriptscriptstyle{*}\scriptscriptstyle{*}}\)} & 0.937 \\
\textit{RFL-3} & Opac. & 506 & 342 & 0.058 & \makebox[\widthof{\(^{\scriptscriptstyle{*}\scriptscriptstyle{*}\scriptscriptstyle{*}}\)}][c]{}0.784\makebox[\widthof{\(^{\scriptscriptstyle{*}\scriptscriptstyle{*}\scriptscriptstyle{*}}\)}][l]{\(^{ns}\)} & \makebox[\widthof{\(^{\scriptscriptstyle{*}\scriptscriptstyle{*}\scriptscriptstyle{*}}\)}][c]{}0.782\makebox[\widthof{\(^{\scriptscriptstyle{*}\scriptscriptstyle{*}\scriptscriptstyle{*}}\)}][l]{\(^{ns}\)} & \makebox[\widthof{\(^{\scriptscriptstyle{*}\scriptscriptstyle{*}\scriptscriptstyle{*}}\)}][c]{}0.772\makebox[\widthof{\(^{\scriptscriptstyle{*}\scriptscriptstyle{*}\scriptscriptstyle{*}}\)}][l]{\(^{\scriptscriptstyle{*}}\)} & \makebox[\widthof{\(^{\scriptscriptstyle{*}\scriptscriptstyle{*}\scriptscriptstyle{*}}\)}][c]{}\textbf{0.798}\makebox[\widthof{\(^{\scriptscriptstyle{*}\scriptscriptstyle{*}\scriptscriptstyle{*}}\)}][l]{\(^{ns}\)} & \makebox[\widthof{\(^{\scriptscriptstyle{*}\scriptscriptstyle{*}\scriptscriptstyle{*}}\)}][c]{}0.786\makebox[\widthof{\(^{\scriptscriptstyle{*}\scriptscriptstyle{*}\scriptscriptstyle{*}}\)}][l]{\(^{ns}\)} & \makebox[\widthof{\(^{\scriptscriptstyle{*}\scriptscriptstyle{*}\scriptscriptstyle{*}}\)}][c]{}0.796\makebox[\widthof{\(^{\scriptscriptstyle{*}\scriptscriptstyle{*}\scriptscriptstyle{*}}\)}][l]{\(^{ns}\)} & 0.796 \\
NIH & Effus. & 200 & 60 & 0.024 & \makebox[\widthof{\(^{\scriptscriptstyle{*}\scriptscriptstyle{*}\scriptscriptstyle{*}}\)}][c]{}0.855\makebox[\widthof{\(^{\scriptscriptstyle{*}\scriptscriptstyle{*}\scriptscriptstyle{*}}\)}][l]{\(^{\scriptscriptstyle{*}\scriptscriptstyle{*}}\)} & \makebox[\widthof{\(^{\scriptscriptstyle{*}\scriptscriptstyle{*}\scriptscriptstyle{*}}\)}][c]{}0.873\makebox[\widthof{\(^{\scriptscriptstyle{*}\scriptscriptstyle{*}\scriptscriptstyle{*}}\)}][l]{\(^{\scriptscriptstyle{*}\scriptscriptstyle{*}}\)} & - & \makebox[\widthof{\(^{\scriptscriptstyle{*}\scriptscriptstyle{*}\scriptscriptstyle{*}}\)}][c]{}0.915\makebox[\widthof{\(^{\scriptscriptstyle{*}\scriptscriptstyle{*}\scriptscriptstyle{*}}\)}][l]{\(^{ns}\)} & \makebox[\widthof{\(^{\scriptscriptstyle{*}\scriptscriptstyle{*}\scriptscriptstyle{*}}\)}][c]{}0.902\makebox[\widthof{\(^{\scriptscriptstyle{*}\scriptscriptstyle{*}\scriptscriptstyle{*}}\)}][l]{\(^{\scriptscriptstyle{*}}\)} & \makebox[\widthof{\(^{\scriptscriptstyle{*}\scriptscriptstyle{*}\scriptscriptstyle{*}}\)}][c]{}0.914\makebox[\widthof{\(^{\scriptscriptstyle{*}\scriptscriptstyle{*}\scriptscriptstyle{*}}\)}][l]{\(^{\scriptscriptstyle{*}}\)} & \textbf{0.959} \\
MIMIC & Effus. & 350 & 134 & 0.100 & \makebox[\widthof{\(^{\scriptscriptstyle{*}\scriptscriptstyle{*}\scriptscriptstyle{*}}\)}][c]{}0.925\makebox[\widthof{\(^{\scriptscriptstyle{*}\scriptscriptstyle{*}\scriptscriptstyle{*}}\)}][l]{\(^{\scriptscriptstyle{*}}\)} & \makebox[\widthof{\(^{\scriptscriptstyle{*}\scriptscriptstyle{*}\scriptscriptstyle{*}}\)}][c]{}0.916\(^{\scriptscriptstyle{*}\scriptscriptstyle{*}\scriptscriptstyle{*}}\) & \makebox[\widthof{\(^{\scriptscriptstyle{*}\scriptscriptstyle{*}\scriptscriptstyle{*}}\)}][c]{}0.928\makebox[\widthof{\(^{\scriptscriptstyle{*}\scriptscriptstyle{*}\scriptscriptstyle{*}}\)}][l]{\(^{\scriptscriptstyle{*}}\)} & \makebox[\widthof{\(^{\scriptscriptstyle{*}\scriptscriptstyle{*}\scriptscriptstyle{*}}\)}][c]{}0.958\makebox[\widthof{\(^{\scriptscriptstyle{*}\scriptscriptstyle{*}\scriptscriptstyle{*}}\)}][l]{\(^{ns}\)} & \makebox[\widthof{\(^{\scriptscriptstyle{*}\scriptscriptstyle{*}\scriptscriptstyle{*}}\)}][c]{}0.942\makebox[\widthof{\(^{\scriptscriptstyle{*}\scriptscriptstyle{*}\scriptscriptstyle{*}}\)}][l]{\(^{ns}\)} & \makebox[\widthof{\(^{\scriptscriptstyle{*}\scriptscriptstyle{*}\scriptscriptstyle{*}}\)}][c]{}0.951\makebox[\widthof{\(^{\scriptscriptstyle{*}\scriptscriptstyle{*}\scriptscriptstyle{*}}\)}][l]{\(^{ns}\)} & \textbf{0.962} \\
NIH & PTX & 200 & 26 & 0.014 & \makebox[\widthof{\(^{\scriptscriptstyle{*}\scriptscriptstyle{*}\scriptscriptstyle{*}}\)}][c]{}0.912\makebox[\widthof{\(^{\scriptscriptstyle{*}\scriptscriptstyle{*}\scriptscriptstyle{*}}\)}][l]{\(^{ns}\)} & \makebox[\widthof{\(^{\scriptscriptstyle{*}\scriptscriptstyle{*}\scriptscriptstyle{*}}\)}][c]{}0.926\makebox[\widthof{\(^{\scriptscriptstyle{*}\scriptscriptstyle{*}\scriptscriptstyle{*}}\)}][l]{\(^{ns}\)} & - & \makebox[\widthof{\(^{\scriptscriptstyle{*}\scriptscriptstyle{*}\scriptscriptstyle{*}}\)}][c]{}\textbf{0.981}\makebox[\widthof{\(^{\scriptscriptstyle{*}\scriptscriptstyle{*}\scriptscriptstyle{*}}\)}][l]{\(^{ns}\)} & \makebox[\widthof{\(^{\scriptscriptstyle{*}\scriptscriptstyle{*}\scriptscriptstyle{*}}\)}][c]{}0.897\makebox[\widthof{\(^{\scriptscriptstyle{*}\scriptscriptstyle{*}\scriptscriptstyle{*}}\)}][l]{\(^{ns}\)} & \makebox[\widthof{\(^{\scriptscriptstyle{*}\scriptscriptstyle{*}\scriptscriptstyle{*}}\)}][c]{}0.929\makebox[\widthof{\(^{\scriptscriptstyle{*}\scriptscriptstyle{*}\scriptscriptstyle{*}}\)}][l]{\(^{ns}\)} & 0.881 \\
\textit{RFL-3} & PTX & 506 & 16 & 0.000 & \makebox[\widthof{\(^{\scriptscriptstyle{*}\scriptscriptstyle{*}\scriptscriptstyle{*}}\)}][c]{}0.258\makebox[\widthof{\(^{\scriptscriptstyle{*}\scriptscriptstyle{*}\scriptscriptstyle{*}}\)}][l]{\(^{\scriptscriptstyle{*}}\)} & \makebox[\widthof{\(^{\scriptscriptstyle{*}\scriptscriptstyle{*}\scriptscriptstyle{*}}\)}][c]{}0.429\makebox[\widthof{\(^{\scriptscriptstyle{*}\scriptscriptstyle{*}\scriptscriptstyle{*}}\)}][l]{\(^{ns}\)} & \makebox[\widthof{\(^{\scriptscriptstyle{*}\scriptscriptstyle{*}\scriptscriptstyle{*}}\)}][c]{}0.333\makebox[\widthof{\(^{\scriptscriptstyle{*}\scriptscriptstyle{*}\scriptscriptstyle{*}}\)}][l]{\(^{ns}\)} & \makebox[\widthof{\(^{\scriptscriptstyle{*}\scriptscriptstyle{*}\scriptscriptstyle{*}}\)}][c]{}0.333\makebox[\widthof{\(^{\scriptscriptstyle{*}\scriptscriptstyle{*}\scriptscriptstyle{*}}\)}][l]{\(^{ns}\)} & \makebox[\widthof{\(^{\scriptscriptstyle{*}\scriptscriptstyle{*}\scriptscriptstyle{*}}\)}][c]{}0.383\makebox[\widthof{\(^{\scriptscriptstyle{*}\scriptscriptstyle{*}\scriptscriptstyle{*}}\)}][l]{\(^{ns}\)} & \makebox[\widthof{\(^{\scriptscriptstyle{*}\scriptscriptstyle{*}\scriptscriptstyle{*}}\)}][c]{}\textbf{0.529}\makebox[\widthof{\(^{\scriptscriptstyle{*}\scriptscriptstyle{*}\scriptscriptstyle{*}}\)}][l]{\(^{ns}\)} & 0.486 \\
\textit{PTX} & PTX & 24709 & 2912 & 0.408 & \makebox[\widthof{\(^{\scriptscriptstyle{*}\scriptscriptstyle{*}\scriptscriptstyle{*}}\)}][c]{}0.758\makebox[\widthof{\(^{\scriptscriptstyle{*}\scriptscriptstyle{*}\scriptscriptstyle{*}}\)}][l]{\(^{ns}\)} & \makebox[\widthof{\(^{\scriptscriptstyle{*}\scriptscriptstyle{*}\scriptscriptstyle{*}}\)}][c]{}0.778\(^{\scriptscriptstyle{*}\scriptscriptstyle{*}\scriptscriptstyle{*}}\) & - & \makebox[\widthof{\(^{\scriptscriptstyle{*}\scriptscriptstyle{*}\scriptscriptstyle{*}}\)}][c]{}\textbf{0.782}\(^{\scriptscriptstyle{*}\scriptscriptstyle{*}\scriptscriptstyle{*}}\) & \makebox[\widthof{\(^{\scriptscriptstyle{*}\scriptscriptstyle{*}\scriptscriptstyle{*}}\)}][c]{}0.734\(^{\scriptscriptstyle{*}\scriptscriptstyle{*}\scriptscriptstyle{*}}\) & \makebox[\widthof{\(^{\scriptscriptstyle{*}\scriptscriptstyle{*}\scriptscriptstyle{*}}\)}][c]{}0.770\(^{\scriptscriptstyle{*}\scriptscriptstyle{*}\scriptscriptstyle{*}}\) & 0.756 \\
\midrule
NIH & - & - & - & - & \makebox[\widthof{\(^{\scriptscriptstyle{*}\scriptscriptstyle{*}\scriptscriptstyle{*}}\)}][c]{}0.868\makebox[\widthof{\(^{\scriptscriptstyle{*}\scriptscriptstyle{*}\scriptscriptstyle{*}}\)}][l]{\(^{\scriptscriptstyle{*}\scriptscriptstyle{*}}\)} & \makebox[\widthof{\(^{\scriptscriptstyle{*}\scriptscriptstyle{*}\scriptscriptstyle{*}}\)}][c]{}0.847\(^{\scriptscriptstyle{*}\scriptscriptstyle{*}\scriptscriptstyle{*}}\) & - & \makebox[\widthof{\(^{\scriptscriptstyle{*}\scriptscriptstyle{*}\scriptscriptstyle{*}}\)}][c]{}0.906\makebox[\widthof{\(^{\scriptscriptstyle{*}\scriptscriptstyle{*}\scriptscriptstyle{*}}\)}][l]{\(^{ns}\)} & \makebox[\widthof{\(^{\scriptscriptstyle{*}\scriptscriptstyle{*}\scriptscriptstyle{*}}\)}][c]{}0.857\(^{\scriptscriptstyle{*}\scriptscriptstyle{*}\scriptscriptstyle{*}}\) & \makebox[\widthof{\(^{\scriptscriptstyle{*}\scriptscriptstyle{*}\scriptscriptstyle{*}}\)}][c]{}0.888\makebox[\widthof{\(^{\scriptscriptstyle{*}\scriptscriptstyle{*}\scriptscriptstyle{*}}\)}][l]{\(^{\scriptscriptstyle{*}}\)} & \textbf{0.914} \\
MIMIC & - & - & - & - & \makebox[\widthof{\(^{\scriptscriptstyle{*}\scriptscriptstyle{*}\scriptscriptstyle{*}}\)}][c]{}0.860\(^{\scriptscriptstyle{*}\scriptscriptstyle{*}\scriptscriptstyle{*}}\) & \makebox[\widthof{\(^{\scriptscriptstyle{*}\scriptscriptstyle{*}\scriptscriptstyle{*}}\)}][c]{}0.902\(^{\scriptscriptstyle{*}\scriptscriptstyle{*}\scriptscriptstyle{*}}\) & \makebox[\widthof{\(^{\scriptscriptstyle{*}\scriptscriptstyle{*}\scriptscriptstyle{*}}\)}][c]{}0.896\(^{\scriptscriptstyle{*}\scriptscriptstyle{*}\scriptscriptstyle{*}}\) & \makebox[\widthof{\(^{\scriptscriptstyle{*}\scriptscriptstyle{*}\scriptscriptstyle{*}}\)}][c]{}\textbf{0.939}\makebox[\widthof{\(^{\scriptscriptstyle{*}\scriptscriptstyle{*}\scriptscriptstyle{*}}\)}][l]{\(^{ns}\)} & \makebox[\widthof{\(^{\scriptscriptstyle{*}\scriptscriptstyle{*}\scriptscriptstyle{*}}\)}][c]{}0.903\(^{\scriptscriptstyle{*}\scriptscriptstyle{*}\scriptscriptstyle{*}}\) & \makebox[\widthof{\(^{\scriptscriptstyle{*}\scriptscriptstyle{*}\scriptscriptstyle{*}}\)}][c]{}0.922\(^{\scriptscriptstyle{*}\scriptscriptstyle{*}\scriptscriptstyle{*}}\) & 0.936 \\
\textit{RFL-3} & - & - & - & - & \makebox[\widthof{\(^{\scriptscriptstyle{*}\scriptscriptstyle{*}\scriptscriptstyle{*}}\)}][c]{}0.707\makebox[\widthof{\(^{\scriptscriptstyle{*}\scriptscriptstyle{*}\scriptscriptstyle{*}}\)}][l]{\(^{\scriptscriptstyle{*}\scriptscriptstyle{*}}\)} & \makebox[\widthof{\(^{\scriptscriptstyle{*}\scriptscriptstyle{*}\scriptscriptstyle{*}}\)}][c]{}0.725\makebox[\widthof{\(^{\scriptscriptstyle{*}\scriptscriptstyle{*}\scriptscriptstyle{*}}\)}][l]{\(^{\scriptscriptstyle{*}}\)} & \makebox[\widthof{\(^{\scriptscriptstyle{*}\scriptscriptstyle{*}\scriptscriptstyle{*}}\)}][c]{}0.701\(^{\scriptscriptstyle{*}\scriptscriptstyle{*}\scriptscriptstyle{*}}\) & \makebox[\widthof{\(^{\scriptscriptstyle{*}\scriptscriptstyle{*}\scriptscriptstyle{*}}\)}][c]{}0.739\makebox[\widthof{\(^{\scriptscriptstyle{*}\scriptscriptstyle{*}\scriptscriptstyle{*}}\)}][l]{\(^{ns}\)} & \makebox[\widthof{\(^{\scriptscriptstyle{*}\scriptscriptstyle{*}\scriptscriptstyle{*}}\)}][c]{}0.732\makebox[\widthof{\(^{\scriptscriptstyle{*}\scriptscriptstyle{*}\scriptscriptstyle{*}}\)}][l]{\(^{ns}\)} & \makebox[\widthof{\(^{\scriptscriptstyle{*}\scriptscriptstyle{*}\scriptscriptstyle{*}}\)}][c]{}0.740\makebox[\widthof{\(^{\scriptscriptstyle{*}\scriptscriptstyle{*}\scriptscriptstyle{*}}\)}][l]{\(^{ns}\)} & \textbf{0.743} \\
\textit{Human} & - & - & - & - & \makebox[\widthof{\(^{\scriptscriptstyle{*}\scriptscriptstyle{*}\scriptscriptstyle{*}}\)}][c]{}0.708\(^{\scriptscriptstyle{*}\scriptscriptstyle{*}\scriptscriptstyle{*}}\) & \makebox[\widthof{\(^{\scriptscriptstyle{*}\scriptscriptstyle{*}\scriptscriptstyle{*}}\)}][c]{}0.731\(^{\scriptscriptstyle{*}\scriptscriptstyle{*}\scriptscriptstyle{*}}\) & - & \makebox[\widthof{\(^{\scriptscriptstyle{*}\scriptscriptstyle{*}\scriptscriptstyle{*}}\)}][c]{}0.731\makebox[\widthof{\(^{\scriptscriptstyle{*}\scriptscriptstyle{*}\scriptscriptstyle{*}}\)}][l]{\(^{\scriptscriptstyle{*}\scriptscriptstyle{*}}\)} & \makebox[\widthof{\(^{\scriptscriptstyle{*}\scriptscriptstyle{*}\scriptscriptstyle{*}}\)}][c]{}0.708\(^{\scriptscriptstyle{*}\scriptscriptstyle{*}\scriptscriptstyle{*}}\) & \makebox[\widthof{\(^{\scriptscriptstyle{*}\scriptscriptstyle{*}\scriptscriptstyle{*}}\)}][c]{}0.737\makebox[\widthof{\(^{\scriptscriptstyle{*}\scriptscriptstyle{*}\scriptscriptstyle{*}}\)}][l]{\(^{ns}\)} & \textbf{0.740} \\
All-W & - & - & - & - & \makebox[\widthof{\(^{\scriptscriptstyle{*}\scriptscriptstyle{*}\scriptscriptstyle{*}}\)}][c]{}0.778\(^{\scriptscriptstyle{*}\scriptscriptstyle{*}\scriptscriptstyle{*}}\) & \makebox[\widthof{\(^{\scriptscriptstyle{*}\scriptscriptstyle{*}\scriptscriptstyle{*}}\)}][c]{}0.803\(^{\scriptscriptstyle{*}\scriptscriptstyle{*}\scriptscriptstyle{*}}\) & - & \makebox[\widthof{\(^{\scriptscriptstyle{*}\scriptscriptstyle{*}\scriptscriptstyle{*}}\)}][c]{}0.822\makebox[\widthof{\(^{\scriptscriptstyle{*}\scriptscriptstyle{*}\scriptscriptstyle{*}}\)}][l]{\(^{ns}\)} & \makebox[\widthof{\(^{\scriptscriptstyle{*}\scriptscriptstyle{*}\scriptscriptstyle{*}}\)}][c]{}0.792\(^{\scriptscriptstyle{*}\scriptscriptstyle{*}\scriptscriptstyle{*}}\) & \makebox[\widthof{\(^{\scriptscriptstyle{*}\scriptscriptstyle{*}\scriptscriptstyle{*}}\)}][c]{}0.818\(^{\scriptscriptstyle{*}\scriptscriptstyle{*}\scriptscriptstyle{*}}\) & \textbf{0.827} \\
All-M & - & - & - & - & \makebox[\widthof{\(^{\scriptscriptstyle{*}\scriptscriptstyle{*}\scriptscriptstyle{*}}\)}][c]{}0.698\(^{\scriptscriptstyle{*}\scriptscriptstyle{*}\scriptscriptstyle{*}}\) & \makebox[\widthof{\(^{\scriptscriptstyle{*}\scriptscriptstyle{*}\scriptscriptstyle{*}}\)}][c]{}0.740\(^{\scriptscriptstyle{*}\scriptscriptstyle{*}\scriptscriptstyle{*}}\) & - & \makebox[\widthof{\(^{\scriptscriptstyle{*}\scriptscriptstyle{*}\scriptscriptstyle{*}}\)}][c]{}0.767\(^{\scriptscriptstyle{*}\scriptscriptstyle{*}\scriptscriptstyle{*}}\) & \makebox[\widthof{\(^{\scriptscriptstyle{*}\scriptscriptstyle{*}\scriptscriptstyle{*}}\)}][c]{}0.754\(^{\scriptscriptstyle{*}\scriptscriptstyle{*}\scriptscriptstyle{*}}\) & \makebox[\widthof{\(^{\scriptscriptstyle{*}\scriptscriptstyle{*}\scriptscriptstyle{*}}\)}][c]{}0.766\(^{\scriptscriptstyle{*}\scriptscriptstyle{*}\scriptscriptstyle{*}}\) & \textbf{0.803} \\

\bottomrule
  \end{tabular}
\end{table*}

In tests, we compared MAPLEZ against six labelers: 
\begin{itemize}
\item CheXpert~\citep{chexpert} and CheXbert~\citep{chexbert}: we ran these labelerers on the MIMIC-CXR dataset~\citep{mimiccxr} and the test set of the NIH dataset~\citep{nihdataset}. They were only compared in the categorical presence task.
\item Medical-Diff-VQA~\citep{vqa}: the annotations only included the MIMIC-CXR dataset~\citep{mimiccxr}, so this labeler does not have results for some of the experiments. The labeler provides categorical presence, probability expressions, location expressions, and severity words. Probability expressions were converted to percentages with a conversion table. Severity words were converted to severity classes using the rules from our manual labeling of the MIMIC-CXR and NIH ChestXray14 ground truths. We also grouped a few abnormalities of the dataset using the rules described in \cref{sec:merging}. The grouped abnormalities and the probability and severity conversion rules are presented in \cref{sec:vqachanges}.
\item Vicuna~\citep{vicunanih}: we employed its annotations for the test sets of the MIMIC-CXR~\citep{mimiccxr} and NIH~\citep{nihdataset} datasets. It was only compared in the categorical presence task. We employed the SOLAR-0-70b-16bit \gls*{LLM} model instead of the less powerful Vicuna model to allow for a fair comparison between prompt systems.
\item Template~\citep{opentemplate}: we reimplemented the prompt system to the best of our ability. The employed repetition penalty had to be changed to 0.01 since the \gls*{LLM} outputs were non-sensical with any repetition penalty lower or equal to 0.0001. From the \glspl*{LLM} for which this prompt was tested in the original paper, we chose to use the meta-llama/Llama-2-70b-hf~\citep{llama2} model so that it was comparable to the \gls*{LLM} used for MAPLEZ.
\item MAPLEZ-Generic: MAPLEZ-Generic is a version of our labeler using simpler abnormality denominations without synonyms or subtypes of each abnormality. The abnormality denomination strings for MAPLEZ and MAPLEZ-Generic are presented in \cref{sec:labelprompts}. 
\end{itemize}

For all four types of annotations, we computed precision, recall, and the F1 score for eight abnormalities common for the five labelers. More details about these calculations and complete tables are presented in \cref{sec:expanded}. We calculated a weighted average for score aggregations using the minimum variance unbiased estimator. We provide more weighted average calculation details in \cref{sec:mvue} and the employed weights in the respective result tables. We also provide macro scores in all tables. They were calculated by weighting each aggregated abnormality row by the same weight in an average calculation. These macro scores have wider confidence intervals, but they balance the disproportions of each abnormality's test set size since some had lower prevalence or lower label availability.

\begin{table}[!htb]
  \caption{F1 scores ($\uparrow$) for categorical presence annotations for two radiologists (Rad.) and our proposed method in phases 1 and 2 of the \textit{REFLACX} dataset~\citep{reflacx}, with $N=109$. The ground truth was the majority vote of the other three radiologists. \Cref{tab:humansfull} is a complete version of this table, with precision, recall, and confidence intervals. Refer to \cref{tab:labels} for a list of abbreviation meanings.}
  \label{tab:humans}
  \centering
  \begin{tabular}{@{}lccrccc@{}}
\toprule
 Abn. & $N$ & $N^+$ & W & Rad. & MAPLEZ \\
\midrule
Card. & 109 & 30 & 0.10 & \makebox[\widthof{\(^{\scriptscriptstyle{*}\scriptscriptstyle{*}\scriptscriptstyle{*}}\)}][c]{}0.459\makebox[\widthof{\(^{\scriptscriptstyle{*}\scriptscriptstyle{*}\scriptscriptstyle{*}}\)}][l]{\(^{ns}\)} & \textbf{0.600} \\
Cons. & 109 & 33 & 0.09 & \makebox[\widthof{\(^{\scriptscriptstyle{*}\scriptscriptstyle{*}\scriptscriptstyle{*}}\)}][c]{}0.448\makebox[\widthof{\(^{\scriptscriptstyle{*}\scriptscriptstyle{*}\scriptscriptstyle{*}}\)}][l]{\(^{ns}\)} & \textbf{0.491} \\
Edema & 109 & 13 & 0.03 & \makebox[\widthof{\(^{\scriptscriptstyle{*}\scriptscriptstyle{*}\scriptscriptstyle{*}}\)}][c]{}0.219\makebox[\widthof{\(^{\scriptscriptstyle{*}\scriptscriptstyle{*}\scriptscriptstyle{*}}\)}][l]{\(^{ns}\)} & \textbf{0.350} \\
Opac. & 109 & 65 & 0.79 & \makebox[\widthof{\(^{\scriptscriptstyle{*}\scriptscriptstyle{*}\scriptscriptstyle{*}}\)}][c]{}0.730\makebox[\widthof{\(^{\scriptscriptstyle{*}\scriptscriptstyle{*}\scriptscriptstyle{*}}\)}][l]{\(^{ns}\)} & \textbf{0.785} \\
\midrule
All-W & - & - & - & \makebox[\widthof{\(^{\scriptscriptstyle{*}\scriptscriptstyle{*}\scriptscriptstyle{*}}\)}][c]{}0.664\makebox[\widthof{\(^{\scriptscriptstyle{*}\scriptscriptstyle{*}\scriptscriptstyle{*}}\)}][l]{\(^{ns}\)} & \textbf{0.729} \\
All-M & - & - & - & \makebox[\widthof{\(^{\scriptscriptstyle{*}\scriptscriptstyle{*}\scriptscriptstyle{*}}\)}][c]{}0.464\makebox[\widthof{\(^{\scriptscriptstyle{*}\scriptscriptstyle{*}\scriptscriptstyle{*}}\)}][l]{\(^{\scriptscriptstyle{*}}\)} & \textbf{0.556} \\

\bottomrule
  \end{tabular}
\end{table}

Results for the categorical presence annotations are presented in \cref{tab:labels}. When processing labeler outputs and ground truths, we considered ``Uncertain" as ``Present" and ``Stable" as ``Absent". To compare the F1 scores against humans, we present \cref{tab:humans}, which contains scores of one radiologist against the majority vote of 3 radiologists.

To evaluate probability annotations, we used the probability labels set by radiologists on the REFLACX dataset~\citep{reflacx}. We calculated the \gls*{MAE} to get numerical results between the predicted probabilities and the radiologist ground truth. We considered ``Stable" probabilities from MAPLEZ as 0\%. Results are presented in \cref{tab:probs}. Radiologists' performance is presented in \cref{tab:humansprobfull}.

\begin{table*}[!htb]
  \caption{ Aggregated results for three of the tested tasks. The full results, with precision, recall, confidence intervals, and subdivision by abnormality, are presented in   \cref{tab:probsfull,tab:locationfull,tab:severityfull}. The location and severity tasks had a varying $N$ depending on the evaluated abnormality. The caption of \cref{tab:labels} presents the meaning of abbreviations and symbols.}
  \label{tab:probs}
  \centering
  \begin{tabular}{@{}llllcccc@{}}
\toprule
Task & Score & Type & Dataset & N & VQA~\citep{vqa} & MAPLEZ-G & MAPLEZ (Ours) \\
\midrule
Probability & MAE ($\downarrow$) & All-W & \textit{RFL-3} &   506 & \makebox[\widthof{\(^{\scriptscriptstyle{*}\scriptscriptstyle{*}\scriptscriptstyle{*}}\)}][c]{}25.2 [23.6,26.7]\(^{\scriptscriptstyle{*}\scriptscriptstyle{*}\scriptscriptstyle{*}}\) & \makebox[\widthof{\(^{\scriptscriptstyle{*}\scriptscriptstyle{*}\scriptscriptstyle{*}}\)}][c]{}22.8 [21.4,24.3]\makebox[\widthof{\(^{\scriptscriptstyle{*}\scriptscriptstyle{*}\scriptscriptstyle{*}}\)}][l]{\(^{\scriptscriptstyle{*}\scriptscriptstyle{*}}\)} & \textbf{21.9 [20.5,23.3]} \\
Probability & MAE ($\downarrow$) & All-M & \textit{RFL-3} &   506  & \makebox[\widthof{\(^{\scriptscriptstyle{*}\scriptscriptstyle{*}\scriptscriptstyle{*}}\)}][c]{}21 [19.7,22.3]\(^{\scriptscriptstyle{*}\scriptscriptstyle{*}\scriptscriptstyle{*}}\) & \makebox[\widthof{\(^{\scriptscriptstyle{*}\scriptscriptstyle{*}\scriptscriptstyle{*}}\)}][c]{}19.4 [18.2,20.6]\makebox[\widthof{\(^{\scriptscriptstyle{*}\scriptscriptstyle{*}\scriptscriptstyle{*}}\)}][l]{\(^{\scriptscriptstyle{*}\scriptscriptstyle{*}}\)} & \textbf{18.6 [17.4,19.8]} \\
Location & F1 ($\uparrow$) & All-W & MIMIC & - & \makebox[\widthof{\(^{\scriptscriptstyle{*}\scriptscriptstyle{*}\scriptscriptstyle{*}}\)}][c]{}0.558 [0.514,0.603]\(^{\scriptscriptstyle{*}\scriptscriptstyle{*}\scriptscriptstyle{*}}\) & \makebox[\widthof{\(^{\scriptscriptstyle{*}\scriptscriptstyle{*}\scriptscriptstyle{*}}\)}][c]{}0.787 [0.751,0.825]\makebox[\widthof{\(^{\scriptscriptstyle{*}\scriptscriptstyle{*}\scriptscriptstyle{*}}\)}][l]{\(^{\scriptscriptstyle{*}\scriptscriptstyle{*}}\)} & \textbf{0.842 [0.812,0.867]} \\
Location & F1 ($\uparrow$) & All-M & MIMIC & - & \makebox[\widthof{\(^{\scriptscriptstyle{*}\scriptscriptstyle{*}\scriptscriptstyle{*}}\)}][c]{}0.431 [0.362,0.512]\(^{\scriptscriptstyle{*}\scriptscriptstyle{*}\scriptscriptstyle{*}}\) & \makebox[\widthof{\(^{\scriptscriptstyle{*}\scriptscriptstyle{*}\scriptscriptstyle{*}}\)}][c]{}\textbf{0.725 [0.651,0.795]}\makebox[\widthof{\(^{\scriptscriptstyle{*}\scriptscriptstyle{*}\scriptscriptstyle{*}}\)}][l]{\(^{ns}\)} & 0.684 [0.637,0.717] \\
Severity & F1 ($\uparrow$) & All-W & MIMIC & - & \makebox[\widthof{\(^{\scriptscriptstyle{*}\scriptscriptstyle{*}\scriptscriptstyle{*}}\)}][c]{}\textbf{0.776 [0.690,0.845]}\makebox[\widthof{\(^{\scriptscriptstyle{*}\scriptscriptstyle{*}\scriptscriptstyle{*}}\)}][l]{\(^{ns}\)} & \makebox[\widthof{\(^{\scriptscriptstyle{*}\scriptscriptstyle{*}\scriptscriptstyle{*}}\)}][c]{}0.756 [0.670,0.833]\makebox[\widthof{\(^{\scriptscriptstyle{*}\scriptscriptstyle{*}\scriptscriptstyle{*}}\)}][l]{\(^{ns}\)} & 0.716 [0.625,0.791] \\
Severity & F1 ($\uparrow$) & All-M & MIMIC & - & \makebox[\widthof{\(^{\scriptscriptstyle{*}\scriptscriptstyle{*}\scriptscriptstyle{*}}\)}][c]{}0.662 [0.542,0.789]\makebox[\widthof{\(^{\scriptscriptstyle{*}\scriptscriptstyle{*}\scriptscriptstyle{*}}\)}][l]{\(^{ns}\)} & \makebox[\widthof{\(^{\scriptscriptstyle{*}\scriptscriptstyle{*}\scriptscriptstyle{*}}\)}][c]{}\textbf{0.737 [0.625,0.830]}\makebox[\widthof{\(^{\scriptscriptstyle{*}\scriptscriptstyle{*}\scriptscriptstyle{*}}\)}][l]{\(^{ns}\)} & 0.726 [0.610,0.818] \\

\bottomrule
  \end{tabular}
\end{table*}
 
To evaluate the labeling of the location of abnormalities, we considered F1 scores for the presence of location keywords instead of full location expressions. These results are presented in \cref{tab:probs}. Keywords and replacement words are listed in \cref{sec:evalwords}. Severity annotations were evaluated by considering any severity present as a positive label. 
Details about the location and severity score calculations and further results are presented in \cref{sec:locationseveritydetails,sec:expanded}, in \cref{tab:locationfull,tab:locationall,tab:locationsimple,tab:severityfull,tab:severityall}. Severity and location annotations were evaluated only on the MIMIC-CXR dataset to allow the comparison with the Medical-Diff-VQA method.

To evaluate the adaptation of the prompt system to other modalities, we labeled 40 CT, 40 MRI, and 39 PET reports for categorical presence and, except for PET, location. A senior radiologist specializing in abdominal imaging chose the abnormality categories employed for each modality. The location keywords employed for this evaluation are presented in \cref{sec:evalwords}, and the replacement list was the same as for the \gls*{CXR} reports. The results are presented in \cref{tab:othermodalities}.

\begin{table}[!htb]
  \caption{F1 scores ($\uparrow$) for the MAPLEZ annotations for reports of modalities other than \gls*{CXR}. For precision, recall, confidence intervals, and subdivision by abnormality, check \cref{tab:othermodalitiesfull,tab:othermodalitieslocationfull}.}
  \label{tab:othermodalities}
  \centering
  \begin{tabular}{@{}lccc|cc@{}}
\toprule
Data & CT & MRI & PET & All-W & All-M\\
\midrule
Presence & 0.887 & 0.884 & 0.827 & 0.871 &  0.825 \\
Location & 0.864 & 0.812 & - & 0.850 & 0.836 \\
\bottomrule
  \end{tabular}
\end{table}

\subsection{Classifier evaluation}

We compared a model trained with MAPLEZ annotations against three baselines: one trained with the Medical-Diff-VQA labeler~\citep{vqa} annotations, one trained with the CheXpert labeler~\citep{chexpert} annotations, and one trained with the CheXbert~\citep{chexbert} labeler. All used annotations and reports were from the MIMIC-CXR dataset~\citep{mimiccxr}. A comparison of the statistics of the annotations from each labeler is provided in \cref{sec:datasetstatistics}. A complete list of tested hyperparameters and employed training parameters and architectures for all methods is presented in \cref{sec:trainingexplained}.

We evaluated our classification results only on datasets for which the ground truth annotations were labeled by radiologists directly from the \gls*{CXR} for a more robust evaluation. In addition to the datasets presented in \cref{sec:labelereval}, we also employed the test set of the \textit{CheXpert}~\citep{chexpert} dataset, which contains 500 \gls*{CXR} studies labeled each by majority vote from 5 radiologists among a set of 8 radiologists. Three of the four employed test datasets contain images from other \gls*{CXR} datasets not seen during training: NIH ChestXray14~\citep{nihdataset} and CheXpert~\citep{chexpert}. We also did an ablation study to evaluate the impact of modifications proposed in this paper: the use of probability annotations instead of categorical presence labels, the multi-task use of location labels, the ignoring of ``Stable" abnormality labels, and the inclusion of synonyms in abnormality denominations. These results are presented in \cref{tab:classifier}.

\begin{table*}[!htb]
  \caption{\gls*{AUC} scores ($\uparrow$) in four radiologist-labeled datasets for the classifiers we trained. To the left, we compare the classifier trained with annotations from MAPLEZ or its competing labelers. To the right, we show the results of the ablation study, where `$\lambda_{loc}=0$' is trained without the multi-task loss, `Cat. Labels' is trained with categorical labels instead of probability annotations, `Use ``Stable"' does not ignore the  ``Stable" abnormalities, setting them to a 50\% probability, `MAPLEZ-G' is the MAPLEZ-Generic model, with simplified abnormality denominations in the prompts, and ``All Changes" is the model with all four modifications, and considering ``Stable" as ``Present". \Cref{tab:classifierfull} is a complete version of this table, subdivided by abnormalities and containing confidence intervals. \textit{CXt}=\textit{CheXpert} dataset. Refer to \cref{tab:labels} for the meaning of other abbreviations and symbols.}
  \label{tab:classifier}
  \centering
  \begin{tabular}{@{}lcccc|ccccc@{}}
  \toprule
\footnotesize{Data} & \footnotesize{CheXpert} & \footnotesize{VQA} & \footnotesize{CheXbert} & \makecell{\footnotesize{MAPLEZ}\\\footnotesize{(Ours)}} & \footnotesize{$\lambda_{loc}=0$} & \footnotesize{Cat. Labels} & \footnotesize{Use ``Stable"} & \makecell
{\footnotesize{MAPLEZ}\\\footnotesize{Generic}} & \footnotesize{All Changes} \\
\midrule
\textit{PNA} & \makebox[\widthof{\(^{\scriptscriptstyle{*}\scriptscriptstyle{*}\scriptscriptstyle{*}}\)}][c]{}0.795\(^{\scriptscriptstyle{*}\scriptscriptstyle{*}\scriptscriptstyle{*}}\) & \makebox[\widthof{\(^{\scriptscriptstyle{*}\scriptscriptstyle{*}\scriptscriptstyle{*}}\)}][c]{}0.783\(^{\scriptscriptstyle{*}\scriptscriptstyle{*}\scriptscriptstyle{*}}\) & \makebox[\widthof{\(^{\scriptscriptstyle{*}\scriptscriptstyle{*}\scriptscriptstyle{*}}\)}][c]{}0.799\(^{\scriptscriptstyle{*}\scriptscriptstyle{*}\scriptscriptstyle{*}}\) & 0.842 & \makebox[\widthof{\(^{\scriptscriptstyle{*}\scriptscriptstyle{*}\scriptscriptstyle{*}}\)}][c]{}0.834\(^{\scriptscriptstyle{*}\scriptscriptstyle{*}\scriptscriptstyle{*}}\) & \makebox[\widthof{\(^{\scriptscriptstyle{*}\scriptscriptstyle{*}\scriptscriptstyle{*}}\)}][c]{}0.834\(^{\scriptscriptstyle{*}\scriptscriptstyle{*}\scriptscriptstyle{*}}\) & \makebox[\widthof{\(^{\scriptscriptstyle{*}\scriptscriptstyle{*}\scriptscriptstyle{*}}\)}][c]{}0.840\makebox[\widthof{\(^{\scriptscriptstyle{*}\scriptscriptstyle{*}\scriptscriptstyle{*}}\)}][l]{\(^{ns}\)} & \makebox[\widthof{\(^{\scriptscriptstyle{*}\scriptscriptstyle{*}\scriptscriptstyle{*}}\)}][c]{}\textbf{0.845}\(^{\scriptscriptstyle{*}\scriptscriptstyle{*}\scriptscriptstyle{*}}\) & \makebox[\widthof{\(^{\scriptscriptstyle{*}\scriptscriptstyle{*}\scriptscriptstyle{*}}\)}][c]{}0.821\(^{\scriptscriptstyle{*}\scriptscriptstyle{*}\scriptscriptstyle{*}}\) \\
\textit{PTX} & \makebox[\widthof{\(^{\scriptscriptstyle{*}\scriptscriptstyle{*}\scriptscriptstyle{*}}\)}][c]{}0.920\(^{\scriptscriptstyle{*}\scriptscriptstyle{*}\scriptscriptstyle{*}}\) & \makebox[\widthof{\(^{\scriptscriptstyle{*}\scriptscriptstyle{*}\scriptscriptstyle{*}}\)}][c]{}0.927\(^{\scriptscriptstyle{*}\scriptscriptstyle{*}\scriptscriptstyle{*}}\) & \makebox[\widthof{\(^{\scriptscriptstyle{*}\scriptscriptstyle{*}\scriptscriptstyle{*}}\)}][c]{}0.937\makebox[\widthof{\(^{\scriptscriptstyle{*}\scriptscriptstyle{*}\scriptscriptstyle{*}}\)}][l]{\(^{\scriptscriptstyle{*}}\)} & \textbf{0.940} & \makebox[\widthof{\(^{\scriptscriptstyle{*}\scriptscriptstyle{*}\scriptscriptstyle{*}}\)}][c]{}0.934\(^{\scriptscriptstyle{*}\scriptscriptstyle{*}\scriptscriptstyle{*}}\) & \makebox[\widthof{\(^{\scriptscriptstyle{*}\scriptscriptstyle{*}\scriptscriptstyle{*}}\)}][c]{}0.937\(^{\scriptscriptstyle{*}\scriptscriptstyle{*}\scriptscriptstyle{*}}\) & \makebox[\widthof{\(^{\scriptscriptstyle{*}\scriptscriptstyle{*}\scriptscriptstyle{*}}\)}][c]{}0.935\(^{\scriptscriptstyle{*}\scriptscriptstyle{*}\scriptscriptstyle{*}}\) & \makebox[\widthof{\(^{\scriptscriptstyle{*}\scriptscriptstyle{*}\scriptscriptstyle{*}}\)}][c]{}0.933\(^{\scriptscriptstyle{*}\scriptscriptstyle{*}\scriptscriptstyle{*}}\) & \makebox[\widthof{\(^{\scriptscriptstyle{*}\scriptscriptstyle{*}\scriptscriptstyle{*}}\)}][c]{}0.936\(^{\scriptscriptstyle{*}\scriptscriptstyle{*}\scriptscriptstyle{*}}\) \\
\textit{RFL-3} & \makebox[\widthof{\(^{\scriptscriptstyle{*}\scriptscriptstyle{*}\scriptscriptstyle{*}}\)}][c]{}0.842\makebox[\widthof{\(^{\scriptscriptstyle{*}\scriptscriptstyle{*}\scriptscriptstyle{*}}\)}][l]{\(^{\scriptscriptstyle{*}}\)} & \makebox[\widthof{\(^{\scriptscriptstyle{*}\scriptscriptstyle{*}\scriptscriptstyle{*}}\)}][c]{}0.850\makebox[\widthof{\(^{\scriptscriptstyle{*}\scriptscriptstyle{*}\scriptscriptstyle{*}}\)}][l]{\(^{ns}\)} & \makebox[\widthof{\(^{\scriptscriptstyle{*}\scriptscriptstyle{*}\scriptscriptstyle{*}}\)}][c]{}0.850\makebox[\widthof{\(^{\scriptscriptstyle{*}\scriptscriptstyle{*}\scriptscriptstyle{*}}\)}][l]{\(^{ns}\)} & \textbf{0.857} & \makebox[\widthof{\(^{\scriptscriptstyle{*}\scriptscriptstyle{*}\scriptscriptstyle{*}}\)}][c]{}0.856\makebox[\widthof{\(^{\scriptscriptstyle{*}\scriptscriptstyle{*}\scriptscriptstyle{*}}\)}][l]{\(^{ns}\)} & \makebox[\widthof{\(^{\scriptscriptstyle{*}\scriptscriptstyle{*}\scriptscriptstyle{*}}\)}][c]{}0.857\makebox[\widthof{\(^{\scriptscriptstyle{*}\scriptscriptstyle{*}\scriptscriptstyle{*}}\)}][l]{\(^{ns}\)} & \makebox[\widthof{\(^{\scriptscriptstyle{*}\scriptscriptstyle{*}\scriptscriptstyle{*}}\)}][c]{}0.856\makebox[\widthof{\(^{\scriptscriptstyle{*}\scriptscriptstyle{*}\scriptscriptstyle{*}}\)}][l]{\(^{ns}\)} & \makebox[\widthof{\(^{\scriptscriptstyle{*}\scriptscriptstyle{*}\scriptscriptstyle{*}}\)}][c]{}0.857\makebox[\widthof{\(^{\scriptscriptstyle{*}\scriptscriptstyle{*}\scriptscriptstyle{*}}\)}][l]{\(^{ns}\)} & \makebox[\widthof{\(^{\scriptscriptstyle{*}\scriptscriptstyle{*}\scriptscriptstyle{*}}\)}][c]{}0.846\makebox[\widthof{\(^{\scriptscriptstyle{*}\scriptscriptstyle{*}\scriptscriptstyle{*}}\)}][l]{\(^{\scriptscriptstyle{*}\scriptscriptstyle{*}}\)} \\
\textit{CXt} & \makebox[\widthof{\(^{\scriptscriptstyle{*}\scriptscriptstyle{*}\scriptscriptstyle{*}}\)}][c]{}0.903\(^{\scriptscriptstyle{*}\scriptscriptstyle{*}\scriptscriptstyle{*}}\) & \makebox[\widthof{\(^{\scriptscriptstyle{*}\scriptscriptstyle{*}\scriptscriptstyle{*}}\)}][c]{}0.918\makebox[\widthof{\(^{\scriptscriptstyle{*}\scriptscriptstyle{*}\scriptscriptstyle{*}}\)}][l]{\(^{ns}\)} & \makebox[\widthof{\(^{\scriptscriptstyle{*}\scriptscriptstyle{*}\scriptscriptstyle{*}}\)}][c]{}0.907\(^{\scriptscriptstyle{*}\scriptscriptstyle{*}\scriptscriptstyle{*}}\) & 0.928 & \makebox[\widthof{\(^{\scriptscriptstyle{*}\scriptscriptstyle{*}\scriptscriptstyle{*}}\)}][c]{}0.924\makebox[\widthof{\(^{\scriptscriptstyle{*}\scriptscriptstyle{*}\scriptscriptstyle{*}}\)}][l]{\(^{ns}\)} & \makebox[\widthof{\(^{\scriptscriptstyle{*}\scriptscriptstyle{*}\scriptscriptstyle{*}}\)}][c]{}0.927\makebox[\widthof{\(^{\scriptscriptstyle{*}\scriptscriptstyle{*}\scriptscriptstyle{*}}\)}][l]{\(^{ns}\)} & \makebox[\widthof{\(^{\scriptscriptstyle{*}\scriptscriptstyle{*}\scriptscriptstyle{*}}\)}][c]{}0.925\makebox[\widthof{\(^{\scriptscriptstyle{*}\scriptscriptstyle{*}\scriptscriptstyle{*}}\)}][l]{\(^{ns}\)} & \makebox[\widthof{\(^{\scriptscriptstyle{*}\scriptscriptstyle{*}\scriptscriptstyle{*}}\)}][c]{}\textbf{0.928}\makebox[\widthof{\(^{\scriptscriptstyle{*}\scriptscriptstyle{*}\scriptscriptstyle{*}}\)}][l]{\(^{ns}\)} & \makebox[\widthof{\(^{\scriptscriptstyle{*}\scriptscriptstyle{*}\scriptscriptstyle{*}}\)}][c]{}0.918\makebox[\widthof{\(^{\scriptscriptstyle{*}\scriptscriptstyle{*}\scriptscriptstyle{*}}\)}][l]{\(^{\scriptscriptstyle{*}\scriptscriptstyle{*}}\)} \\
\midrule
All-W & \makebox[\widthof{\(^{\scriptscriptstyle{*}\scriptscriptstyle{*}\scriptscriptstyle{*}}\)}][c]{}0.899\(^{\scriptscriptstyle{*}\scriptscriptstyle{*}\scriptscriptstyle{*}}\) & \makebox[\widthof{\(^{\scriptscriptstyle{*}\scriptscriptstyle{*}\scriptscriptstyle{*}}\)}][c]{}0.905\(^{\scriptscriptstyle{*}\scriptscriptstyle{*}\scriptscriptstyle{*}}\) & \makebox[\widthof{\(^{\scriptscriptstyle{*}\scriptscriptstyle{*}\scriptscriptstyle{*}}\)}][c]{}0.912\(^{\scriptscriptstyle{*}\scriptscriptstyle{*}\scriptscriptstyle{*}}\) & \textbf{0.923} & \makebox[\widthof{\(^{\scriptscriptstyle{*}\scriptscriptstyle{*}\scriptscriptstyle{*}}\)}][c]{}0.916\(^{\scriptscriptstyle{*}\scriptscriptstyle{*}\scriptscriptstyle{*}}\) & \makebox[\widthof{\(^{\scriptscriptstyle{*}\scriptscriptstyle{*}\scriptscriptstyle{*}}\)}][c]{}0.919\(^{\scriptscriptstyle{*}\scriptscriptstyle{*}\scriptscriptstyle{*}}\) & \makebox[\widthof{\(^{\scriptscriptstyle{*}\scriptscriptstyle{*}\scriptscriptstyle{*}}\)}][c]{}0.918\(^{\scriptscriptstyle{*}\scriptscriptstyle{*}\scriptscriptstyle{*}}\) & \makebox[\widthof{\(^{\scriptscriptstyle{*}\scriptscriptstyle{*}\scriptscriptstyle{*}}\)}][c]{}0.918\(^{\scriptscriptstyle{*}\scriptscriptstyle{*}\scriptscriptstyle{*}}\) & \makebox[\widthof{\(^{\scriptscriptstyle{*}\scriptscriptstyle{*}\scriptscriptstyle{*}}\)}][c]{}0.915\(^{\scriptscriptstyle{*}\scriptscriptstyle{*}\scriptscriptstyle{*}}\) \\
All-M & \makebox[\widthof{\(^{\scriptscriptstyle{*}\scriptscriptstyle{*}\scriptscriptstyle{*}}\)}][c]{}0.853\(^{\scriptscriptstyle{*}\scriptscriptstyle{*}\scriptscriptstyle{*}}\) & \makebox[\widthof{\(^{\scriptscriptstyle{*}\scriptscriptstyle{*}\scriptscriptstyle{*}}\)}][c]{}0.858\(^{\scriptscriptstyle{*}\scriptscriptstyle{*}\scriptscriptstyle{*}}\) & \makebox[\widthof{\(^{\scriptscriptstyle{*}\scriptscriptstyle{*}\scriptscriptstyle{*}}\)}][c]{}0.861\makebox[\widthof{\(^{\scriptscriptstyle{*}\scriptscriptstyle{*}\scriptscriptstyle{*}}\)}][l]{\(^{\scriptscriptstyle{*}\scriptscriptstyle{*}}\)} & 0.876 & \makebox[\widthof{\(^{\scriptscriptstyle{*}\scriptscriptstyle{*}\scriptscriptstyle{*}}\)}][c]{}0.872\makebox[\widthof{\(^{\scriptscriptstyle{*}\scriptscriptstyle{*}\scriptscriptstyle{*}}\)}][l]{\(^{\scriptscriptstyle{*}\scriptscriptstyle{*}}\)} & \makebox[\widthof{\(^{\scriptscriptstyle{*}\scriptscriptstyle{*}\scriptscriptstyle{*}}\)}][c]{}0.874\makebox[\widthof{\(^{\scriptscriptstyle{*}\scriptscriptstyle{*}\scriptscriptstyle{*}}\)}][l]{\(^{ns}\)} & \makebox[\widthof{\(^{\scriptscriptstyle{*}\scriptscriptstyle{*}\scriptscriptstyle{*}}\)}][c]{}0.873\makebox[\widthof{\(^{\scriptscriptstyle{*}\scriptscriptstyle{*}\scriptscriptstyle{*}}\)}][l]{\(^{\scriptscriptstyle{*}}\)} & \makebox[\widthof{\(^{\scriptscriptstyle{*}\scriptscriptstyle{*}\scriptscriptstyle{*}}\)}][c]{}\textbf{0.877}\makebox[\widthof{\(^{\scriptscriptstyle{*}\scriptscriptstyle{*}\scriptscriptstyle{*}}\)}][l]{\(^{ns}\)} & \makebox[\widthof{\(^{\scriptscriptstyle{*}\scriptscriptstyle{*}\scriptscriptstyle{*}}\)}][c]{}0.863\(^{\scriptscriptstyle{*}\scriptscriptstyle{*}\scriptscriptstyle{*}}\) \\

\bottomrule
  \end{tabular}
\end{table*}

\section{Discussion}

Our proposed method showed a better overall performance than the competing methods. In the categorical labels task, presented in~\cref{tab:labels}, the MAPLEZ labeler had the significantly highest macro score over all compared competing labelers (all \textit{P}\textless.001). For the weighted overall scores, the scores of MAPLEZ were also the highest, even though not significantly against CheXbert. The MAPLEZ prompt system was the only tested \gls*{LLM} based method that surpassed the established CheXbert method. 

Some precision and recall scores might seem relatively low for a medical task. For example, the ``Pneumothorax'' label has a recall of 0.738 and a precision of 0.786 for the MAPLEZ labeler. This lower score for the \textit{Human} datasets, in which separate radiologists annotated CXR reports and image labels, probably happened because of inter-observer variability. For example, we show in \cref{tab:humans} that the macro scores of our method, which annotates \glspl*{CXR} from a radiology report from the MIMIC-CXR dataset, are significantly better than the macro inter-observer scores of individual radiologists annotating \glspl*{CXR} directly from the image (\textit{P}=.037). The low score of the ``Pneumothorax'' label probably happened because of inter-rater variability originating from the \textit{Pneumothorax} dataset, which dominated the score of the ``Pneumothorax" label, representing 98\% of the score. There was no inter- or intra-observer variability for the NIH ChestXray14 and MIMIC-CXR datasets since manual annotations and extracted labels came from the same report, and F1 scores were approximately 0.9 for those datasets, except for the ``Edema'' label. 

We reviewed MAPLEZ's ``Edema" outputs in 20 test mistakes to further understand the lower scores for this label. False negatives seemed to happen from a combination of the presence of alternative abnormality wording in the reports (\say{vascular congestion}, \say{vascular engorgement}) and the presence of low probability/severity adjectives in the reports (\say{less likely}, \say{minimal if any}, \say{without other evidence of}). ``Edema" false positives happened because the \gls*{LLM} confused \say{enlarged heart} with ``Edema", because of complex wording (\say{patient history of edema}, \say{interval resolution of edema}) and because there were a few incorrect annotations in the test set ground truth. 

\Cref{tab:probs} shows that the probabilities outputted by the MAPLEZ method from a radiologist report significantly (\textit{P}\textless.001; \textit{P}\textless.001) conform better to probabilities assigned by other radiologists directly to the same \gls*{CXR} than the outputs of the Medical-Diff-VQA method. Even though radiologists are not consistent in their language to express probability~\citep{radprobabilities}, the \gls*{LLM} is still able to extract meaningful probabilities and even outperform (\textit{P}\textless.001; \textit{P}\textless.001) other radiologists assigning probabilities directly to the \gls*{CXR}, as shown in \cref{tab:humansprobfull}.

Our method significantly outperformed the Medical-Diff-VQA method for location annotations (all \textit{P}\textless.001), achieving an F1 score of more than 0.2 higher over weighted and macro scores, as shown in \cref{tab:probs,tab:locationfull,tab:locationall,tab:locationsimple}. This superiority even happened when we limited the evaluated vocabulary to only words included in the manual rules of the Medical-Diff-VQA dataset (\cref{tab:locationsimple}). Rule-based location extraction is probably in early development, and its rule set could still be expanded. It does not identify several location expressions (``lateral", ``perihilar" or ``fissure", for example), leading to a low recall. Using an \gls*{LLM} for location expression extraction seems more generalizable in a short development time. 

The MAPLEZ labeler achieved an F1 score lower than the Medical-Diff-VQA method for severity, with the macro scores from \cref{tab:severityall} significantly lower (\textit{P}=.024). This result might show one of our method's deficiencies. We evaluated severity outputs in 20 test mistakes. False negative errors were caused by a combination of alternative wording not included in the prompt (\say{small}, \say{extensive}, \say{subtle}) and because some severity adjectives were not adjacent to the abnormality mention. False positives were caused by the presence of nearby adjectives characterizing another abnormality and by a few incorrect test set ground truth annotations. 

\Cref{tab:labelsfull,tab:locationfull,tab:locationall,tab:locationsimple} demonstrate that MAPLEZ outperformed other labelers in the categorical presence and location annotations mainly because of a higher recall. \Cref{tab:severityfull,tab:severityall} show that MAPLEZ's relatively lower performance for the severity annotations was caused mainly by a lower precision than other methods.

The MAPLEZ method scored better than the MAPLEZ-Generic method in the location and categorical presence annotation tasks. These results show that adding rule-based aspects to the prompts -- the multiple ways of mentioning the same abnormality -- can positively impact the labeler. This enhancement likely occurred because the short answers prevented the model from identifying synonyms in some cases, which was remediated by including those in the prompts. However, the fact that the performance of MAPLEZ-Generic, in most cases, was closer to MAPLEZ than to CheXpert, Vicuna, or Medical-Diff-VQA shows that the main advantage of the proposed MAPLEZ method comes from the use of a performant \gls*{LLM} and of a well-validated extensive prompt system. It also shows that the technique will likely perform well when adapted to other abnormalities, even if a careful manual definition of abnormality denominations is not performed.

The scores of our adaptation to other medical modalities achieved in \cref{tab:othermodalities} are comparable to the scores reached by the MAPLEZ and MAPLEZ-Generic model in \cref{tab:locationall,tab:labels}. These results show the potential and accessibility of such a tool in facilitating research in various medical imaging projects in other modality types or for a different label set. 

As shown in \cref{tab:classifier}, the classifier trained with the annotations from the MAPLEZ labeler performed better than the classifiers trained with the annotations of the CheXpert labeler, the CheXbert labeler, or the Medical-Diff-VQA labeler  (\textit{P}\textless.001, \textit{P}\textless.001, \textit{P}\textless.001; \textit{P}\textless.001, \textit{P}\textless.001, \textit{P}=.002). These results show that the annotations we extracted are more useful in a downstream classification task. For example, in the \textit{Pneumothorax} dataset, the CheXbert labeler performed significantly better than MAPLEZ (\textit{P}\textless.001) whereas the classifier using MAPLEZ's annotations was significantly better than the CheXbert classifier (\textit{P}=.029). This contradictory result might have happened because MAPLEZ 's mistakes happened when the pneumothorax finding was borderline or because the multi-type annotations and the labels of other abnormalities compensated with additional supervision information. Additionally, for the weighted aggregated scores, CheXbert was not significantly different from the MAPLEZ labeler in the report labeling task, but it was in the classification task (\textit{P}\textless.001). This difference might happen because of the difference in report labeling performance for the \textit{Human} datasets, for which MAPLEZ shows a significant superiority (\textit{P}=.003). The performance on the \textit{Human} datasets in \cref{tab:labels} might be predictive of the classifier performance because the classifier test datasets were also labeled by humans directly from the \glspl*{CXR}.

The ablation study from \cref{tab:classifier} shows that the method choices of how to employ the data from the MAPLEZ labeler were, together, significantly beneficial to the classifier (\textit{P}\textless.001; \textit{P}\textless.001). Individually, each change, except the use of probability labels and the use of synonyms in the abnormality denominations in the macro score, also significantly improved the classifier (\textit{P}\textless.001, \textit{P}\textless.001, \textit{P}\textless.001, \textit{P}\textless.001; \textit{P}=.005, \textit{P}=.026). The use of the extracted anatomical location through the location loss ($\lambda_{loc}=0.01$) provides the model with additional supervision, possibly teaching it to focus on the correct area of the \gls*{CXR} when a finding is present. This hypothesis could be tested in future work. The optimal $\lambda_{loc}$ is 0.01 for MAPLEZ's labels but only 0.001 for the Medical-Diff-VQA dataset. This fact corroborates the proposed labeler's superiority and lower noise labels against its compared baseline. Employing probability labels instead of categorical labels leads to a better \gls*{AUC} probably because the model has a more forgiving loss when the radiologist is unsure about an abnormality in complex or dubious cases. Ignoring the cases labeled as ``Stable" is probably beneficial because those cases have very noisy labels. For those cases, the information about the abnormality presence is inaccessible to the labeler because it is only listed in a previous \gls*{CXR} report of the same patient. Having fewer training cases (Ignore ``Stable") showed benefits against having more noisy data (Use ``Stable"). As shown in the weighted scores of \cref{tab:labels,tab:locationfull}, the MAPLEZ labeler is more accurate than MAPLEZ-Generic, so the use of abnormality synonyms in the \gls*{LLM} prompts leads to a better classifier through less noisy training labels.

\subsection{Limitations}
\label{sec:limitations}

Our classifier did not achieve state-of-the-art performance in some tasks. For example, the model trained with radiologist-labeled annotations by Hallinan \textit{et~al.}~\citep{pneumothorax} achieved an \gls*{AUC} of 0.943 [0.939, 0.946], and the performance of our model was slightly lower than the lower end of that confidence interval. The classifier has also not been trained or tested for clinical uses. Indeed, our study focused on showing the advantage of using the \gls*{MAPLEZ} labeler instead of achieving the best classifier. For example, we did not use lateral images as inputs or the resolution and bit depth from raw DICOMs during training. There are also many unexplored ways of integrating the provided annotations into classification models and losses that were not proposed in this paper. Additionally, we did not show any positive impact of using the severity annotations, likely because the annotations for severity were too noisy to be used in supervision.

When other researchers try to adapt the \gls*{MAPLEZ} method to their work needs, defining the appropriate abnormality denominations and local hardware requirements may pose deployment difficulty. However, we showed that simple adaptations to other modalities can achieve results comparable to results that MAPLEZ-Generic achieved for \glspl*{CXR}. For deployment in the future, it is unclear if the same prompt system will be adaptable to a more sophisticated next-generation \glspl*{LLM} since we did not evaluate prompt transferability between \glspl*{LLM}. A new round of prompt engineering with a validation set might be necessary for the full potential use of a different \gls*{LLM}. However, the difference between the results of different prompt systems might become smaller as the understanding power and the medical knowledge of the \glspl*{LLM} grows. The procedure of extracting labels from reports and then using them for training a classifier might also become obsolete as there will likely be further development of large multimodal models that learn end-to-end medical tasks involving language and vision.

In future work, when trying to improve the MAPLEZ's performance in the categorical presence of ``Edema" and in severity tasks, alternative wording mistakes could potentially be solved with longer prompts. For example, the abnormality denomination \say{lung edema (CHF or vascular congestion or vascular prominence or indistinctness)} could be validated for ``Edema", and including abnormality size or other synonyms could be validated for the severity. Other types of mistakes would probably need a more powerful \gls*{LLM} or a chain-of-thought answer prompt~\citep{cot}. Other potential improvements for the prompt system are optimizing its computational speed and including a ``Normal'' category output to fully match the functionalities of the CheXpert labeler.
 
\section{Conclusion}

We showed that \glspl*{LLM} can improve the quality of annotation of medical reports and still be run locally without sharing potentially confidential data. We also showed that the answers given by the \glspl*{LLM} can have high quality even if chain-of-thought reasoning is not used. \Glspl*{LLM} can also help estimate the uncertainty expressed by radiologists in their reports, which can reduce noise in annotations. The use of \glspl*{LLM} has the potential to expedite medical research by facilitating the extraction of textual information. Compared to rule-based systems, \glspl*{LLM} enable the fast development of strategies for extracting data from texts and, as our findings show, provide annotations with superior quality. Finally, we showed that training modifications made possible by the \gls*{MAPLEZ} method led to improved classification scores for a \gls*{CXR} abnormality detection model.

\section{Data and code availability}
Part of the anonymized datasets and labelers we used are publicly available: CheXpert labeler~\citep{chexpert}, \textit{CheXpert} test set~\citep{chexpert}, \glspl*{CXR} of the NIH ChestXray 14 dataset~\citep{nihdataset}, and the images/reports of MIMIC-CXR~\citep{mimiccxr}. Other datasets and baseline methods are private and were obtained after anonymization and analyzed with IRB approval. The code for producing these results and the \gls*{MAPLEZ} annotations, the annotations we processed for the complete MIMIC-CXR and NIH ChestXray14 datasets, and the ground truth manual annotations used for part of our evaluation are available at \url{https://github.com/rsummers11/CADLab/tree/master/MAPLEZ_LLM_report_labeler/}.

\section*{Acknowledgments}
 Nicholas K Lee and Abhi Suri participated in anonymizing reports for dataset sharing. This work was supported by the Intramural Research Programs of the NIH Clinical Center. This work utilized the computational resources of the NIH HPC Biowulf cluster. (\url{http://hpc.nih.gov})

\section*{Declaration of generative AI and AI-assisted technologies in the writing process}

During the preparation of this work the authors
used ChatGPT and Grammarly in order to improve writing quality. After using this tool/service, the authors reviewed and edited the content as needed and take full responsibility for the content of the publication.

% \section*{References}
%%Harvard
\bibliographystyle{model2-names.bst}\biboptions{authoryear}
\bibliography{refs}

% \textbf{Ricardo Bigolin Lanfredi} received his BS degree in Electrical Engineering from the Federal University of Rio Grande do Sul, his M.Eng. degree from CentraleSupélec, his PhD degree from the University of UTah, and is in 2024 a postdoctoral fellow  at the National Institutes of Health working with medical image analysis and deep learning at the Imaging Biomarkers and Computer-Aided Diagnosis Laboratory.

% \textbf{Ronald M. Summers, M.D., Ph.D.}, is a tenured Senior Investigator and Staff Radiologist in the Radiology and Imaging Sciences Department at the NIH Clinical Center in Bethesda, MD.  He is a Fellow of the Society of Abdominal Radiologists, the American Institute for Medical and Biological Engineering and SPIE. His awards include the Presidential Early Career Award for Scientists and Engineers, the NIH Director’s Award, and the NIH Ruth L. Kirschstein Mentoring Award. He is a member of the editorial boards of Radiology: Artificial Intelligence and Academic Radiology and a past member of the editorial boards of Radiology and the Journal of Medical Imaging. He has co-authored over 650 journal, review and conference proceedings articles and is a co-inventor on 17 patents. His research interests include thoracic and abdominal imaging, large radiology image databases, and artificial intelligence.

\clearpage
\appendix
% \section{Appendix}

\section{Experimental details}
\subsection{Prompts}
\label{sec:prompts}

\begin{figure*}[t]
  \centering
   \includegraphics[width=0.95\linewidth]{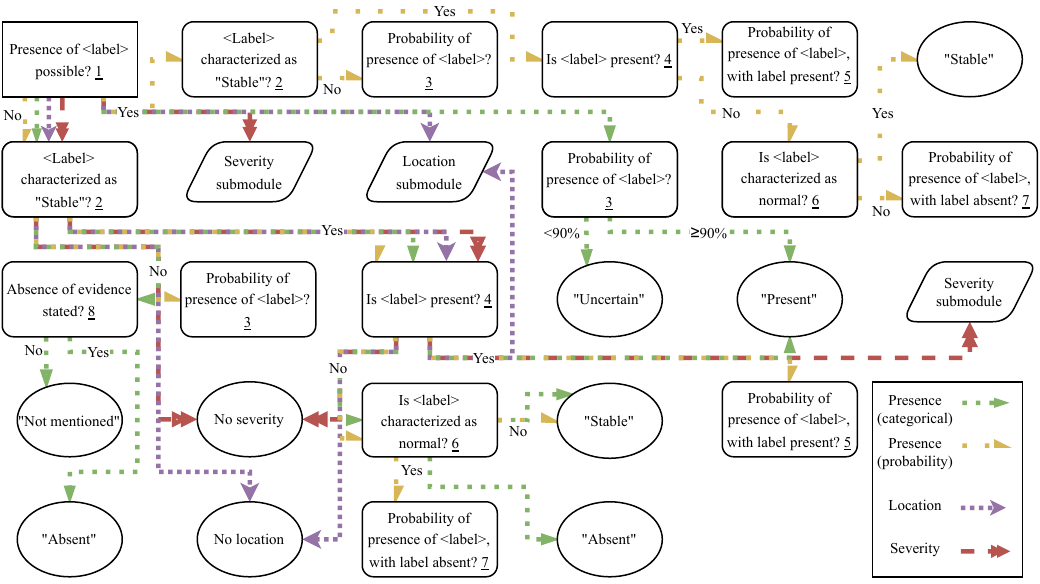}
   \caption{Full representation of the knowledge-driven decision tree prompt system, except for the exact prompts used. Different arrow styles represent the decision tree and simplified prompts for each of the four types of annotations. Some blocks are submodules whose prompt system is represented in \cref{fig:submodules}. Questions/prompts are numbered, and their complete respective text prompts are presented in \cref{sec:prompttexts}. For simplicity, probability outputs are not indicated, but they are the \gls*{LLM} answer to each of the probability questions (\underline{3}, \underline{5}, and \underline{7}). }
   \label{fig:systemfull}
\end{figure*}

\begin{figure}
  \centering
  \begin{subfigure}{0.56\linewidth}
       \includegraphics[width=0.99\linewidth]{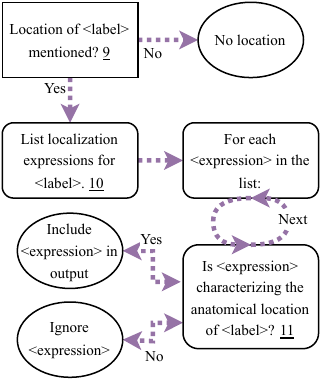}
    \caption{Location submodule}
    \label{fig:locationsubmodule}
  \end{subfigure}
  \hfill
  \begin{subfigure}{0.43\linewidth}
    \includegraphics[width=0.99\linewidth]{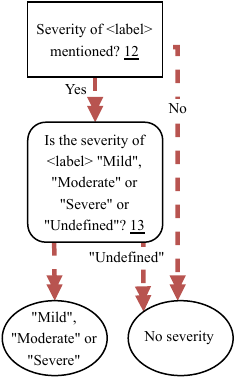}
    \caption{Severity submodule}
    \label{fig:severitysubmodule}
  \end{subfigure}
  \caption{Submodules that are indicated in \cref{fig:systemfull}.}
  \label{fig:submodules}
\end{figure}

A complete representation of the knowledge-drive decision tree prompt system is given in \cref{fig:systemfull,fig:submodules}. As seen in \cref{fig:systemfull}, the \gls*{LLM} interactions for getting the four types of annotations start with common prompts and branch out to prompts specific to each annotation type. This configuration saves computational time when annotating for all four annotation types simultaneously.

\subsubsection{Prompt texts}
\label{sec:prompttexts}

These are the complete prompts used in our knowledge-driven decision tree prompt system:

\begin{itemize}
\item System prompt (included before all MAPLEZ prompts as an introduction): \say{A chat between a radiologist and an artificial intelligence assistant trained to understand radiology reports and any synonyms and word equivalency of findings and medical terms that may appear in the report. The assistant gives helpful structured answers to the radiologist.}

\item \underline{1}:
\say{Given the full report \texttt{"}\textless report\textgreater\texttt{"}, use a one sentence logical deductive reasoning to infer if the radiologist observed possible presence of evidence of \texttt{"}\textless label\textgreater \texttt{"}. Respond only with \texttt{"}Yes\texttt{"} or \texttt{"}No\texttt{"}.}

\item \underline{2}:
For ``Enlarged cardiomediastinum" and ``Cardiomegaly", the prompt was:
\say{Given the full report \texttt{"}\textless report\textgreater \texttt{"}, use a one sentence logical deductive reasoning to infer if the radiologist characterized \texttt{"}\textless label\textgreater \texttt{"} as stable or unchanged. Respond only with \texttt{"}Yes\texttt{"} or \texttt{"}No\texttt{"}.}. For other labels, the prompt was:
\say{Given the full report \texttt{"}\textless report\textgreater \texttt{"}, use a one sentence logical deductive reasoning to infer if the radiologist characterized specifically \texttt{"}\textless label\textgreater \texttt{"} as stable or unchanged. Respond only with \texttt{"}Yes\texttt{"} or \texttt{"}No\texttt{"}.}

\item \underline{3}:
\say{Consider in your answer: 1) radiologists might skip some findings because of their low priority 2) explore all range of probabilities, giving preference to non-round probabilities 3) medical wording synonyms, subtypes of abnormalities 4) radiologists might express their uncertainty using words such as \texttt{"}or\texttt{"}, \texttt{"}possibly\texttt{"}, \texttt{"}can\textquotesingle t exclude\texttt{"}, etc... Given the complete report \texttt{"}\textless report\textgreater \texttt{"}, estimate from the report wording how likely another radiologist is to observe the presence of any type of \texttt{"}\textless label\textgreater \texttt{"} in the same imaging. Respond with the template \texttt{"}\_\_\_\% likely.\texttt{"} and no other words.}

\item \underline{4}:
For ``Support devices", the prompt was:
\say{Say \texttt{"}Yes\texttt{"}.}. For other labels, the prompt was:
\say{Given the full report \texttt{"}\textless report\textgreater \texttt{"}, use a one sentence logical deductive reasoning to infer if \texttt{"}\textless label\textgreater \texttt{"} might be present. Respond only with \texttt{"}Yes\texttt{"} or \texttt{"}No\texttt{"}.}

\item \underline{5}:
\say{Consider in your answer: 1) radiologists might skip some findings because of their low priority 2) explore all range of probabilities, giving preference to non-round probabilities 3) medical wording synonyms, subtypes of abnormalities 4) radiologists might express their uncertainty using words such as \texttt{"}or\texttt{"}, \texttt{"}possibly\texttt{"}, \texttt{"}can\textquotesingle t exclude\texttt{"}, etc... Given the complete report \texttt{"}\textless report\textgreater \texttt{"}, consistent with the radiologist observing \texttt{"}\textless label\textgreater \texttt{"}, estimate from the report wording how likely another radiologist is to observe the presence of any type of \texttt{"}\textless label\textgreater \texttt{"} in the same imaging. Respond with the template \texttt{"}\_\_\_\% likely.\texttt{"} and no other words.}

\item \underline{6}:
For ``Enlarged cardiomediastinum" and ``Cardiomegaly", the prompt was:
\say{Given the full report \texttt{"}\textless report\textgreater \texttt{"}, use a one sentence logical deductive reasoning to infer if the radiologist characterized  \texttt{"}\textless label\textgreater \texttt{"} as normal. Respond only with \texttt{"}Yes\texttt{"} or \texttt{"}No\texttt{"}.}. For other labels, the prompt was:
\say{Given the full report \texttt{"}\textless report\textgreater \texttt{"}, use a one sentence logical deductive reasoning to infer if the radiologist characterized specifically \texttt{"}\textless label\textgreater \texttt{"} as normal. Respond only with \texttt{"}Yes\texttt{"} or \texttt{"}No\texttt{"}.}

\item \underline{7}:
\say{Consider in your answer: 1) radiologists might skip some findings because of their low priority 2) explore all range of probabilities, giving preference to non-round probabilities 3) medical wording synonyms, subtypes of abnormalities 4) radiologists might express their uncertainty using words such as \texttt{"}or\texttt{"}, \texttt{"}possibly\texttt{"}, \texttt{"}can\textquotesingle t exclude\texttt{"}, etc... Given the complete report \texttt{"}\textless report\textgreater \texttt{"}, consistent with the radiologist stating the absence of evidence \texttt{"}\textless label\textgreater \texttt{"}, estimate from the report wording how likely another radiologist is to observe the presence of any type of \texttt{"}\textless label\textgreater \texttt{"} in the same imaging. Respond with the template \texttt{"}\_\_\_\% likely.\texttt{"} and no other words.}

\item \underline{8}:
\say{Given the full report \texttt{"}\textless report\textgreater \texttt{"}, use a one sentence logical deductive reasoning to infer if the radiologist stated the absence of evidence of \texttt{"}\textless label\textgreater \texttt{"}. Respond only with \texttt{"}Yes\texttt{"} or \texttt{"}No\texttt{"}.}

\item \underline{9}:
\say{Given the complete report \texttt{"}\textless report\textgreater \texttt{"}, does it mention a location for specifically \texttt{"}\textless label\textgreater \texttt{"}? Respond only with \texttt{"}Yes\texttt{"} or \texttt{"}No\texttt{"}.}

\item \underline{10}:
\say{Given the report \texttt{"}\textless report\textgreater \texttt{"}, list the localizing expressions characterizing specifically the \texttt{"}\textless label\textgreater \texttt{"} finding. Each adjective expression should be between quotes, broken down into each and every one of the localizing adjectives and each independent localiziation prepositional phrase, and separated by comma. Output an empty list (\texttt{"}[]\texttt{"} is an empty list) if there are 0 locations mentioned for \texttt{"}\textless label\textgreater \texttt{"}. Do not mention the central nouns identified as \texttt{"}\textless label\textgreater \texttt{"}. Do not mention any nouns that are not part of an adjective. Only include in your answer location adjectives adjacent to the mention of the \texttt{"}\textless label\textgreater \texttt{"} finding. Exclude from your answer adjectives for other findings. Use very short answers without complete sentences. Start the list (0+ elements) of only localizing adjectives or localizing expressions (preposition + noun) right here: [}

\item \underline{11}:
\say{Consider in your answer: 1) medical wording synonyms, subtypes of abnormalities 2) abreviations of the medical vocabulary. Given the complete report \texttt{"}\textless report\textgreater \texttt{"}, is the isolated adjective \texttt{"}\textless expression\textgreater \texttt{"}, on its own, characterizing a medical finding in what way? Respond only with \texttt{"}\_\texttt{"} where \_ is the number corresponding to the correct answer.
(1) Anatomical location of \texttt{"}\textless label\textgreater \texttt{"}
(2) Comparison with a previous report for \texttt{"}\textless label\textgreater \texttt{"}
(3) Severity of \texttt{"}\textless label\textgreater \texttt{"}
(4) Size of \texttt{"}\textless label\textgreater \texttt{"}
(5) Probability of presence of \texttt{"}\textless label\textgreater \texttt{"}
(6) Visual texture description of \texttt{"}\textless label\textgreater \texttt{"}
(7) It is not characterizing the \texttt{"}\textless label\textgreater \texttt{"} mention noun
(8) A type of support device
Answer:\texttt{"}}

\item \underline{12}:
\say{Given the complete report \texttt{"}\textless report\textgreater \texttt{"}, would you be able to characterize the severity of \texttt{"}\textless label\textgreater \texttt{"}, as either \texttt{"}Mild\texttt{"}, \texttt{"}Moderate\texttt{"} or \texttt{"}Severe\texttt{"} only from the words of the report? Respond only with \texttt{"}Yes\texttt{"} or \texttt{"}No\texttt{"}.}

\item \underline{13}:
\say{Given the complete report \texttt{"}\textless report\textgreater \texttt{"}, characterize the severity of \texttt{"}\textless label\textgreater \texttt{"} as either \texttt{"}Mild\texttt{"}, \texttt{"}Moderate\texttt{"} or \texttt{"}Severe\texttt{"} or \texttt{"}Undefined\texttt{"} only from the words of the report, and not from comparisons or changes. Do not add extra words to your answer and exclusively use the words from one of those four options.}
\end{itemize}

The underlined number in each item connects the full prompts to the simplified prompts used in \cref{fig:systemfull,fig:submodules}. For the listed prompts, \textless report\textgreater~is replaced with the report being analyzed, \textless label\textgreater~is replaced with the abnormality denominations listed in \cref{sec:labelprompts}, and \textless expression\textgreater~is replaced with one of the expressions listed by the \gls*{LLM} as an answer to question \underline{9}. In the reports that were input to the prompt, anonymized words, usually indicated by \say{\_\_\_\_}, were replaced by \say{thing}. For the MIMIC-CXR dataset, the findings, comparison, and impression sections were included in the input report. Only the findings section was included for the CT, MRI, and PET modalities.

\subsubsection{Abnormalities denominations}
\label{sec:labelprompts} 
We followed the definitions of abnormalities from the CheXpert labeler GitHub repository~\citep{chexpertdefinitions}. The abnormality denominations included a subset of those terms and were chosen with validation experiments after being compared to other CheXpert term subsets. For all 13 abnormalities that our prompt system evaluates, these are the denominations used inside the prompts:

\begin{itemize}
\item ``Enlarged cardiomediastinum": \say{enlarged cardiomediastinum (enlarged heart silhouette or large heart vascularity or cardiomegaly or abnormal mediastinal contour)}
\item ``Cardiomegaly": \say{cardiomegaly (enlarged cardiac/heart contours)}
\item ``Atelectasis": \say{atelectasis (collapse of the lung)}
\item ``Consolidation": \say{consolidation or infiltrate}
\item ``Edema": \say{lung edema (congestive heart failure)}
\item ``Fracture": \say{fracture (bone)}
\item ``Lung lesion": \say{lung lesion (mass or nodule)}
\item ``Pleural effusion": \say{pleural effusion (pleural fluid) or hydrothorax/hydropneumothorax}
\item ``Pneumonia": \say{pneumonia (infection)}
\item ``Pneumothorax": \say{pneumothorax (or pneumomediastinum or hydropneumothorax)}
\item ``Support devices": \say{medical equipment or medical support devices (lines or tubes or pacers or apparatus)}
\item ``Lung opacity": \say{lung opacity (or decreased lucency or lung scarring or bronchial thickening or infiltration or reticulation or interstitial lung)}
\item ``Pleural other": \say{pleural abnormalities other than pleural effusion (pleural thickening, fibrosis, fibrothorax, pleural scaring)}
\end{itemize}

For prompts that are exclusively evaluating probability, location, or severity, the following abnormalities had their denominations slightly modified:
\begin{itemize}
\item ``Support devices": \say{medical equipment or support device (line or tube or pacer or apparatus or valve or catheter)}
\item ``Pleural other": \say{fibrothorax (not lung fibrosis) or pleural thickening or abnormalities in the pleura (not pleural effusion)}
\end{itemize}

In \cref{sec:labelereval}, the MAPLEZ-Generic labeler abnormality denominations are simplified to evaluate the importance of adding the synonyms and subtypes to them. These are the simplified denominations used in that case:
\begin{itemize}
\item ``Enlarged cardiomediastinum": \say{enlarged cardiomediastinum}
\item ``Cardiomegaly": \say{cardiomegaly}
\item ``Atelectasis": \say{atelectasis}
\item ``Consolidation": \say{consolidation}
\item ``Edema": \say{lung edema}
\item ``Fracture": \say{fracture}
\item ``Lung lesion": \say{lung lesion}
\item ``Pleural effusion": \say{pleural effusion}
\item ``Pneumonia": \say{pneumonia}
\item ``Pneumothorax": \say{pneumothorax}
\item ``Support devices": \say{medical equipment or support device}
\item ``Lung opacity": \say{lung opacity}
\item ``Pleural other": \say{abnormalities in the pleura (not pleural effusion)}
\end{itemize}

In \cref{sec:adapt}, we mentioned the MAPLEZ labeler's adaptation to other medical reports modalities. A senior radiologist specializing in abdominal imaging selected The CT/MRI/PET labels. These are the denominations that we used in our prompts and the chosen abnormalities for each modality, listed in parenthesis:
\begin{itemize}
\item ``Lung lesion" (CT): \say{lung lesion};
\item ``Liver lesion" (CT, MRI): \say{liver lesion};
\item ``Kidney lesion" (CT, MRI): \say{kidney lesion};
\item ``Adrenal gland abnormality" (CT, MRI): \say{adrenal gland abnormality};
\item ``Pleural effusion" (CT): \say{pleural effusion};
\item ``Hypermetabolic abnormality in the thorax'' (PET): \say{hypermetabolic abnormality in the thorax};
\item ``Hypermetabolic abnormality in the abdomen'' (PET): \say{hypermetabolic abnormality in the abdomen};
\item ``Hypermetabolic abnormality in the pelvis'' (PET): \say{hypermetabolic abnormality in the pelvis}.
\end{itemize}

\subsection{Location keywords}

\subsubsection{Location keywords for training a classifier}
\label{sec:classifierwords}

The following words were all of the location category keywords that were selected for employment during the training process for each abnormality:
%keywords
\begin{itemize}
\item ``Cardiomegaly": \say{right}, \say{left}, \say{upper}, \say{lower}, \say{base}, \say{ventricle}, \say{atrium}    
\item ``Lung Opacity":\say{apical}, \say{middle}, \say{right}, \say{left}, \say{upper}, \say{lower}, \say{base}, \say{lateral}, \say{perihilar}, \say{retrocardiac},
\item ``Edema":\say{apical}, \say{middle}, \say{right}, \say{left}, \say{upper}, \say{lower}, \say{base}, \say{lateral}, \say{perihilar}, \say{retrocardiac},
\item ``Consolidation":\say{apical}, \say{middle}, \say{right}, \say{left}, \say{upper}, \say{lower}, \say{base}, \say{lateral}, \say{perihilar}, \say{retrocardiac},
\item ``Atelectasis":\say{apical}, \say{middle}, \say{right}, \say{left}, \say{upper}, \say{lower}, \say{base}, \say{lateral}, \say{perihilar}, \say{retrocardiac},
\item ``Pneumothorax":\say{apical}, \say{middle}, \say{right}, \say{left}, \say{upper}, \say{lower}, \say{base}, \say{lateral}, \say{perihilar}, \say{retrocardiac},
\item ``Pleural Effusion":\say{apical}, \say{middle}, \say{right}, \say{left}, \say{upper}, \say{lower}, \say{base}, \say{lateral}, \say{perihilar}, \say{retrocardiac},
\item ``Fracture":\say{middle}, \say{right}, \say{left}, \say{upper}, \say{lower}, \say{lateral}, \say{posterior}, \say{anterior}, \say{rib}, \say{third}, \say{fourth}, \say{fifth}, \say{sixth}, \say{seventh}, \say{eighth}, \say{ninth}, \say{clavicular}, \say{spine}
\end{itemize}

These are the words that, when present in the list of free-form text annotated by the model, made each of the keyword categories positive (replacement list):
\begin{itemize}
\item \say{right}: \say{bilateral}, \say{bilaterally}, \say{lungs}, \say{biapical}, \say{apices}, \say{apexes}, \say{bases}, \say{bibasilar}, \say{chest walls}, \say{ventricles}, \say{atriums}, \say{clavicles}
\item \say{left}: \say{bilateral}, \say{bilaterally}, \say{lungs}, \say{biapical}, \say{apices}, \say{apexes}, \say{retrocardiac}, \say{bases}, \say{bibasilar}, \say{chest walls}, \say{ventricles}, \say{atriums}, \say{clavicles}
\item \say{apical}:\say{biapical}, \say{apices}, \say{apexes}, \say{apical}
\item \say{lower}:\say{retrocardiac}
\item \say{middle}:\say{mid}
\item \say{base}:\say{basilar}, \say{bases}, \say{bibasilar}
\item \say{ventricle}:\say{ventricles}
\item \say{atrium}:\say{atriums}
\item \say{posterior}:\say{posterolateral}
\item \say{lateral}:\say{posterolateral}
\item \say{rib}:\say{ribs}
\item \say{third}:\say{3rd}, \say{3}
\item \say{fourth}:\say{4th}, \say{4}, \say{four}
\item \say{fifth}:\say{5th}, \say{5}, \say{five}
\item \say{sixth}:\say{6th}, \say{6}, \say{six}
\item \say{seventh}:\say{7th}, \say{7}, \say{seven}
\item \say{eighth}:\say{8th}, \say{8}, \say{eight}
\item \say{ninth}:\say{9th}, \say{9}, \say{nine}
\item \say{clavicular}:\say{clavicle}, \say{clavicles}
\item \say{spine}:\say{vertebrae}, \say{vertebra}, \say{vertebral}
\end{itemize}

The items of the following list are location keywords that, when positive, prevented the keywords in the associated list on the right from being negative:
\begin{itemize}
\item \say{apical}: \say{upper},
\item \say{middle}: \say{upper}, \say{lower},
\item \say{upper}: \say{apical}, \say{middle},
\item \say{lower}: \say{base}, \say{retrocardiac}, \say{middle}, \say{base}:\ say{lower}, \say{retrocardiac}, 
\item \say{perihilar}: \say{atrium}, \say{ventricle},
\item \say{ventricle}: \say{retrocardiac}, \say{perihilar}, \say{lower}, \say{left}, 
\item \say{atrium}: \say{retrocardiac}, \say{perihilar}, \say{lower}, 
\item \say{posterior}: \say{rib}, 
\item \say{anterior}: \say{rib}
\item \say{third}: \say{rib}, 
\item \say{fourth}: \say{rib}, 
\item \say{fifth}: \say{rib}, 
\item \say{sixth}: \say{rib}, 
\item \say{seventh}: \say{rib}, 
\item \say{eighth}: \say{rib}, 
\item \say{ninth}: \say{rib},
\item \say{clavicular}: \say{upper}, \say{apical}, 
\item \say{spine}: \say{perihilar}, 
\item \say{retrocardiac}: \say{left}, \say{lower}, \say{perihilar}, \say{base}, \say{atrium}, \say{ventricle},
\end{itemize}

The items of the following list are location keywords that, when positive, made the keywords in the associated list on the right negative unless the associated keyword was prevented from being negative (as shown in the list above):
\begin{itemize}
\item \say{apical}: \say{lower}, \say{base}, \say{retrocardiac}, \say{ventricle}, \say{atrium} 
\item \say{middle}: \say{apical}, \say{base}, \say{ventricle}, \say{atrium} 
\item \say{right}: \say{left}, \say{retrocardiac}, \say{atrium}, \say{ventricle}, 
\item \say{left}: \say{right}, \say{atrium}, 
\item \say{upper}: \say{lower}, \say{base}, \say{retrocardiac}, \say{ventricle}, \say{atrium}, 
\item \say{lower}: \say{upper}, \say{apical}, \say{clavicular}, 
\item \say{base}: \say{upper}, \say{apical}, \say{middle}, 
\item \say{lateral}: \say{perihilar}, \say{retrocardiac}, \say{atrium}, \say{spine}, 
\item \say{perihilar}: \say{lateral},
\item \say{ventricle}: \say{right}, \say{atrium}, \say{upper}, \say{apical},
\item \say{atrium}: \say{ventricle}, \say{upper}, \say{apical}, \say{lateral},
\item \say{posterior}: \say{clavicular}, \say{anterior}, \say{spine},         
\item \say{anterior}: \say{clavicular}, \say{posterior}, \say{spine}, 
\item \say{rib}: \say{clavicular}, \say{spine},
\item \say{third}: \say{fourth}, \say{fifth}, \say{sixth}, \say{seventh}, \say{eighth}, \say{ninth}, \say{clavicular}, \say{spine}, 
\item \say{fourth}: \say{third}, \say{fifth}, \say{sixth}, \say{seventh}, \say{eighth}, \say{ninth}, \say{clavicular}, \say{spine}, 
\item \say{fifth}: \say{third}, \say{fourth}, \say{sixth}, \say{seventh}, \say{eighth}, \say{ninth}, \say{clavicular}, \say{spine}, 
\item \say{sixth}: \say{third}, \say{fourth}, \say{fifth}, \say{seventh}, \say{eighth}, \say{ninth}, \say{clavicular}, \say{spine}, 
\item \say{seventh}: \say{third}, \say{fourth}, \say{fifth}, \say{sixth}, \say{eighth}, \say{ninth}, \say{clavicular}, \say{spine}, 
\item \say{eighth}: \say{third}, \say{fourth}, \say{fifth}, \say{sixth}, \say{seventh}, \say{ninth}, \say{clavicular}, \say{spine}, 
\item \say{ninth}: \say{third}, \say{fourth}, \say{fifth}, \say{sixth}, \say{seventh}, \say{eighth}, \say{clavicular}, \say{spine},
\item \say{clavicular}: \say{third}, \say{fourth}, \say{fifth}, \say{sixth}, \say{seventh}, \say{eighth}, \say{ninth}, \say{spine}, \say{rib}, \say{lower},
\item \say{spine}: \say{third}, \say{fourth}, \say{fifth}, \say{sixth}, \say{seventh}, \say{eighth}, \say{ninth}, \say{clavicular}, \say{rib}, \say{lateral}, 
\item \say{retrocardiac}:  \say{right}, \say{lateral}, \say{upper}, \say{apical},
\end{itemize}

\subsubsection{Location keywords for labeler evaluation}
\label{sec:evalwords}

The selection and use of keywords and replacement words are performed similarly to \cref{sec:methods2}, but also analyzing the ground truth annotations to decide what keywords to use. Furthermore, the vocabulary of location expressions provided by the Medical-Diff-VQA labeler is limited. We built a list of frequent keywords in that dataset and performed evaluation only for those, without allowing for the replacement of words. The considered keywords were: \say{right}, \say{left},\say{upper}, \say{lower}, \say{base}, \say{apical}, \say{retrocardiac}, \say{rib},\say{middle}, and results in \cref{tab:locationsimple}.  

The following list presents the keywords that were considered location categorical labels when the evaluation of the location outputs of the labelers, presented in \cref{tab:probs}, was calculated:
\begin{itemize}
\item ``Cardiomegaly": \say{right}, \say{left}, \say{upper}, \say{lower}, \say{base}, \say{ventricle}, \say{atrium}
\item ``Lung opacity": \say{apical}, \say{middle}, \say{right}, \say{left}, \say{upper}, \say{lower}, \say{base}, \say{lateral}, \say{perihilar}, \say{retrocardiac}
\item ``Edema": \say{apical}, \say{middle}, \say{right}, \say{left}, \say{upper}, \say{lower}, \say{base}, \say{lateral}, \say{perihilar}, \say{retrocardiac}
\item ``Consolidation": \say{apical}, \say{middle}, \say{right}, \say{left}, \say{upper}, \say{lower}, \say{base}, \say{lateral}, \say{perihilar}, \say{retrocardiac}
\item ``Atelectasis": \say{apical}, \say{middle}, \say{right}, \say{left}, \say{upper}, \say{lower}, \say{base}, \say{lateral}, \say{perihilar}, \say{retrocardiac}
\item ``Pneumothorax": \say{apical}, \say{middle}, \say{right}, \say{left}, \say{upper}, \say{lower}, \say{base}, \say{lateral}, \say{perihilar}, \say{retrocardiac}
\item ``Pleural effusion": \say{apical}, \say{middle}, \say{right}, \say{left}, \say{upper}, \say{lower}, \say{base}, \say{lateral}, \say{perihilar}, \say{retrocardiac}, \say{fissure}
\item ``Fracture": \say{middle}, \say{right}, \say{left}, \say{upper}, \say{lower}, \say{lateral}, \say{posterior}, \say{anterior}, \say{rib}, \say{third}, \say{fourth}, \say{fifth}, \say{sixth}, \say{seventh}, \say{eighth}, \say{ninth}, \say{clavicular}, \say{spine}
\end{itemize}

These are keywords and the respective words from the text that, when present, also made the anatomical location annotation for that keyword to be present (replacement list):

\begin{itemize}
\item \say{left}: \say{leftward}, \say{left-sided}, \say{left-side}, \say{lingula}, \say{bilateral}, \say{bilaterally}, \say{lungs}, \say{biapical}, \say{apices}, \say{apexes}, \say{retrocardiac}, \say{bases}, \say{bibasilar}, \say{ventricles}, \say{atriums}, \say{clavicles}
\item \say{right}: \say{right-sided}, \say{right-side}, \say{rightward}, \say{bilateral}, \say{bilaterally}, \say{lungs}, \say{biapical}, \say{apices}, \say{apexes}, \say{bases}, \say{bibasilar}, \say{ventricles}, \say{atriums}, \say{clavicles}
\item \say{lower}: \say{infrahilar}, \say{lingula}, \say{retrocardiac}
\item \say{upper}: \say{suprahilar}
\item \say{fissure}: \say{fissural}, \say{fissures}
\item \say{perihilar}: \say{central}, \say{medial}, \say{medially}
\item \say{base}: \say{costrophenic}, \say{basilar}, \say{bases}, \say{bibasilar}
\item \say{apical}: \say{biapical}, \say{apices}, \say{apexes}, \say{apex}, \say{apically}, \say{apicolateral}
\item \say{middle}: \say{mid}
\item \say{ventricle}: \say{ventricular}, \say{ventricles}
\item \say{atrium}: \say{atriums}
\item \say{spine}: \say{vertebrae}, \say{vertebra}, \say{vertebral}, \say{spinal}
\item \say{clavicular}: \say{clavicles}, \say{clavicle}
\item \say{posterior}: \say{posterolateral}
\item \say{lateral}: \say{posterolateral}, \say{apicolateral}
\item \say{rib}: \say{ribs}
\item \say{third}: \say{3rd}, \say{3}
\item: \say{fourth}: \say{4th}, \say{4}, \say{four}
\item \say{fifth}: \say{5th}, \say{5}, \say{five}
\item \say{sixth}: \say{6th}, \say{6}, \say{six}
\item \say{seventh}: \say{7th}, \say{7}, \say{seven}
\item \say{eighth}: \say{8th}, \say{8}, \say{eight}
\item \say{ninth}: \say{9th}, \say{9}, \say{nine}
\end{itemize}       

The location keywords employed for the evaluation of the adaptability of the prompt system to other modalities were:
\begin{itemize}
\item ``Lung lesion": \say{right}, \say{left}, \say{upper}, \say{lower}, \say{middle},
\item ``Liver lesion": \say{right}, \say{left},
\item ``Kidney lesion": \say{right}, \say{left},
\item ``Adrenal gland abnormality": \say{right}, \say{left},
\item ``Pleural effusion": \say{right}, \say{left},
   \end{itemize}     

\subsection{Dataset Annotations}
\label{sec:datasetstatistics}

Reports from the non-public datasets for which we calculated MAPLEZ annotations were obtained retrospectively, and consent was waived. We automatically annotated 227,827 reports from the MIMIC-CXR dataset~\citep{mimiccxr,mimiccxrjpg,physionet,mimicxrjpgphysionet,mimiccxrphysionet} and 112,120 reports from the NIH ChestXray14 dataset~\citep{nihdataset}. We did not use the annotations of the NIH dataset for training a classifier in this paper, but we made them publicly available.

We trained our classifiers with all of the images with a \say{ViewPosition} value of \say{AP} or \say{PA} from the training set of the MIMIC-CXR dataset~\citep{mimiccxr,mimiccxrjpg,physionet,mimicxrjpgphysionet,mimiccxrphysionet}. The statistics of the employed annotations are presented in \cref{tab:statisticsannotations,tab:statisticsannotations2}.

\begin{table*}[!htbp]
  \caption{The number of cases fitting each category for the categorical labels and probabilities for the annotations used to train the classifier with the MIMIC-CXR dataset. The percentages are given related to the total number of cases, 237,973. Pr.=``Present"; Abs.=``Absent"; Unc.=``Uncertain"; NM=``Not mentioned"; St.=``Stable"; P.$\mu$=probability average, when not ``Stable"; P.St.=Probability ``Stable"; P.$\leq$10\%=Cases with probability less than 10\%; P.$\leq$90\%=Cases with probability greater than 90\%; C.St.=combined cases for ``Stable" from categorical presence annotations and probability annotations; Abn.=Abnormality; Atel.=``Atelectasis"; Card.=``Cardiomegaly"; Cons.=``Consolidation"; Fract.=``Fracture"; Opac.=``Lung opacity"; Effus.=``Pleural Effusion"; PTX=``Pneumothorax".}
  \label{tab:statisticsannotations}
  \centering
  \begin{tabular}{@{}llrrrrrrrrrr@{}}
\toprule
Abn. & Labeler & Pr. & Ab. & Unc. & NM & St. & P.$\mu$ & P.St. & P.$\leq$10\% & P.$\geq$90\% & C.St. \\
\midrule
Atel. & MAPLEZ & 25\% & 19\% & 9.6\% & 46\% & 0.3\% & 33\% & 0.5\% & 63\% & 22\% & 0.5\% \\
Atel. & VQA & 24\% & 0.5\% & 8.1\% & 67\% & - & 36\% & - & 68\% & 24\% & - \\
Atel. & CheXpert & 20\% & 0.7\% & 4.5\% & 75\% & - & - & - & - & - & - \\
Atel. & CheXbert & 27\% & 0.4\% & 5.2\% & 68\% & - & - & - & - & - & - \\
Card. & MAPLEZ & 21\% & 44\% & 5.0\% & 18\% & 12\% & 27\% & 13\% & 61\% & 20\% & 13\% \\
Card. & VQA & 16\% & 0.0\% & 0.2\% & 84\% & - & 24\% & - & 84\% & 16\% & - \\
Card. & CheXpert & 19\% & 6.7\% & 2.6\% & 71\% & - & - & - & - & - & - \\
Card. & CheXbert & 26\% & 25\% & 6.3\% & 43\% & - & - & - & - & - & - \\
Cons. & MAPLEZ & 9.4\% & 38\% & 12\% & 40\% & 0.8\% & 21\% & 0.9\% & 75\% & 11\% & 0.9\% \\
Cons. & VQA & 11\% & 32\% & 7.0\% & 50\% & - & 20\% & - & 82\% & 11\% & - \\
Cons. & CheXpert & 4.7\% & 3.5\% & 1.9\% & 90\% & - & - & - & - & - & - \\
Cons. & CheXbert & 5.3\% & 24\% & 2.9\% & 68\% & - & - & - & - & - & - \\
Edema & MAPLEZ & 9.3\% & 30\% & 9.6\% & 50\% & 1.0\% & 16\% & 1.4\% & 80\% & 6.2\% & 1.4\% \\
Edema & VQA & 16\% & 0.0\% & 2.6\% & 81\% & - & 24\% & - & 81\% & 16\% & - \\
Edema & CheXpert & 12\% & 11\% & 5.8\% & 72\% & - & - & - & - & - & - \\
Edema & CheXbert & 13\% & 20\% & 5.5\% & 61\% & - & - & - & - & - & - \\
Fract. & MAPLEZ & 3.2\% & 9.2\% & 0.3\% & 87\% & 0.2\% & 2.2\% & 0.2\% & 98\% & 2.0\% & 0.2\% \\
Fract. & VQA & 3.2\% & 2.3\% & 0.2\% & 94\% & - & 13\% & - & 97\% & 3.2\% & - \\
Fract. & CheXpert & 2.0\% & 0.4\% & 0.2\% & 97\% & - & - & - & - & - & - \\
Fract. & CheXbert & 3.6\% & 2.2\% & 0.2\% & 94\% & - & - & - & - & - & - \\
Opac. & MAPLEZ & 49\% & 31\% & 9.4\% & 8.1\% & 2.5\% & 54\% & 3.8\% & 38\% & 46\% & 3.8\% \\
Opac. & VQA & 54\% & 0.0\% & 2.4\% & 44\% & - & 57\% & - & 44\% & 54\% & - \\
Opac. & CheXpert & 22\% & 1.3\% & 1.6\% & 75\% & - & - & - & - & - & - \\
Opac. & CheXbert & 27\% & 2.4\% & 0.1\% & 71\% & - & - & - & - & - & - \\
Effus. & MAPLEZ & 20\% & 52\% & 9.0\% & 17\% & 1.3\% & 27\% & 1.8\% & 69\% & 21\% & 1.8\% \\
Effus. & VQA & 24\% & 46\% & 4.8\% & 24\% & - & 30\% & - & 71\% & 24\% & - \\
Effus. & CheXpert & 24\% & 11\% & 2.5\% & 62\% & - & - & - & - & - & - \\
Effus. & CheXbert & 27\% & 48\% & 3.3\% & 22\% & - & - & - & - & - & - \\
PTX & MAPLEZ & 4.1\% & 67\% & 0.6\% & 28\% & 0.4\% & 5.2\% & 0.5\% & 95\% & 3.8\% & 0.5\% \\
PTX & VQA & 4.2\% & 57\% & 0.4\% & 39\% & - & 8.3\% & - & 95\% & 4.2\% & - \\
PTX & CheXpert & 4.6\% & 18\% & 0.5\% & 76\% & - & - & - & - & - & - \\
PTX & CheXbert & 4.0\% & 60\% & 0.7\% & 35\% & - & - & - & - & - & - \\
\bottomrule
  \end{tabular}
\end{table*}

\begin{table*}[!htbp]
  \caption{The number of cases fitting each category for the severity and location annotations used to train the classifier with the MIMIC-CXR dataset. Most percentages are given as a percentage of the total number of cases, 237,973, but Loc.+ and Loc.- are given as a percentage of how many location labels were possible for the cases with at least one location present. Sevs.=Severities present; Locs.=cases with at least one location present; Loc.+=number of positive locations in cases with at least one location present; Loc.-=number of negative locations in cases with at least one location present; Abn.=Abnormality; Atel.=``Atelectasis"; Card.=``Cardiomegaly"; Cons.=``Consolidation"; Fract.=``Fracture"; Opac.=``Lung opacity"; Effus.=``Pleural Effusion"; PTX=``Pneumothorax".}
  \label{tab:statisticsannotations2}
  \centering
  \begin{tabular}{@{}llrrrrrrrrrrr@{}}
\toprule
Abn. & Labeler & Sevs. & Mild & Moderate & Severe & Locs. & Loc.+ & Loc.- \\
\midrule
Atel. & MAPLEZ & 8.9\% & 6.9\% & 1.7\% & 0.3\% & 45\% & 27\% & 27\% \\
Atel. & VQA & 5.8\% & 5.0\% & 0.4\% & 0.3\% & 29\% & 24\% & 28\% \\
Card. & MAPLEZ & 16\% & 8.3\% & 7.0\% & 1.2\% & 0.3\% & 21\% & 34\% \\
Card. & VQA & 10\% & 3.7\% & 5.3\% & 1.3\% & 0.0\% & 17\% & 36\% \\
Cons. & MAPLEZ & 1.8\% & 0.6\% & 0.6\% & 0.6\% & 45\% & 27\% & 27\% \\
Cons. & VQA & 0.5\% & 0.2\% & 0.0\% & 0.3\% & 29\% & 24\% & 28\% \\
Edema & MAPLEZ & 8.1\% & 5.3\% & 2.4\% & 0.4\% & 45\% & 27\% & 27\% \\
Edema & VQA & 9.9\% & 7.6\% & 1.8\% & 0.5\% & 29\% & 24\% & 28\% \\
Fract. & MAPLEZ & 0.1\% & 0.1\% & 0.0\% & 0.0\% & 53\% & 12\% & 9.2\% \\
Fract. & VQA & 0.1\% & 0.1\% & 0.0\% & 0.0\% & 44\% & 10\% & 8.2\% \\
Opac. & MAPLEZ & 21\% & 13\% & 6.2\% & 1.5\% & 45\% & 27\% & 27\% \\
Opac. & VQA & 17\% & 14\% & 2.2\% & 1.2\% & 29\% & 24\% & 28\% \\
Effus. & MAPLEZ & 13\% & 6.2\% & 6.7\% & 0.4\% & 53\% & 26\% & 25\% \\
Effus. & VQA & 17\% & 13\% & 3.6\% & 0.3\% & 43\% & 22\% & 23\% \\
PTX & MAPLEZ & 2.1\% & 1.4\% & 0.6\% & 0.1\% & 47\% & 27\% & 27\% \\
PTX & VQA & 2.1\% & 1.8\% & 0.3\% & 0.0\% & 32\% & 23\% & 28\% \\

\bottomrule
  \end{tabular}
\end{table*}

\subsubsection{Manual Ground Truth Annotations}
\label{sec:handannotations}

The annotation ground truth for the MIMIC-CXR and NIH ChestXray14 datasets was performed by an image analysis researcher with supervision from a radiologist. The researcher and the radiologist had an initial meeting to discuss the initial annotation of six random reports. The researcher was in constant communication with the radiologist to clarify annotation questions. Reports from the non-public datasets were obtained retrospectively, and consent was waived. 

During annotation, location prepositional phrases are annotated from the preposition to the end of the prepositional phrase. Locations used to compare the intensity of abnormality, for example, \say{right greater than left}, were omitted. The report was annotated for severity with the maximum of all the severities mentioned for that abnormality. Probabilities were not annotated. The statistics of the ground truth annotations we labeled are presented in \cref{tab:statisticsgts,tab:statisticsgts2}. Severities were not annotated for the CT, MRI, and PET reports. For the PET reports, locations were not annotated, and only ``Present" and ``Absent" categorical labels were used.

\begin{table*}[!htpb]
  \caption{Statistics of the test sets we hand-annotated, related to categorical labels. NM=``Not mentioned"; \#=number of reports }
  \label{tab:statisticsgts}
  \centering
  \begin{tabular}{@{}llrrrrrr@{}}
\toprule
Abnormality & Data & ``Present" & ``Absent" & ``Uncertain" & NM & ``Stable" & \# \\
\midrule
Atelectasis & NIH & 9 (4.5\%) & 26 (13\%) & 18 (9.0\%) & 138 (69\%) & 9 (4.5\%) & 200 \\
Atelectasis & MIMIC & 63 (18\%) & 52 (15\%) & 41 (12\%) & 180 (51\%) & 14 (4.0\%) & 350 \\
Cardiomegaly & NIH & 13 (6.5\%) & 40 (20\%) & 8 (4.0\%) & 92 (46\%) & 47 (24\%) & 200 \\
Cardiomegaly & MIMIC & 117 (33\%) & 91 (26\%) & 15 (4.3\%) & 68 (19\%) & 59 (17\%) & 350 \\
Consolidation & NIH & 42 (21\%) & 44 (22\%) & 28 (14\%) & 77 (38\%) & 9 (4.5\%) & 200 \\
Consolidation & MIMIC & 31 (8.9\%) & 117 (33\%) & 59 (17\%) & 126 (36\%) & 17 (4.9\%) & 350 \\
Enlarged cardiomediastinum & NIH & 22 (11\%) & 33 (16\%) & 6 (3.0\%) & 92 (46\%) & 47 (24\%) & 200 \\
Enlarged cardiomediastinum & MIMIC & 124 (35\%) & 58 (17\%) & 8 (2.3\%) & 94 (27\%) & 66 (19\%) & 350 \\
Fracture & NIH & 1 (0.5\%) & 21 (10\%) & 0 (0.0\%) & 172 (86\%) & 6 (3.0\%) & 200 \\
Fracture & MIMIC & 18 (5.1\%) & 39 (11\%) & 2 (0.6\%) & 282 (81\%) & 9 (2.6\%) & 350 \\
Edema & NIH & 3 (1.5\%) & 26 (13\%) & 12 (6.0\%) & 150 (75\%) & 9 (4.5\%) & 200 \\
Edema & MIMIC & 96 (27\%) & 104 (30\%) & 15 (4.3\%) & 123 (35\%) & 12 (3.4\%) & 350 \\
Lung lesion & NIH & 17 (8.5\%) & 29 (14\%) & 5 (2.5\%) & 142 (71\%) & 7 (3.5\%) & 200 \\
Lung lesion & MIMIC & 18 (5.1\%) & 58 (17\%) & 6 (1.7\%) & 255 (73\%) & 13 (3.7\%) & 350 \\
Lung lesion & CT & 23 (57\%) & 5 (12\%) & 0 (0.0\%) & 10 (25\%) & 2 (5.0\%) & 40 \\
Lung opacity & NIH & 105 (52\%) & 23 (12\%) & 17 (8.5\%) & 52 (26\%) & 3 (1.5\%) & 200 \\
Lung opacity & MIMIC & 258 (74\%) & 38 (11\%) & 4 (1.1\%) & 39 (11\%) & 11 (3.1\%) & 350 \\
Pleural effusion & NIH & 43 (22\%) & 36 (18\%) & 17 (8.5\%) & 95 (48\%) & 9 (4.5\%) & 200 \\
Pleural effusion & MIMIC & 106 (30\%) & 135 (39\%) & 28 (8.0\%) & 67 (19\%) & 14 (4.0\%) & 350 \\
Pleural effusion & CT & 8 (20\%) & 21 (52\%) & 1 (2.5\%) & 10 (25\%) & 0 (0.0\%) & 40 \\
Pleural other & NIH & 8 (4.0\%) & 17 (8.5\%) & 4 (2.0\%) & 165 (82\%) & 6 (3.0\%) & 200 \\
Pleural other & MIMIC & 14 (4.0\%) & 42 (12\%) & 2 (0.6\%) & 281 (80\%) & 11 (3.1\%) & 350 \\
Pneumothorax & NIH & 25 (12\%) & 51 (26\%) & 1 (0.5\%) & 114 (57\%) & 9 (4.5\%) & 200 \\
Pneumothorax & MIMIC & 9 (2.6\%) & 206 (59\%) & 1 (0.3\%) & 126 (36\%) & 8 (2.3\%) & 350 \\
Support device & NIH & 151 (76\%) & 8 (4.0\%) & 1 (0.5\%) & 38 (19\%) & 2 (1.0\%) & 200 \\
Support device & MIMIC & 161 (46\%) & 1 (0.3\%) & 0 (0.0\%) & 184 (53\%) & 4 (1.1\%) & 350 \\
Liver lesion & CT & 20 (50\%) & 11 (28\%) & 0 (0.0\%) & 3 (7.5\%) & 6 (15\%) & 40 \\
Liver lesion & MRI & 23 (57\%) & 9 (22\%) & 0 (0.0\%) & 5 (12\%) & 3 (7.5\%) & 40 \\
Kidney lesion & CT & 8 (20\%) & 19 (48\%) & 1 (2.5\%) & 10 (25\%) & 2 (5.0\%) & 40 \\
Kidney lesion & MRI & 21 (52\%) & 6 (15\%) & 1 (2.5\%) & 11 (28\%) & 1 (2.5\%) & 40 \\
Adrenal gland abnormality & CT & 8 (20\%) & 21 (52\%) & 1 (2.5\%) & 8 (20\%) & 2 (5.0\%) & 40 \\
Adrenal gland abnormality & MRI & 9 (22\%) & 18 (45\%) & 0 (0.0\%) & 12 (30\%) & 1 (2.5\%) & 40 \\
Hypermet. thorax & PET & 25 (64\%) & 14 (36\%) & - & - & - & 39 \\
Hypermet. abdomen & PET & 16 (41\%) & 23 (59\%) & - & - & - & 39 \\
Hypermet. pelvis & PET & 13 (33\%) & 26 (67\%) & - & - & - & 39 \\
\bottomrule
  \end{tabular}
\end{table*}

\begin{table*}[!htpb]
  \caption{ Statistics of the test sets that were hand-annotated by us, related to severity and location. Sevs.=total of severities present; Loc.+=number of positive location keywords; Loc.\#=total possible number of positive location keywords.}
  \label{tab:statisticsgts2}
  \centering
  \begin{tabular}{@{}llrrrrrr@{}}
\toprule
Abnormality & Data & Sevs. & Mild & Moderate & Severe & Loc.+ & Loc.\# \\
\midrule
Atelectasis & NIH & 6 (3.0\%) & 6 (3.0\%) & 0 (0.0\%) & 0 (0.0\%) & 60 (3.0\%) & 2000 \\
Atelectasis & MIMIC & 24 (6.9\%) & 18 (5.1\%) & 3 (0.9\%) & 3 (0.9\%) & 255 (7.3\%) & 3500 \\
Cardiomegaly & NIH & 7 (3.5\%) & 4 (2.0\%) & 2 (1.0\%) & 1 (0.5\%) & 0 (0.0\%) & 1400 \\
Cardiomegaly & MIMIC & 76 (22\%) & 35 (10\%) & 22 (6.3\%) & 19 (5.4\%) & 0 (0.0\%) & 2450 \\
Consolidation & NIH & 8 (4.0\%) & 5 (2.5\%) & 3 (1.5\%) & 0 (0.0\%) & 131 (6.6\%) & 2000 \\
Consolidation & MIMIC & 9 (2.6\%) & 5 (1.4\%) & 1 (0.3\%) & 3 (0.9\%) & 197 (5.6\%) & 3500 \\
Enlarged cardiomediastinum & NIH & 8 (4.0\%) & 5 (2.5\%) & 2 (1.0\%) & 1 (0.5\%) & 7 (0.5\%) & 1400 \\
Enlarged cardiomediastinum & MIMIC & 70 (20\%) & 33 (9.4\%) & 19 (5.4\%) & 18 (5.1\%) & 7 (0.3\%) & 2450 \\
Fracture & NIH & 0 (0.0\%) & 0 (0.0\%) & 0 (0.0\%) & 0 (0.0\%) & 3 (0.1\%) & 3600 \\
Fracture & MIMIC & 1 (0.3\%) & 1 (0.3\%) & 0 (0.0\%) & 0 (0.0\%) & 40 (0.6\%) & 6300 \\
Edema & NIH & 0 (0.0\%) & 0 (0.0\%) & 0 (0.0\%) & 0 (0.0\%) & 24 (1.2\%) & 2000 \\
Edema & MIMIC & 67 (19\%) & 47 (13\%) & 16 (4.6\%) & 4 (1.1\%) & 65 (1.9\%) & 3500 \\
Lung lesion & NIH & 2 (1.0\%) & 1 (0.5\%) & 0 (0.0\%) & 1 (0.5\%) & 42 (2.1\%) & 2000 \\
Lung lesion & MIMIC & 2 (0.6\%) & 0 (0.0\%) & 0 (0.0\%) & 2 (0.6\%) & 45 (1.3\%) & 3500 \\
Lung lesion & CT & - & - & - & - & 61 (30\%) & 200 \\
Lung opacity & NIH & 19 (9.5\%) & 14 (7.0\%) & 3 (1.5\%) & 2 (1.0\%) & 259 (13\%) & 2000 \\
Lung opacity & MIMIC & 85 (24\%) & 59 (17\%) & 15 (4.3\%) & 11 (3.1\%) & 576 (16\%) & 3500 \\
Pleural effusion & NIH & 25 (12\%) & 17 (8.5\%) & 7 (3.5\%) & 1 (0.5\%) & 85 (3.9\%) & 2200 \\
Pleural effusion & MIMIC & 92 (26\%) & 59 (17\%) & 21 (6.0\%) & 12 (3.4\%) & 223 (5.8\%) & 3850 \\
Pleural effusion & CT & - & - & - & - & 12 (15\%) & 80 \\
Pleural other & NIH & 2 (1.0\%) & 1 (0.5\%) & 0 (0.0\%) & 1 (0.5\%) & 23 (1.0\%) & 2200 \\
Pleural other & MIMIC & 3 (0.9\%) & 3 (0.9\%) & 0 (0.0\%) & 0 (0.0\%) & 31 (0.8\%) & 3850 \\
Pneumothorax & NIH & 16 (8.0\%) & 14 (7.0\%) & 2 (1.0\%) & 0 (0.0\%) & 44 (2.2\%) & 2000 \\
Pneumothorax & MIMIC & 5 (1.4\%) & 4 (1.1\%) & 0 (0.0\%) & 1 (0.3\%) & 17 (0.5\%) & 3500 \\
Support device & NIH & 1 (0.5\%) & 1 (0.5\%) & 0 (0.0\%) & 0 (0.0\%) & 396 (8.6\%) & 4600 \\
Support device & MIMIC & 0 (0.0\%) & 0 (0.0\%) & 0 (0.0\%) & 0 (0.0\%) & 475 (5.9\%) & 8050 \\
Liver lesion & CT & - & - & - & - & 11 (14\%) & 80 \\
Liver lesion & MRI & - & - & - & - & 15 (19\%) & 80 \\
Kidney lesion & CT & - & - & - & - & 10 (12\%) & 80 \\
Kidney lesion & MRI & - & - & - & - & 34 (42\%) & 80 \\
Adrenal gland abnormality & CT & - & - & - & - & 8 (10\%) & 80 \\
Adrenal gland abnormality & MRI & - & - & - & - & 9 (11\%) & 80 \\
\bottomrule
  \end{tabular}
\end{table*}

\subsubsection{External datasets annotation adaptations}
\paragraph{REFLACX adaptations}
\label{sec:reflacxchanges}
For the analyses using the \textit{REFLACX} dataset, these were the merges done to labels for including subtypes of abnormalities:
\begin{itemize}
\item ``Lung opacity" (phase 2 and 3): \say{Interstitial lung disease}, \say{Pulmonary edema},\say{Consolidation}, \say{Atelectasis},\say{Enlarged hilum}, \say{Groundglass opacity}, \say{Lung nodule or mass}
\item ``Lung opacity" (phase 1): \say{Fibrosis}, \say{Pulmonary edema}, \say{Consolidation}, \say{Atelectasis}, \say{Groundglass opacity}, \say{Mass},\say{Nodule}
\end{itemize}

\paragraph{Medical-Diff-VQA adaptations}
\label{sec:vqachanges}
For the analyses using the Medical-Diff-VQA labeler, these were the merges done to the annotations:
\begin{itemize}
\item ``Cardiomegaly": \say{cardiomegaly} (primary), \say{enlargement of the cardiac silhouette};
\item ``Edema": \say{edema} (primary), \say{vascular congestion}, \say{heart failure}, \say{hilar congestion};
\item ``Lung opacity": \say{lung opacity} (primary), \say{consolidation}, \say{edema}, \say{vascular congestion}, \say{atelectasis}, \say{heart failure}, \say{hilar congestion}, \say{pneumonia};
\item ``Consolidation": \say{consolidation} (primary), \say{pneumonia}.
\end{itemize}

For the analyses using the Medical-Diff-VQA labeler, these were the rules used to convert expressions to probabilities:

\begin{itemize}
\item 100\%: \say{positive}, \say{change in}
\item 70\%: \say{probable},\say{probabl}, \say{likely}, \say{may}, \say{could}, \say{potential}
\item 50\%: \say{might}, \say{possibl}, \say{possible}
\item 30\%: \say{unlikely},\say{not exclude}, \say{cannot be verified}, \say{difficult rule out}, \say{not ruled out}, \say{cannot be accurately assessed}, \say{not rule out}, \say{impossible exclude}, \say{cannot accurately assesses}, \say{cannot be assessed}, \say{cannot be identified}, \say{cannot be confirmed}, \say{cannot be evaluated}, \say{difficult exclude};
\item 0\%: \say{no}, \say{without}, \say{negative}, \say{clear of}, \say{exclude}, \say{lack of}, \say{rule out}, \say{ruled out}.
\end{itemize}

For the analyses using the Medical-Diff-VQA labeler, these were the rules used to convert severity words to severity classes:

\begin{itemize}
\item ``Mild": \say{mild},\say{minimal},\say{small}, \say{subtle},\say{minimally}, \say{mildly}, \say{trace}, \say{minor};
\item ``Moderate": \say{moderate}, \say{mild to moderate}, \say{moderately};
\item ``Severe": \say{massive}, \say{severe}, \say{moderate to severe}, \say{moderate to large};
\item No severity: \say{increasing}, \say{decreasing}, \say{acute}.
\end{itemize}

\subsection{LLM inference}

We employed a temperature of 0 for our \gls*{LLM} inference since a higher temperature hurts performance~\citep{vicunanih}. Our inference for roughly 350,000 reports we automatically annotated took about 15 days to finish when using 22 80GB A100 GPUs, corresponding to around 40 s per report with 2 A100 GPUs. We highlight there have been optimizations to LLM inference (e.g., vLLM~\citep{vllm} or llama.cpp~\citep{llamacpp}) with a potential of speeding up our code by $>20\times$. The inference of the model needs about 150 GB of VRAM. We based our \gls*{LLM} code off of the FastChat open platform~\citep{fastchat}.

\subsection{Classifier training}
\label{sec:trainingexplained}

To adapt the \gls*{CNN} to our tasks, we modified its final layer, replacing it with classifier heads, each dedicated to an abnormality and with a single hidden layer. The multi-task implementation was considered in two ways: additional outputs to each abnormality head or new localization and severity heads for each abnormality. For the MAPLEZ method, we employed the probability of presence annotations and $\lambda_{loc}=0.01$ with independent heads for the location outputs. Furthermore, we chose to ignore any abnormalities labeled as ``Stable" for the presence loss. For the Medical-Diff-VQA method, we employed the categorical presence annotations and $\lambda_{loc}=0.001$ with shared heads. We did not use probability or location labels for the CheXpert method because that annotation type is not provided. We performed validation of several hyperparameters for the three compared methods by checking for the highest average \gls*{AUC} on the validation set of the respective method. We considered the following hyperparameters:
\begin{itemize}
\item Models: the ViT-B/32 \gls*{CXR} pre-trained CheXzero~\citep{chexzero} (224$\times$224 input), the resnet50-res512-all \gls*{CXR} pre-trained model from the TorchXRayVision library~\citep{torchxrayvision} (512$\times$512 input), and EfficientNetV2-M~\citep{efficientnet} with ImageNet pre-trained weights (480$\times$480 input);
\item $\lambda_{sev}$: 0, 0.001, 0.01, 0.1, 1, 10;
\item $\lambda_{loc}$: 0, 0.001, 0.01, 0.1, 1, 10;
\item Weight decay: 2e-5, 5e-5, 1e-4, 2e-4, 5e-4, 1e-3;
\item Data augmentation: random erasing~\citep{randomerase}, TrivialAugment~\citep{ta_wide}, AugMix~\citep{augmix},  mixup~\citep{mixup}, CutMix~\citep{cutmix}, AutoAugment~\citep{autoaugment}, and the set of augmentations from the training script of the TorchXRayVision library~\citep{torchxrayvision}. We excluded the ``Equalize" and ``Posterize" transformations from all augmentation methods that included those.
\item exponential-moving-average decay: no decay, 0.9, 0.999, 0.99998
\item Gradient clip (gradient norm maximum): No gradient clipping, 1;
\item Label smoothing~\citep{labelsmoothing}: 0, 0.1;
\item Number of hidden neurons in each classification head: 32, 64, 128, 256, 512, 1024;
\item Annotation type: label, probability.
\end{itemize}

We trained the models with the following hyperparameters:
EfficientNetV2-M, $\lambda_{sev}=0$, no gradient clipping, no exponential-moving-average decay, no label smoothing, a batch size of 16 samples, SGD optimizer, with a momentum of 0.9, a learning rate of 0.0078125 ($0.5/1024\times 16$), weight decay of 5e-5, a total of 30 epochs of training, for which the four first did a linear warmup of the learning rate, the ten first only had the weights of the classifier heads unfrozen, and the 15 last had the learning rate be a tenth of the original value. For models trained with the Medical-Diff-VQA, CheXpert, and CheXbert labelers, the best data augmentation was a combination of the  TorchXRayVision augmentation with the random erasing. For MAPLEZ, it was TrivialAugment with random erasing. We employed 1024 hidden neurons for all classifiers except for the one trained with CheXbert labels, which employed 512 hidden neurons. We based our code and some of the choices of hyperparameters on a training script and training suggestions provided by PyTorch~\citep{pytorchrecipe,pytorch}. We did early stopping by choosing the weights from the epoch with the best \gls*{AUC} score in the respective validation set. We employed bootstrap and permutation tests with 2,000 samples to account for test case sampling variability and to perform two-sided hypothesis testing of the difference in scores. We also calculated the classifier scores with the average score over three random training seeds for the classifiers to account for that variability.

Using an A100 80GB GPU, training a model with the chosen hyperparameters took around one day, and the test set inference, including the initialization of models and datasets, took from three to five minutes. Training required approximately 22GB VRAM.

\subsection{Detailed results for location and severity labels}
\label{sec:locationseveritydetails}
Since the labelers only provide locations for abnormalities they considered ``Present" or ``Uncertain", some of their location mistakes might be due to categorical presence mistakes. To isolate this effect from the results, \cref{tab:probs} present results considering only cases where all labelers and the ground truth annotations agree that the abnormality is present. For completeness of results, we also present the scores for all of the cases in the dataset, without filtering for agreement, in \cref{tab:locationall}. We calculated the location scores by considering all location keywords simultaneously as a batch of binary labels. 

For severity results in \cref{tab:probs}, we only considered cases for which Medical-Diff-VQA, MAPLEZ, MAPLEZ-Generic, and manual ground truth agreed about the presence of the abnormality. Complete scores with precision, recall, confidence intervals, and abnormality subdivision are shown in \cref{tab:severityfull}. We present results for all dataset cases in \cref{tab:severityall}.

\subsection{Calculation of average weights for different abnormalities and datasets}
\label{sec:mvue}

We employed a minimum variance unbiased estimator, sometimes called Markowitz optimization, to combine the scores of individual table rows through a weighted average and to get the average with the least normalized variance. With this calculation, uncertain scores or uncertain table rows have a lower weight in the grouped scores. This calculation assumed that the bootstrap scores followed a Gaussian distribution. We calculated the weights of each row by first normalizing all distributions to have a score of 1. In this case, the new variance is the old one divided by the square of the score. This normalization avoided unfairly giving large weights to lower scores. We then calculated the regular average variance of each row from all of the participating labelers or classifiers. The inverse of the average row variance then gives the unnormalized weight for each row. We present the weights in all tables as a normalization over all non-aggregated rows so they sum to 1.

\section{Expanded Results}
\label{sec:expanded}
In this section, we present the full versions of all the tables in the paper and a few additional tables briefly mentioned in the main text. All table results had 95\% confidence intervals calculated using paired bootstrapping with correction and acceleration with 2,000 samples and p-values with permutation tests with 2,000 samples. P-values are presented to the right of the score in these tables. Our results only employed abnormalities/datasets with more than ten positive cases in their annotation.

\onecolumn

%tab:labelsfull
\begin{landscape}
\begin{longtable}{llrrclcrcrcr}
  \caption{Complete table for the scores of the labelers for the categorical presence annotations. Refer to \cref{tab:labels} for more contextual information. }
  \label{tab:labelsfull}
  \endfirsthead
\endhead
\toprule
Data & Abn. & $N$ & $N^+$ & W & Labeler & Precision ($\uparrow$) & \textit{P} & Recall ($\uparrow$) & \textit{P} & F1 ($\uparrow$) & \textit{P} \\
\midrule
NIH & Atel. & 200 & 27 & 0.009 & CheXpert & \makebox[\widthof{\(^{\scriptscriptstyle{*}\scriptscriptstyle{*}\scriptscriptstyle{*}}\)}][c]{}0.923 [0.750,1.000]\makebox[\widthof{\(^{\scriptscriptstyle{*}\scriptscriptstyle{*}\scriptscriptstyle{*}}\)}][l]{\(^{\scriptscriptstyle{*}\scriptscriptstyle{*}}\)}& .004 & \makebox[\widthof{\(^{\scriptscriptstyle{*}\scriptscriptstyle{*}\scriptscriptstyle{*}}\)}][c]{}0.889 [0.700,0.968]\makebox[\widthof{\(^{\scriptscriptstyle{*}\scriptscriptstyle{*}\scriptscriptstyle{*}}\)}][l]{\(^{ns}\)}& .284 & \makebox[\widthof{\(^{\scriptscriptstyle{*}\scriptscriptstyle{*}\scriptscriptstyle{*}}\)}][c]{}0.906 [0.789,0.970]\makebox[\widthof{\(^{\scriptscriptstyle{*}\scriptscriptstyle{*}\scriptscriptstyle{*}}\)}][l]{\(^{ns}\)}& .097 \\
 &  &  &  &  & Vicuna & \makebox[\widthof{\(^{\scriptscriptstyle{*}\scriptscriptstyle{*}\scriptscriptstyle{*}}\)}][c]{}0.812 [0.640,0.923]\makebox[\widthof{\(^{\scriptscriptstyle{*}\scriptscriptstyle{*}\scriptscriptstyle{*}}\)}][l]{\(^{\scriptscriptstyle{*}}\)}& .030 & \makebox[\widthof{\(^{\scriptscriptstyle{*}\scriptscriptstyle{*}\scriptscriptstyle{*}}\)}][c]{}0.963 [0.802,1.000]\makebox[\widthof{\(^{\scriptscriptstyle{*}\scriptscriptstyle{*}\scriptscriptstyle{*}}\)}][l]{\(^{ns}\)}& .956 & \makebox[\widthof{\(^{\scriptscriptstyle{*}\scriptscriptstyle{*}\scriptscriptstyle{*}}\)}][c]{}0.881 [0.753,0.946]\makebox[\widthof{\(^{\scriptscriptstyle{*}\scriptscriptstyle{*}\scriptscriptstyle{*}}\)}][l]{\(^{ns}\)}& .134 \\
 &  &  &  &  & VQA & - & - & - & - & - & - \\
\bottomrule
\end{longtable}

% \pagebreak
%tab:locationfull
\begin{longtable}{lcccccrcrcr}
  \caption{Location scores for cases of MIMIC-CXR for which all labelers and ground truth agreed about the presence of an abnormality. Check \cref{tab:labels} for the meanings of symbols and abbreviations.}
  \label{tab:locationfull}
  \endfirsthead
\endhead
\toprule
Abn. & $N$ & $N^+$ & W & Labeler & Precision ($\uparrow$) & \textit{P} & Recall ($\uparrow$) & \textit{P} & F1 ($\uparrow$) & \textit{P} \\
\midrule
Atel. & 570 & 30 & 0.041 & VQA & \makebox[\widthof{\(^{\scriptscriptstyle{*}\scriptscriptstyle{*}\scriptscriptstyle{*}}\)}][c]{}0.727 [0.375,1.000]\(^{\scriptscriptstyle{*}\scriptscriptstyle{*}\scriptscriptstyle{*}}\)& \textless .001 & \makebox[\widthof{\(^{\scriptscriptstyle{*}\scriptscriptstyle{*}\scriptscriptstyle{*}}\)}][c]{}0.267 [0.131,0.458]\(^{\scriptscriptstyle{*}\scriptscriptstyle{*}\scriptscriptstyle{*}}\)& \textless .001 & \makebox[\widthof{\(^{\scriptscriptstyle{*}\scriptscriptstyle{*}\scriptscriptstyle{*}}\)}][c]{}0.390 [0.207,0.584]\(^{\scriptscriptstyle{*}\scriptscriptstyle{*}\scriptscriptstyle{*}}\)& \textless .001 \\
 &  &  &  & MAPLEZ-G & \makebox[\widthof{\(^{\scriptscriptstyle{*}\scriptscriptstyle{*}\scriptscriptstyle{*}}\)}][c]{}\textbf{0.893 [0.727,0.968]}\makebox[\widthof{\(^{\scriptscriptstyle{*}\scriptscriptstyle{*}\scriptscriptstyle{*}}\)}][l]{\(^{ns}\)}& .581 & \makebox[\widthof{\(^{\scriptscriptstyle{*}\scriptscriptstyle{*}\scriptscriptstyle{*}}\)}][c]{}\textbf{0.833 [0.654,0.943]}\makebox[\widthof{\(^{\scriptscriptstyle{*}\scriptscriptstyle{*}\scriptscriptstyle{*}}\)}][l]{\(^{ns}\)}& .630 & \makebox[\widthof{\(^{\scriptscriptstyle{*}\scriptscriptstyle{*}\scriptscriptstyle{*}}\)}][c]{}\textbf{0.862 [0.741,0.939]}\makebox[\widthof{\(^{\scriptscriptstyle{*}\scriptscriptstyle{*}\scriptscriptstyle{*}}\)}][l]{\(^{ns}\)}& .655 \\
 &  &  &  & MAPLEZ & 0.885 [0.693,0.971]&  & 0.767 [0.591,0.896]&  & 0.821 [0.696,0.913]&  \\
 &  &  &  \\
\bottomrule
\end{longtable}

\pagebreak

%tab:locationall
\begin{longtable}{lcccccrcrcr}
  \caption{Location scores for all annotated cases in the MIMIC-CXR dataset. Refer to \cref{tab:labels} for explanations of the symbols and abbreviations}
  \label{tab:locationall}
  \endfirsthead
\endhead
\toprule
Abn. & $N$ & $N^+$ & W & Labeler & Precision ($\uparrow$) & \textit{P} & Recall ($\uparrow$) & \textit{P} & F1 ($\uparrow$) & \textit{P} \\
\midrule
Atel. & 3500 & 255 & 0.196 & VQA & \makebox[\widthof{\(^{\scriptscriptstyle{*}\scriptscriptstyle{*}\scriptscriptstyle{*}}\)}][c]{}\textbf{0.942 [0.892,0.975]}\makebox[\widthof{\(^{\scriptscriptstyle{*}\scriptscriptstyle{*}\scriptscriptstyle{*}}\)}][l]{\(^{ns}\)}& .304 & \makebox[\widthof{\(^{\scriptscriptstyle{*}\scriptscriptstyle{*}\scriptscriptstyle{*}}\)}][c]{}0.443 [0.382,0.504]\(^{\scriptscriptstyle{*}\scriptscriptstyle{*}\scriptscriptstyle{*}}\)& \textless .001 & \makebox[\widthof{\(^{\scriptscriptstyle{*}\scriptscriptstyle{*}\scriptscriptstyle{*}}\)}][c]{}0.603 [0.547,0.662]\(^{\scriptscriptstyle{*}\scriptscriptstyle{*}\scriptscriptstyle{*}}\)& \textless .001 \\
 &  &  &  & MAPLEZ-G & \makebox[\widthof{\(^{\scriptscriptstyle{*}\scriptscriptstyle{*}\scriptscriptstyle{*}}\)}][c]{}0.901 [0.849,0.938]\(^{\scriptscriptstyle{*}\scriptscriptstyle{*}\scriptscriptstyle{*}}\)& \textless .001 & \makebox[\widthof{\(^{\scriptscriptstyle{*}\scriptscriptstyle{*}\scriptscriptstyle{*}}\)}][c]{}0.608 [0.547,0.663]\(^{\scriptscriptstyle{*}\scriptscriptstyle{*}\scriptscriptstyle{*}}\)& \textless .001 & \makebox[\widthof{\(^{\scriptscriptstyle{*}\scriptscriptstyle{*}\scriptscriptstyle{*}}\)}][c]{}0.726 [0.675,0.770]\(^{\scriptscriptstyle{*}\scriptscriptstyle{*}\scriptscriptstyle{*}}\)& \textless .001 \\
 &  &  &  & MAPLEZ & 0.919 [0.876,0.950]&  & \textbf{0.753 [0.698,0.806]}&  & \textbf{0.828 [0.788,0.864]}&  \\
 &  &  &  \\
\bottomrule
\end{longtable}

% \pagebreak

%tab:locationsimple
\begin{longtable}{lcccccrcrcr}
  \caption{Location scores for cases of the MIMIC-CXR dataset for which labelers and ground truth agreed about the presence of an abnormality. These scores were computed with no replacement of keywords allowed and only a  limited set of keywords frequently employed in location annotations by the Medical-Diff-VQA labeler.}
  \label{tab:locationsimple}
  \endfirsthead
\endhead
\toprule
Abn. & $N$ & $N^+$ & W & Labeler & Precision ($\uparrow$) & \textit{P} & Recall ($\uparrow$) & \textit{P} & F1 ($\uparrow$) & \textit{P} \\
\midrule
Atel. & 456 & 14 & 0.039 & VQA & \makebox[\widthof{\(^{\scriptscriptstyle{*}\scriptscriptstyle{*}\scriptscriptstyle{*}}\)}][c]{}0.750 [0.000,1.000]\makebox[\widthof{\(^{\scriptscriptstyle{*}\scriptscriptstyle{*}\scriptscriptstyle{*}}\)}][l]{\(^{\scriptscriptstyle{*}\scriptscriptstyle{*}}\)}& .003 & \makebox[\widthof{\(^{\scriptscriptstyle{*}\scriptscriptstyle{*}\scriptscriptstyle{*}}\)}][c]{}0.214 [0.053,0.500]\makebox[\widthof{\(^{\scriptscriptstyle{*}\scriptscriptstyle{*}\scriptscriptstyle{*}}\)}][l]{\(^{\scriptscriptstyle{*}\scriptscriptstyle{*}}\)}& .004 & \makebox[\widthof{\(^{\scriptscriptstyle{*}\scriptscriptstyle{*}\scriptscriptstyle{*}}\)}][c]{}0.333 [0.105,0.667]\makebox[\widthof{\(^{\scriptscriptstyle{*}\scriptscriptstyle{*}\scriptscriptstyle{*}}\)}][l]{\(^{\scriptscriptstyle{*}\scriptscriptstyle{*}}\)}& .006 \\
 &  &  &  & MAPLEZ-G & \makebox[\widthof{\(^{\scriptscriptstyle{*}\scriptscriptstyle{*}\scriptscriptstyle{*}}\)}][c]{}0.917 [0.539,1.000]\makebox[\widthof{\(^{\scriptscriptstyle{*}\scriptscriptstyle{*}\scriptscriptstyle{*}}\)}][l]{\(^{ns}\)}& 1.000 & \makebox[\widthof{\(^{\scriptscriptstyle{*}\scriptscriptstyle{*}\scriptscriptstyle{*}}\)}][c]{}0.786 [0.455,0.952]\makebox[\widthof{\(^{\scriptscriptstyle{*}\scriptscriptstyle{*}\scriptscriptstyle{*}}\)}][l]{\(^{ns}\)}& 1.000 & \makebox[\widthof{\(^{\scriptscriptstyle{*}\scriptscriptstyle{*}\scriptscriptstyle{*}}\)}][c]{}0.846 [0.615,0.963]\makebox[\widthof{\(^{\scriptscriptstyle{*}\scriptscriptstyle{*}\scriptscriptstyle{*}}\)}][l]{\(^{ns}\)}& 1.000 \\
 &  &  &  & MAPLEZ & \textbf{0.923 [0.600,1.000]}&  & \textbf{0.857 [0.559,1.000]}&  & \textbf{0.889 [0.667,0.980]}&  \\
 &  &  &  \\
\bottomrule
\end{longtable}

%tab:severityfull
\begin{longtable}{lcccccrcrcr}
 \caption{Scores for severity annotations in the MIMIC-CXR dataset for cases where labelers and ground truth agreed the abnormality was present. \Cref{tab:labels} contains explanations of the symbols and abbreviations.}
  \label{tab:severityfull}
  \endfirsthead
\endhead
\toprule
Abn. & $N$ & $N^+$ & W & Labeler & Precision ($\uparrow$) & \textit{P} & Recall ($\uparrow$) & \textit{P} & F1 ($\uparrow$) & \textit{P}\\
\midrule
Card. & 74 & 12 & 0.044 & VQA & \makebox[\widthof{\(^{\scriptscriptstyle{*}\scriptscriptstyle{*}\scriptscriptstyle{*}}\)}][c]{}\textbf{1.000 [1.000,1.000]}\makebox[\widthof{\(^{\scriptscriptstyle{*}\scriptscriptstyle{*}\scriptscriptstyle{*}}\)}][l]{\(^{ns}\)}& 1.000 & \makebox[\widthof{\(^{\scriptscriptstyle{*}\scriptscriptstyle{*}\scriptscriptstyle{*}}\)}][c]{}0.250 [0.059,0.600]\makebox[\widthof{\(^{\scriptscriptstyle{*}\scriptscriptstyle{*}\scriptscriptstyle{*}}\)}][l]{\(^{ns}\)}& .140 & \makebox[\widthof{\(^{\scriptscriptstyle{*}\scriptscriptstyle{*}\scriptscriptstyle{*}}\)}][c]{}0.400 [0.111,0.744]\makebox[\widthof{\(^{\scriptscriptstyle{*}\scriptscriptstyle{*}\scriptscriptstyle{*}}\)}][l]{\(^{ns}\)}& .124 \\
 &  &  &  & MAPLEZ-G & \makebox[\widthof{\(^{\scriptscriptstyle{*}\scriptscriptstyle{*}\scriptscriptstyle{*}}\)}][c]{}0.875 [0.476,1.000]\makebox[\widthof{\(^{\scriptscriptstyle{*}\scriptscriptstyle{*}\scriptscriptstyle{*}}\)}][l]{\(^{ns}\)}& 1.000 & \makebox[\widthof{\(^{\scriptscriptstyle{*}\scriptscriptstyle{*}\scriptscriptstyle{*}}\)}][c]{}\textbf{0.583 [0.286,0.857]}\makebox[\widthof{\(^{\scriptscriptstyle{*}\scriptscriptstyle{*}\scriptscriptstyle{*}}\)}][l]{\(^{ns}\)}& 1.000 & \makebox[\widthof{\(^{\scriptscriptstyle{*}\scriptscriptstyle{*}\scriptscriptstyle{*}}\)}][c]{}0.700 [0.400,0.909]\makebox[\widthof{\(^{\scriptscriptstyle{*}\scriptscriptstyle{*}\scriptscriptstyle{*}}\)}][l]{\(^{ns}\)}& 1.000 \\
 &  &  &  & MAPLEZ & \textbf{1.000 [1.000,1.000]}&  & \textbf{0.583 [0.250,0.833]}&  & \textbf{0.737 [0.400,0.909]}&  \\
 &  &  &  \\
\bottomrule

\end{longtable}

%tab:severityall
\begin{longtable}{@{}lcccccrcrcr@{}}
 \caption{Severity scores on the MIMIC-CXR dataset for all cases in the dataset, including cases for which labelers and ground truth disagreed about the presence of the abnormality. The scores evaluate the presence of a severity annotation without evaluating the severity level itself. \Cref{tab:probs,tab:labels} contain the meaning of symbols and abbreviations.}
  \label{tab:severityall}
  \endfirsthead
\endhead
\toprule
Abn. & $N$ & $N^+$ & W & Labeler & Precision ($\uparrow$) & \textit{P} & Recall ($\uparrow$) & \textit{P} & F1 ($\uparrow$) & \textit{P} \\
\midrule
Atel. & 350 & 24 & 0.065 & VQA & \makebox[\widthof{\(^{\scriptscriptstyle{*}\scriptscriptstyle{*}\scriptscriptstyle{*}}\)}][c]{}\textbf{1.000 [1.000,1.000]}\makebox[\widthof{\(^{\scriptscriptstyle{*}\scriptscriptstyle{*}\scriptscriptstyle{*}}\)}][l]{\(^{\scriptscriptstyle{*}\scriptscriptstyle{*}}\)}& .005 & \makebox[\widthof{\(^{\scriptscriptstyle{*}\scriptscriptstyle{*}\scriptscriptstyle{*}}\)}][c]{}\textbf{0.708 [0.478,0.885]}\makebox[\widthof{\(^{\scriptscriptstyle{*}\scriptscriptstyle{*}\scriptscriptstyle{*}}\)}][l]{\(^{ns}\)}& 1.000 & \makebox[\widthof{\(^{\scriptscriptstyle{*}\scriptscriptstyle{*}\scriptscriptstyle{*}}\)}][c]{}\textbf{0.829 [0.667,0.930]}\makebox[\widthof{\(^{\scriptscriptstyle{*}\scriptscriptstyle{*}\scriptscriptstyle{*}}\)}][l]{\(^{\scriptscriptstyle{*}}\)}& .031 \\
 &  &  &  & MAPLEZ-G & \makebox[\widthof{\(^{\scriptscriptstyle{*}\scriptscriptstyle{*}\scriptscriptstyle{*}}\)}][c]{}0.773 [0.551,0.913]\makebox[\widthof{\(^{\scriptscriptstyle{*}\scriptscriptstyle{*}\scriptscriptstyle{*}}\)}][l]{\(^{\scriptscriptstyle{*}}\)}& .046 & \makebox[\widthof{\(^{\scriptscriptstyle{*}\scriptscriptstyle{*}\scriptscriptstyle{*}}\)}][c]{}\textbf{0.708 [0.474,0.870]}\makebox[\widthof{\(^{\scriptscriptstyle{*}\scriptscriptstyle{*}\scriptscriptstyle{*}}\)}][l]{\(^{ns}\)}& 1.000 & \makebox[\widthof{\(^{\scriptscriptstyle{*}\scriptscriptstyle{*}\scriptscriptstyle{*}}\)}][c]{}0.739 [0.564,0.868]\makebox[\widthof{\(^{\scriptscriptstyle{*}\scriptscriptstyle{*}\scriptscriptstyle{*}}\)}][l]{\(^{ns}\)}& .121 \\
 &  &  &  & MAPLEZ & 0.630 [0.423,0.792]&  & \textbf{0.708 [0.500,0.875]}&  & 0.667 [0.519,0.814]&  \\
 &  &  &  \\
\bottomrule

\end{longtable}

\pagebreak

%tab:humansprob
\begin{longtable}{@{}lccccccrcrcr@{}}
 \caption{Full table for the comparison of the categorical presence labeling performed by the MAPLEZ method and individual radiologists on a majority vote between 3 radiologists from phase 1 and 2 of the \textit{REFLACX} dataset (\textit{RFL-12}). Check \cref{tab:humans,tab:labels} for the meanings of symbols and abbreviations.}
  \label{tab:humansfull}
  \endfirsthead
\endhead
\toprule
Data & Abn. & $N$ & $N^+$ & W & Labeler & Precision ($\uparrow$) & \textit{P} & Recall ($\uparrow$) & \textit{P} & F1 ($\uparrow$) & \textit{P} \\
\midrule
\textit{RFL-12} & Card. & 109 & 30 & 0.10 & Rad. & \makebox[\widthof{\(^{\scriptscriptstyle{*}\scriptscriptstyle{*}\scriptscriptstyle{*}}\)}][c]{}0.451 [0.295,0.618]\makebox[\widthof{\(^{\scriptscriptstyle{*}\scriptscriptstyle{*}\scriptscriptstyle{*}}\)}][l]{\(^{ns}\)}& .138 & \makebox[\widthof{\(^{\scriptscriptstyle{*}\scriptscriptstyle{*}\scriptscriptstyle{*}}\)}][c]{}0.467 [0.311,0.643]\makebox[\widthof{\(^{\scriptscriptstyle{*}\scriptscriptstyle{*}\scriptscriptstyle{*}}\)}][l]{\(^{ns}\)}& .406 & \makebox[\widthof{\(^{\scriptscriptstyle{*}\scriptscriptstyle{*}\scriptscriptstyle{*}}\)}][c]{}0.459 [0.319,0.609]\makebox[\widthof{\(^{\scriptscriptstyle{*}\scriptscriptstyle{*}\scriptscriptstyle{*}}\)}][l]{\(^{ns}\)}& .208 \\
 &  &  &  &  & MAPLEZ & \textbf{0.600 [0.407,0.771]}&  & \textbf{0.600 [0.418,0.773]}&  & \textbf{0.600 [0.433,0.733]}&  \\
 &  &  &  &  \\
\textit{RFL-12} & Cons. & 109 & 33 & 0.09 & Rad. & \makebox[\widthof{\(^{\scriptscriptstyle{*}\scriptscriptstyle{*}\scriptscriptstyle{*}}\)}][c]{}0.431 [0.295,0.576]\makebox[\widthof{\(^{\scriptscriptstyle{*}\scriptscriptstyle{*}\scriptscriptstyle{*}}\)}][l]{\(^{ns}\)}& .104 & \makebox[\widthof{\(^{\scriptscriptstyle{*}\scriptscriptstyle{*}\scriptscriptstyle{*}}\)}][c]{}\textbf{0.470 [0.345,0.594]}\makebox[\widthof{\(^{\scriptscriptstyle{*}\scriptscriptstyle{*}\scriptscriptstyle{*}}\)}][l]{\(^{ns}\)}& .826 & \makebox[\widthof{\(^{\scriptscriptstyle{*}\scriptscriptstyle{*}\scriptscriptstyle{*}}\)}][c]{}0.448 [0.332,0.564]\makebox[\widthof{\(^{\scriptscriptstyle{*}\scriptscriptstyle{*}\scriptscriptstyle{*}}\)}][l]{\(^{ns}\)}& .630 \\
 &  &  &  &  & MAPLEZ & \textbf{0.583 [0.364,0.773]}&  & 0.424 [0.258,0.600]&  & \textbf{0.491 [0.318,0.653]}&  \\
 &  &  &  &  \\
\bottomrule

\end{longtable}

%tab:othermodalitiesfull
\begin{longtable}{@{}llccrccc@{}}
 \caption{Categorical presence scores for the annotations from the MAPLEZ labeler for three medical imaging modalities for which the labeler did not go through a validation process. Hypermetab. thorax= Hypermetabolic abnormality in the thorax; Hypermetab. abdomen= Hypermetabolic abnormality in the abdomen; Hypermetab. pelvis= Hypermetabolic abnormality in the pelvis.}
  \label{tab:othermodalitiesfull}
  \endfirsthead
\endhead
\toprule
Data & Abn. & $N$ & $N^+$ & W & Precision ($\uparrow$) & Recall ($\uparrow$) & F1 ($\uparrow$)\\
\midrule
CT & Lung lesion & 40 & 23 & 0.264 & 0.846 [0.661,0.959] & 0.957 [0.772,1.000] & 0.898 [0.776,0.964] \\
CT & Liver lesion & 40 & 20 & 0.128 & 0.941 [0.713,1.000] & 0.800 [0.582,0.947] & 0.865 [0.702,0.955] \\
MRI & Liver lesion & 40 & 23 & 0.194 & 0.909 [0.698,1.000] & 0.870 [0.667,0.962] & 0.889 [0.750,0.962] \\
MRI & Kidney lesion & 40 & 22 & 0.174 & 0.947 [0.707,1.000] & 0.818 [0.609,0.950] & 0.878 [0.733,0.957] \\
PET & Hypermetab. thorax & 39 & 25 & 0.141 & 0.840 [0.643,0.957] & 0.840 [0.652,0.958] & 0.840 [0.681,0.923] \\
PET & Hypermetab. abdomen & 39 & 16 & 0.086 & 0.824 [0.578,0.950] & 0.875 [0.611,1.000] & 0.848 [0.641,0.957] \\
PET & Hypermetab. pelvis & 39 & 13 & 0.014 & 0.583 [0.286,0.857] & 0.538 [0.228,0.800] & 0.560 [0.300,0.778] \\
\midrule
- & Liver lesion & - & - & - & 0.922 [0.785,0.979] & 0.842 [0.703,0.934] & 0.879 [0.790,0.943] \\
CT & - & - & - & - & 0.877 [0.752,0.954] & 0.905 [0.779,0.966] & 0.887 [0.797,0.945] \\
MRI & - & - & - & - & 0.927 [0.802,0.978] & 0.845 [0.714,0.929] & 0.884 [0.789,0.941] \\
PET & - & - & - & - & 0.819 [0.688,0.911] & 0.835 [0.701,0.915] & 0.827 [0.729,0.894] \\
All-W & - & - & - & - & 0.882 [0.820,0.930] & 0.866 [0.801,0.915] & 0.871 [0.826,0.914] \\
All-M & - & - & - & - & 0.842 [0.772,0.895] & 0.814 [0.744,0.872] & 0.825 [0.772,0.876] \\

 \bottomrule

\end{longtable}

\end{landscape}

\begin{table}
 \caption{Scores of the location annotations of the MAPLEZ method after quick adaptation to other medical modalities.}
  \label{tab:othermodalitieslocationfull}
  \centering
\begin{tabular}{@{}llccrccc@{}}
\toprule
Data & Abn. & $N$ & $N^+$ & W & Precision ($\uparrow$) & Recall ($\uparrow$) & F1 ($\uparrow$) \\
\midrule
CT & Lung lesion & 200 & 61 & 0.347 & 0.821 [0.702,0.909] & 0.754 [0.629,0.845] & 0.786 [0.690,0.859] \\
CT & Liver lesion & 80 & 11 & 0.052 & 0.889 [0.400,1.000] & 0.727 [0.364,0.938] & 0.800 [0.526,0.947] \\
MRI & Liver lesion & 80 & 15 & 0.098 & 1.000 [1.000,1.000] & 0.733 [0.462,0.917] & 0.846 [0.621,0.959] \\
MRI & Kidney lesion & 80 & 34 & 0.177 & 0.958 [0.789,1.000] & 0.676 [0.500,0.812] & 0.793 [0.653,0.886] \\
CT & Pleural effusion & 80 & 12 & 0.327 & 1.000 [1.000,1.000] & 0.917 [0.604,1.000] & 0.957 [0.690,1.000] \\
\midrule
- & Liver lesion & - & - & - & 0.961 [0.809,1.000] & 0.731 [0.522,0.891] & 0.830 [0.678,0.927] \\
CT & - & - & - & - & 0.907 [0.849,0.949] & 0.825 [0.678,0.890] & 0.864 [0.783,0.910] \\
MRI & - & - & - & - & 0.973 [0.867,1.000] & 0.697 [0.541,0.817] & 0.812 [0.703,0.894] \\
All-W & - & - & - & - & 0.925 [0.880,0.959] & 0.790 [0.683,0.850] & 0.850 [0.791,0.892] \\
All-M & - & - & - & - & 0.934 [0.846,0.968] & 0.762 [0.660,0.840] & 0.836 [0.764,0.889] \\

 \bottomrule
 \end{tabular}

\end{table}

%tab:probsfull
\begin{table*}
\caption{Full table for comparing annotated probabilities of three labelers against the probability annotations provided by radiologists in phase 3 of the \textit{REFLACX} dataset. The scores represent the MAE ($\uparrow$) between probabilities. The meanings of symbols and abbreviations are presented in \cref{tab:probs,tab:labels}.}
  \label{tab:probsfull}
  \centering
\begin{tabular}{@{}lccccrcrc@{}}
\toprule
Abn. & $N$ & $N^+$ & W & VQA & \textit{P} & MAPLEZ-G & \textit{P} & MAPLEZ \\
\midrule
Card. & 506 & 171 & 0.205 & \makebox[\widthof{\(^{\scriptscriptstyle{*}\scriptscriptstyle{*}\scriptscriptstyle{*}}\)}][c]{}23.3 [20.0,26.6]\makebox[\widthof{\(^{\scriptscriptstyle{*}\scriptscriptstyle{*}\scriptscriptstyle{*}}\)}][l]{\(^{ns}\)}& .458 & \makebox[\widthof{\(^{\scriptscriptstyle{*}\scriptscriptstyle{*}\scriptscriptstyle{*}}\)}][c]{}22.6 [19.3,25.6]\makebox[\widthof{\(^{\scriptscriptstyle{*}\scriptscriptstyle{*}\scriptscriptstyle{*}}\)}][l]{\(^{ns}\)}& .467 & \textbf{22.3 [19.4,25.4]} \\
Cons. & 506 & 154 & 0.279 & \makebox[\widthof{\(^{\scriptscriptstyle{*}\scriptscriptstyle{*}\scriptscriptstyle{*}}\)}][c]{}25.4 [22.4,28.7]\makebox[\widthof{\(^{\scriptscriptstyle{*}\scriptscriptstyle{*}\scriptscriptstyle{*}}\)}][l]{\(^{ns}\)}& .151 & \makebox[\widthof{\(^{\scriptscriptstyle{*}\scriptscriptstyle{*}\scriptscriptstyle{*}}\)}][c]{}24.0 [21.3,26.8]\makebox[\widthof{\(^{\scriptscriptstyle{*}\scriptscriptstyle{*}\scriptscriptstyle{*}}\)}][l]{\(^{ns}\)}& .322 & \textbf{23.6 [21.1,26.6]} \\
Edema & 506 & 115 & 0.215 & \makebox[\widthof{\(^{\scriptscriptstyle{*}\scriptscriptstyle{*}\scriptscriptstyle{*}}\)}][c]{}23.8 [20.6,26.9]\(^{\scriptscriptstyle{*}\scriptscriptstyle{*}\scriptscriptstyle{*}}\)& \textless .001 & \makebox[\widthof{\(^{\scriptscriptstyle{*}\scriptscriptstyle{*}\scriptscriptstyle{*}}\)}][c]{}21.6 [18.8,24.5]\(^{\scriptscriptstyle{*}\scriptscriptstyle{*}\scriptscriptstyle{*}}\)& \textless .001 & \textbf{18.0 [15.6,20.5]} \\
\bottomrule
   \end{tabular}
\end{table*}

%tab:humansprobfull
\begin{table*}
 \caption{Table comparing the MAE scores ($\downarrow$) of the MAPLEZ annotations and the individual radiologists assigning probabilities of presence for several types of abnormalities. They are compared with the probability that is the average of the probability assigned by three other radiologists from phases 1 and 2 of the \textit{REFLACX} dataset (\textit{RFL-12}). We applied linear interpolation to the interval boundaries presented in \cref{sec:labelereval} to calculate the necessary accepted probability interval for the MAE scores. Refer to \cref{tab:humans,tab:classifier} for abbreviations and symbols.}
  \label{tab:humansprobfull}
  \centering
\begin{tabular}{@{}llccccrc@{}}
\toprule
    Data & Abn. & $N$ & $N^+$ & W & Rad. & \textit{P} & MAPLEZ \\
\midrule
\textit{RFL-12} & Card. & 109 & 46 & 0.19 & \makebox[\widthof{\(^{\scriptscriptstyle{*}\scriptscriptstyle{*}\scriptscriptstyle{*}}\)}][c]{}21.3 [16.4,26.9]\makebox[\widthof{\(^{\scriptscriptstyle{*}\scriptscriptstyle{*}\scriptscriptstyle{*}}\)}][l]{\(^{\scriptscriptstyle{*}}\)}& .045 & \textbf{14.3 [10.4,19.6]} \\
\textit{RFL-12} & Cons. & 109 & 54 & 0.29 & \makebox[\widthof{\(^{\scriptscriptstyle{*}\scriptscriptstyle{*}\scriptscriptstyle{*}}\)}][c]{}22.4 [18.7,26.9]\makebox[\widthof{\(^{\scriptscriptstyle{*}\scriptscriptstyle{*}\scriptscriptstyle{*}}\)}][l]{\(^{\scriptscriptstyle{*}\scriptscriptstyle{*}}\)}& .005 & \textbf{14.1 [10.9,18.2]} \\
\textit{RFL-12} & Edema & 109 & 37 & 0.19 & \makebox[\widthof{\(^{\scriptscriptstyle{*}\scriptscriptstyle{*}\scriptscriptstyle{*}}\)}][c]{}13.4 [10.5,17.5]\makebox[\widthof{\(^{\scriptscriptstyle{*}\scriptscriptstyle{*}\scriptscriptstyle{*}}\)}][l]{\(^{ns}\)}& .095 & \textbf{9.8 [7.0,13.2]} \\
\textit{RFL-12} & Opac. & 109 & 82 & 0.33 & \makebox[\widthof{\(^{\scriptscriptstyle{*}\scriptscriptstyle{*}\scriptscriptstyle{*}}\)}][c]{}21.6 [17.9,25.5]\makebox[\widthof{\(^{\scriptscriptstyle{*}\scriptscriptstyle{*}\scriptscriptstyle{*}}\)}][l]{\(^{ns}\)}& .105 & \textbf{16.8 [12.9,21.1]} \\
\midrule
All-W & - & - & - & - & \makebox[\widthof{\(^{\scriptscriptstyle{*}\scriptscriptstyle{*}\scriptscriptstyle{*}}\)}][c]{}20.2 [18.5,22.0]\(^{\scriptscriptstyle{*}\scriptscriptstyle{*}\scriptscriptstyle{*}}\)& \textless .001 & \textbf{14.2 [12.2,16.4]} \\
All-M & - & - & - & - & \makebox[\widthof{\(^{\scriptscriptstyle{*}\scriptscriptstyle{*}\scriptscriptstyle{*}}\)}][c]{}19.7 [17.9,21.4]\(^{\scriptscriptstyle{*}\scriptscriptstyle{*}\scriptscriptstyle{*}}\)& \textless .001 & \textbf{13.8 [11.9,15.8]} \\

 \bottomrule
   \end{tabular}
\end{table*}

\clearpage
%tab:classifierfull
  \centering
\begin{longtable}{@{}llccrlcr@{}}
 \caption{Full table with the AUC scores ($\uparrow$) of classifiers trained with annotations from competing labelers and the ablation study performed to evaluate the impact of the proposed ways of applying the annotations extracted by the MAPLEZ labeler. In addition to the score of the classifiers for a specific abnormality of a specific dataset, we also aggregate the scores by abnormality, by dataset, and for all rows using a weighted average (``All-W") and a simple average over abnromalities (``All-M"). Refer to \cref{tab:labels,tab:classifier} for abbreviations and symbols.}
  \label{tab:classifierfull}
  \endfirsthead
\endhead
\toprule
Data & Abn. & $N$ & $N^+$ & W & Labeler & AUC & \textit{P} \\
\midrule
\textit{RFL-3} & Atel. & 506 & 156 & 0.006 & CheXpert & \makebox[\widthof{\(^{\scriptscriptstyle{*}\scriptscriptstyle{*}\scriptscriptstyle{*}}\)}][c]{}0.794 [0.756,0.831]\makebox[\widthof{\(^{\scriptscriptstyle{*}\scriptscriptstyle{*}\scriptscriptstyle{*}}\)}][l]{\(^{ns}\)}& .325 \\
 &  &  &  &  & VQA & \makebox[\widthof{\(^{\scriptscriptstyle{*}\scriptscriptstyle{*}\scriptscriptstyle{*}}\)}][c]{}0.789 [0.748,0.825]\makebox[\widthof{\(^{\scriptscriptstyle{*}\scriptscriptstyle{*}\scriptscriptstyle{*}}\)}][l]{\(^{ns}\)}& .173 \\
 &  &  &  &  & CheXbert & \makebox[\widthof{\(^{\scriptscriptstyle{*}\scriptscriptstyle{*}\scriptscriptstyle{*}}\)}][c]{}0.784 [0.745,0.823]\makebox[\widthof{\(^{\scriptscriptstyle{*}\scriptscriptstyle{*}\scriptscriptstyle{*}}\)}][l]{\(^{\scriptscriptstyle{*}}\)}& .035 \\
 &  &  &  &  & LLM & \textbf{0.807 [0.766,0.842]}&  \\
 &  &  &  &  & $\lambda_{loc}=0$ & \makebox[\widthof{\(^{\scriptscriptstyle{*}\scriptscriptstyle{*}\scriptscriptstyle{*}}\)}][c]{}0.798 [0.757,0.834]\makebox[\widthof{\(^{\scriptscriptstyle{*}\scriptscriptstyle{*}\scriptscriptstyle{*}}\)}][l]{\(^{ns}\)}& .183 \\
 &  &  &  &  & Cat. Labels & \makebox[\widthof{\(^{\scriptscriptstyle{*}\scriptscriptstyle{*}\scriptscriptstyle{*}}\)}][c]{}0.799 [0.760,0.832]\makebox[\widthof{\(^{\scriptscriptstyle{*}\scriptscriptstyle{*}\scriptscriptstyle{*}}\)}][l]{\(^{ns}\)}& .295 \\
 &  &  &  &  & Use ``Stable" & \makebox[\widthof{\(^{\scriptscriptstyle{*}\scriptscriptstyle{*}\scriptscriptstyle{*}}\)}][c]{}0.800 [0.760,0.833]\makebox[\widthof{\(^{\scriptscriptstyle{*}\scriptscriptstyle{*}\scriptscriptstyle{*}}\)}][l]{\(^{ns}\)}& .222 \\
\bottomrule
\end{longtable}

\twocolumn

\end{document}